\documentclass[aps,prb,twocolumn,showpacs,amsmath,amssymb,superscriptaddress,floatfix]{revtex4-2}

\usepackage[english]{babel}
\usepackage{blindtext}
\usepackage{latexsym}
\usepackage{amssymb}
\usepackage{physics}
\usepackage{amsmath}
\usepackage{bm}
\usepackage{amsfonts}
\usepackage{relsize}
\usepackage{xcolor}
\usepackage{verbatim}
\usepackage{bbold}
\usepackage{slashed}
\usepackage{appendix}
\usepackage{graphicx}
\usepackage{color}
\usepackage[normalem]{ulem}
\usepackage[colorlinks = true,
            linkcolor = blue,
            urlcolor  = blue,
            citecolor = blue,
            anchorcolor = blue]{hyperref}
\usepackage{graphicx} 
\usepackage{dcolumn} 
\newcommand{\approptoinn}[2]{\mathrel{\vcenter{
  \offinterlineskip\halign{\hfil$##$\cr
    #1\propto\cr\noalign{\kern2pt}#1\sim\cr\noalign{\kern-2pt}}}}}

\newcommand{\up}{\uparrow}
\newcommand{\down}{\downarrow}

\definecolor{ao(english)}{rgb}{0.0, 0.5, 0.0}
\definecolor{amaranth}{rgb}{0.9, 0.17, 0.31}
\definecolor{green(html/cssgreen)}{rgb}{0.0, 0.5, 0.0}

\newcommand\greensout{\bgroup\markoverwith{\textcolor{green(html/cssgreen)}{\rule[0.5ex]{2pt}{1.0pt}}}\ULon}

\begin{document}

\title{Classification of Lifshitz invariant in multiband superconductors: an application to Leggett modes in the linear response regime in Kagome lattice models}

\author{Raigo Nagashima}
\affiliation{Department of Physics, The University of Tokyo, Hongo, Tokyo, 113-8656, Japan}
\affiliation{Max Planck Institute for Solid State Research, 70569 Stuttgart, Germany}

\author{Sida Tian}
\affiliation{Max Planck Institute for Solid State Research, 70569 Stuttgart, Germany}
\author{Rafael Haenel}
\affiliation{Max Planck Institute for Solid State Research, 70569 Stuttgart, Germany}
\affiliation{Department of Physics and Astronomy, Quantum Matter Institute, University of British Columbia, Vancouver, British Columbia V6T 1Z4, Canada}
\author{Naoto Tsuji}
\affiliation{Department of Physics, The University of Tokyo, Hongo, Tokyo, 113-8656, Japan}
\affiliation{RIKEN Center for Emergent Matter Science (CEMS), Wako 351-0198, Japan}
\author{Dirk Manske}
\affiliation{Max Planck Institute for Solid State Research, 70569 Stuttgart, Germany}

\date{\today}

\begin{abstract}
Multiband superconductors are sources of rich physics arising from multiple order parameters, which show unique collective dynamics including Leggett mode as relative phase oscillations.
Previously, it has been pointed out that the Leggett mode can be optically excited in the linear response regime, as demonstrated in a one-dimensional model for multiband superconductors [T. Kamatani, et al., Phys. Rev. B \textbf{105}, 094520 (2022)].
Here we identify the linear coupling term in the Ginzburg-Landau free energy to be the so-called Lifshitz invariant, which takes the form of $\boldsymbol{d}\cdot\qty(\Psi^{*}_{i}\nabla\Psi_{j} - \Psi^{*}_{j}\nabla\Psi_{i})$, where $\boldsymbol{d}$ is a constant vector and $\Psi_{i}$ and $\Psi_{j}$ $(i\neq j)$ represent superconducting order parameters.
We classify all pairs of irreducible representations of order parameters in the crystallographic point groups that allow for the existence of the Lifshitz invariant.
We emphasize that the Lifshitz invariant can appear even in systems with inversion symmetry.
The results are applied to a model of $s$-wave superconductors on a Kagome lattice with various bond orders, for which in some cases we confirm that the Leggett mode appears as a resonance peak in a linear optical conductivity spectrum based on microscopic calculations.
We discuss a possible experimental observation of the Leggett mode by a linear optical response in multiband superconductors.

\end{abstract}

\maketitle

\section{\label{sec:level1}Introduction}
\begin{figure}[h]
\centering
\includegraphics[width=86mm]{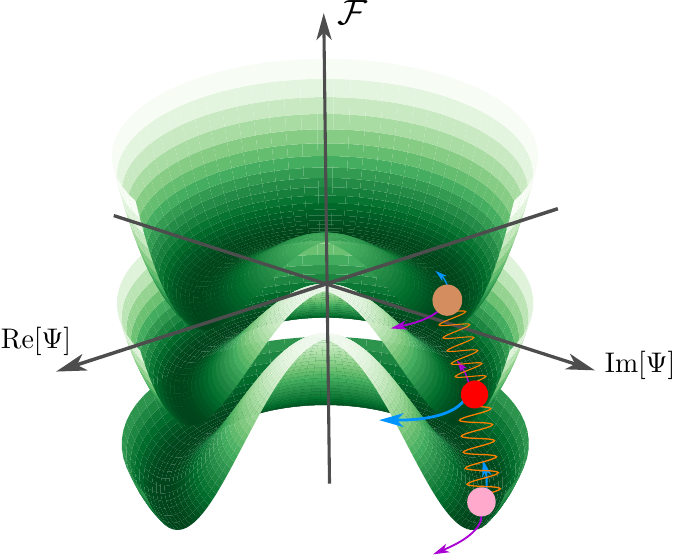}
\caption{A schematic picture of the free energy of a superconductor with three order parameters.
In multiband superconductors, the fluctuations of the relative phase between two order parameters (the Leggett mode) generally constitute a massive mode. 
A three-band superconductor has two such phase modes, with characteristic frequencies $\omega_{_{\mathrm{L}1}}$ and $\omega_{_{\mathrm{L}2}}$. 
In contrast, the overall massless phase mode (Nambu-Goldstone mode) is lifted to the plasma frequency due to the Anderson-Higgs mechanism.
For simplicity only one of the Leggett modes is shown and the other Leggett and Higgs modes are not depicted.
The colored arrows show the oscillation of the Leggett mode (blue and purple arrows indicate oscillating directions at different oscillation phases).}
\label{fig:FreeEnergy}
\end{figure}
Superconductivity embraces rich order parameter dynamics, as shown by the macroscopic Ginzburg-Landau (GL) and microscopic BCS theories.
They tell us that there are typically two types of collective modes in single-band superconductors: one of them is the Higgs (amplitude) mode~\cite{Intro:1_Anderson,Intro:2_Schmid,Intro:3_Littlewood_Varma,Intro:4_Littlewood_Varma,Intro:5_Pekker_Varma,Intro:6_Shimano_Tsuji,Intro:6_2_Tsuji-Ency} and the other is the Nambu-Goldstone (phase) mode~\cite{Intro_NG1,Intro_NG2}, the latter of which is lifted up to the plasma frequency by the Anderson-Higgs mechanism~\cite{Intro:Anderson-Higgs1,Intro:Anderson-Higgs2}.
What remains at low energy is the Higgs mode, which constitutes a massive excitation on top of the continuum of quasiparticle excitations.

Since Higgs mode does not linearly couple to electromagnetic fields in ordinary situations, previous studies have focused on the investigation of nonlinear optical responses of superconductors ~\cite{Intro:HiggsMode1,Intro:HiggsMode2,Intro:HiggsMode3,Intro:HiggsMode4,Intro:HiggsMode5,Intro:HiggsMode6,Intro:HiggsMode7,Intro:HiggsMode8,Intro:HiggsMode9,Intro:HiggsMode10}.
In conventional superconductors, the energy scale of the Higgs mode is usually a few meV, which is in the frequency range of THz lasers. That is why one had to wait for the arrival of high-intensity THz lasers and techniques like THz pump-THz probe experiments and third harmonic generations~\cite{Intro:Ex_PumpProbe_HiggsMode_Matsunaga2013,Intro:Ex_THG_HiggsMode1,Intro:Ex_PumpProbeHiggsMode_Matsunaga2017,Intro:Ex_PumpProbe_HiggsMode2,Intro:Ex_THG_HiggsMode2}.
Another experiment that has observed the Higgs mode is the Raman spectroscopy in $2H$-$\mathrm{NbSe}_{2}$, where the charge density wave (CDW) phase coexists with the superconducting phase. 
In this particular situation the Higgs mode becomes Raman active, and Raman experiments have observed the signal prior to the development of THz lasers~\cite{Intro:Ex_Raman1,Intro:Ex_Raman2} (see also ~\cite{Intro:Ex_Raman3,Intro:Ex_Raman4,Intro:Ex_Raman5,Intro:RamanRelated} for recent research).

The physics of superconductors with multiple order parameters is even richer than the single-band case. 
Many superconductors of interest, such as iron-based superconductors~\cite{Intro:Fe_SC_review}, $\mathrm{MgB}_{2}$~\cite{Intro:MgB2_SC_review}, niobium-based superconductors~\cite{Intro:Niobium_SC_review}, and Kagome superconductors~\cite{Kagome_s_wave,Kagome_StarOfDavid,Intro:Kagome_SC_review, Kagome_Possible_CDW,Kagome_Possible_CDW_3Q,Kagome_Anisotropic_s_wave,Kagome_Preserved_TRS,Kagome_NoTRSB}, are all multiband superconductors, and it is natural to consider those cases. 
Two-band superconductors, for instance, have four real collective modes. 
Two of them are amplitude modes, and the others are phase modes, one of which is just an overall phase and is absorbed into electromagnetic fields due to the Anderson-Higgs mechanism.
The remaining phase mode corresponds to fluctuations of the phase difference between the two order parameters, which is called the Leggett mode~\cite{Intro:Leggett_original}.
So far, there is an example of the observation of the collective phase fluctuations by Raman spectroscopy~\cite{Intro:MgB2RamanLeggett}.
Other examples of collective phase fluctuation have been studied in the nonlinear response regime~\cite{Intro:Higgs_Leggett1,Intro:Higgs_Leggett2,Intro:Higgs_Leggett3,Intro:Leggett_Nonlinear1,Intro:Leggett_Massless,Intro:Leggett_Nonlinear2,Intro:Leggett_Nonlinear3,Intro:Leggett_Nonlinear4,Intro:Leggett_Nonlinear5,Intro:Leggett_Nonlinear6}.
We also note that the phase solitons~\cite{PhaseSoliton_Tanaka,PhaseSoliton_Yerin} arise from the multiband nature of superconductors with a nontrivial geometry like a ring.

Recently, it has been shown that a term containing only a single spatial derivative, responsible for the linear order Leggett-light coupling, could in principle appear in the GL free energy, and its existence was demonstrated in a one-dimensional two-band superconducting model~\cite{Leggett_Kamatani}.
In general, however, it is not clear under what condition the Leggett mode would appear in the linear response regime, or which crystal symmetry could host the linear Leggett mode response. 
Particularly it is not known whether the Leggett mode can appear in the linear response in dimensions higher than one.

In the present work, we point out that the single-derivative term corresponds to the so-called Lifshitz invariant $\bm{d}\cdot\qty(\Psi^{*}_{i}\nabla\Psi_{j}-\Psi^{*}_{j}\nabla\Psi_{i})$ $(i\neq j)$, which is invariant under symmetry operations of the system. 
The possibility of such an antisymmetric term appearing in the GL free energy has been studied by Lifshitz ~\cite{LandauLifshitz} in the context of the stability of second-order phase transitions.
Dzyaloshinskii has also discussed the term considering the helicoidal structure in antiferromagnets~\cite{LI_SymmetryDiscusstion}.
The Lifshitz invariant is linear with respect to the spatial gradient of the order parameter, so it has been supposed to appear in inversion symmetry broken systems.
The Lifshitz invariant is known to emerge, e.g., in noncentrosymmetric superconductors~\cite{II:LI_NCSC_1st,II:LI_NCSC_2nd,II:LI_NCSC_3rd}, parity- and time-reversal symmetry broken superconductors~\cite{II:LI_Yanase_1st,II:LI_Yanase_2nd}, commensurate-incommensurate transitions~\cite{II:LI_com_incom_1st,II:LI_com_incom_2nd}, liquid crystals~\cite{II:LI_LiquidCrystal}, and as the Dzyaloshinskii-Moriya interaction term in magnets~\cite{II:LI_DM_1st,II:LI_DM_2nd}. 
This linear gradient term modifies the free energy and causes various different states such as nonuniform superconducting states in noncentrosymmetric superconductors,  magnetic skyrmion~\cite{MagneticSkyrmion}, and instability around the phase transition point.
Group theoretical classification of this term has been given in the incommensurate phase transition~\cite{II:LI_com_incom_1st} and solids~\cite{LI_Classification_solids}, but not in multiband superconductors.
We here classify all combinations of irreducible representations of order parameters in crystallographic point groups that permit the presence of the Lifshitz invariant.

We also found that the inversion symmetry is not crucial, and, the system could have the Lifshitz invariant even in the presence of the inversion symmetry.
The condition for the Lifshitz invariant to appear in multiband superconductors with sublattice degrees of freedom is determined by the \textit{induced representation} of the crystallographic point group, which is induced by the trivial representation of the site-symmetry group for each Wyckoff position of lattice sites.
As a result, a wide range of multiband superconducting systems are shown to be able to have the Lifshitz invariant. 

As an application, we study a family of two-dimensional microscopic models on the Kagome lattice.
Kagome materials are intensively studied experimentally and theoretically in the context of spin liquids and antiferromagnetism~\cite{Intro:Kagome_SpinLiquid1,Intro:Kagome_SpinLiquid2,Intro:Kagome_SpinLiquid3,Intro:Kagome_antiferro,Intro:Kagome_Breathing}, inherent flat bands~\cite{Intro:Kagome_FlatBand1,Intro:Kagome_FlatBand2,Intro:Kagome_FlatBand3,Intro:Kagome_FlatBand4}, nontrivial topology~\cite{Intro:Kagome_Topology1,Intro:Kagome_Topology2,Intro:Kagome_Topology3,Intro:Kagome_Topology4,Intro:Kagome_Topology5}, as well as  superconductivity~\cite{Kagome_s_wave,Kagome_StarOfDavid,Intro:Kagome_SC_review,Kagome_Possible_CDW,Kagome_Possible_CDW_3Q,Kagome_Anisotropic_s_wave,Kagome_Preserved_TRS,Kagome_NoTRSB}. 
We give insight into the collective mode spectrum and optical response in the case of multi-order parameter superconductivity.

This paper is organized as follows. 
In Sec.~\ref{sec:level2}, we review the GL free energy framework of a two-band superconductor.
While we focus on the two-band case, the argument can be straightforwardly extended to cases of a larger number of bands.
Within the GL approach, we perform a group theoretical analysis of the Lifshitz invariant in Sec.~\ref{sec:level3}. 
We apply the group theoretical techniques to some models in Sec.~\ref{sec:level4} to see that the inversion symmetry is not crucial to discuss the Lifshitz invariant in multiband superconductors. 
In Sec.~\ref{sec:level5} we study a family of Kagome models and explicitly compute signatures of collective phase modes in the linear optical conductivity using an imaginary time path integral approach.
The paper is summarized in Sec.~\ref{sec:level6}.
We set $\hbar=1$ throughout the paper. 

\section{\label{sec:level2}Ginzburg-Landau Free energy}
This section reviews a phenomenological theory of multiband superconductors within the GL free energy framework~\cite{Leggett_Kamatani}.
Fig.~\ref{fig:FreeEnergy} depicts the schematic picture of the free energy of a three-band-superconductor, which we will microscopically consider later in Sec.~\ref{sec:level5}.
To illustrate the physics of the Lifshitz invariant, however, it is sufficient to study the case of two order parameters. 
The argument can be generalized to arbitrary $n$-band superconductors in a straightforward manner.
The single-band case is described in detail, e.g., in the review~\cite{Intro:6_Shimano_Tsuji}.

\subsection{\label{sec:level2-1}Two-band superconductor}
We shall consider the GL free energy density $\mathcal{F}$ for a two-band superconductor given below~\cite{Leggett_Kamatani},
\begin{align}
\mathcal{F} &= \sum_{i=1,2}\qty[ a_{i}\left| \Psi_{i}\right|^{2} + \frac{b_{i}}{2}\left| \Psi_{i} \right|^{4} + \frac{1}{2m^{*}_{i}}\left| \bm{D}\Psi_{i}\right|^{2} ] \notag \\
&\quad + \qty[ \epsilon\Psi^{*}_{1}\Psi_{2} + \eta\qty( \bm{D}^{*}\Psi^{*}_{1})\cdot\qty(\bm{D}\Psi_{2}) + \mathrm{c.c.} ] \notag \\
&\quad + \qty[ \bm{d}\cdot \qty( \Psi^{*}_{1}\bm{D}\Psi_{2} + \Psi^{*}_{2}\bm{D}^{*}\Psi_{1}) 
+ \mathrm{c.c.} ],
\label{eqn:fe0}
\end{align}
where $a_{i}=a_{0, i}\qty(T-T_{\mathrm{c}})$, $a_{0, i}$ and $b_{i}$ are
positive constants, $T$ is the system's temperature, $T_{\mathrm{c}}$ is the transition temperature, $m^{*}_{i}$ is the effective electron mass, and $\bm{D} = -\mathrm{i}\nabla - e^{*}\bm{A}$ is the covariant derivative with an electric charge of a Cooper pair $e^{*}=2e$ and the electromagnetic vector potential $\bm{A}$.
See Appendix.~\ref{Apdx.A} for a microscopic derivation.
The first line in Eq.~\eqref{eqn:fe0} describes the free energy density of two independent single-band superconductors with complex order parameters $\Psi_i$, each describing a Mexican hat potential below the critical temperature $T_{\mathrm{c}}$.

The remaining terms represent couplings between the two order parameters, with coefficients $\epsilon$, $\eta$ and a constant vector $\bm{d}$. 
Here, the term proportional to $\epsilon$ corresponds to the Josephson (proximity) coupling.
The term proportional to $\eta$ is interpreted as a drag effect~\cite{II:FreeEnergy,DragEffect_Doh,DragEffect_Yerin}. 

Of particular relevance in the context of the present study is the term proportional to $\bm{d}$. 
It is responsible for inducing the Leggett mode in the linear response regime as we will see below \cite{Leggett_Kamatani}. 
We will discuss this in Sec.~\ref{sec:level2-2} that the vector $\bm{d}$ may be interpreted as an ``internal field" that induces a flow of the phase.
We note that there is a slight difference in the $\bm{d}$ term between Eq.~\eqref{eqn:fe0} and that of \cite{Leggett_Kamatani}.
However, the difference is only the total derivative, and hence is not physically relevant. 

We now expand the free energy around the mean-field ground state $\Psi_{1,0}$ and $\Psi_{2,0}$,
\begin{align}
\Psi_{1}\qty(\bm{r},t) &= \qty( \Psi_{1,0} + H_{1}\qty(\bm{r}, t)
)e^{\mathrm{i}\theta_{1}\qty(\bm{r},t)} \, , \nonumber\\
\Psi_{2}\qty(\bm{r},t) &= \qty( \Psi_{2,0} + H_{2}\qty(\bm{r}, t) )e^{\mathrm{i}\theta_{2}\qty(\bm{r},t)} \, , 
\end{align}
where $H_{i}$ and $\theta_{i}$ describe amplitude and phase fluctuations, respectively. 
The overall phase $\theta_1 + \theta_2$ can be removed by gauge transformation due to the
Anderson-Higgs mechanism. The only relevant phase degree of freedom will be
$\theta_1 - \theta_2$. In general, one has $n-1$ phase degrees of freedom, where
$n$ is the number of order parameters. In the expansions, we only keep
terms including the electromagnetic vector potential up to the second order.
Additionally, we restrict ourselves to the uniform limit $\nabla H_{i}=\nabla \theta_i=0$.
A uniform solution usually has a lower free energy.
In the presence of the Lifshitz invariant, however, it is not obvious.
As we show in Appendix.~\ref{Apdx.B}, if the magnitude of the vector $\bm{d}$ is sufficiently small, the uniform solution has a lower free energy, and the order parameters are not spatially modulated at the ground state.

We obtain
\begin{align}
\mathcal{F}_{\mathrm{EM}} = \mathcal{F}_{1} + \mathcal{F}_{2} 
+ O\qty(H_{i}^{2}, \theta_{i}^{2}, H_{i}\theta_{i}), 
\end{align}
where
\begin{align}
\mathcal{F}_{1} &= e^{*2}\qty( \frac{\Psi_{1,0}}{m^{*}_{1}} + 2\eta\Psi_{2,0}
)\bm{A}^{2}H_{1} 
\notag \\
&\quad + e^{*2}\qty( \frac{\Psi_{2,0}}{m^{*}_{2}} + 2\eta\Psi_{1,0}
)\bm{A}^{2}H_{2} \,, 
\end{align}
and
\begin{align}
	\mathcal{F}_{2} &= -4e^{*}\bm{A}\cdot \mathrm{Re}  \qty[\bm{d}] \,\qty( \Psi_{1,0}H_{2} + \Psi_{2,0}H_{1} ) \notag \\
	& \quad - 4e^{*}\bm{A}\cdot\mathrm{Im}\qty[\bm{d}] \, \Psi_{1,0}\Psi_{2,0}\qty(\theta_{1}-\theta_{2}). 
	\label{eq:dterm}
\end{align}
The first term, $\mathcal{F}_{1}$, represents the nonlinear coupling between the amplitude fluctuation (Higgs mode) and the external field for each band~\cite{Intro:6_Shimano_Tsuji}. This coupling yields the third harmonic generation responses of Higgs modes in multiband superconductors. 

The second term, $\mathcal{F}_2$, is a linear coupling of the electromagnetic
vector potential to the collective modes. It describes the collective mode
contribution to the linear response. 
The real part induces the linear response of the Higgs modes,
while the imaginary part is responsible for the Leggett mode. 

In superconductors, the particle-hole symmetry is an effective (approximate)
low-energy symmetry, that acts as $\Psi_{i}\rightarrow\Psi^{*}_{i}$, or
\begin{align}
	H_i &\rightarrow H_i \nonumber\\
	\theta_i &\rightarrow -\theta_i \nonumber\\
	e^* &\rightarrow -e^* \,.
\end{align}
It is thus clear that only the Leggett mode linear response contribution is
invariant under the particle-hole symmetry. 
The constant vector $\bm{d}$ is in fact purely imaginary according to the microscopic calculation (see Appendix.~\ref{Apdx.A}).
Consequently, the amplitude contributions are
suppressed in the linear response. 
Potential observations of collective
modes in the linear response therefore require a multiband structure of superconductors. 

The term $\mathcal{F}_2$ may further be restricted by spatial symmetries of the underlying crystal lattice. 
Symmetry requirements that allow for the presence of the linear Leggett coupling will be discussed in Sec.~\ref{sec:level3}.

\subsection{\label{sec:level2-2}The Lifshitz invariant}
The term $\mathcal{F}_2$ in Eq.~\eqref{eq:dterm} originates from the expression
\begin{align}
\bm{d}\cdot\qty(\Psi^{*}_{1}\bm{D}\Psi_{2} + \Psi^{*}_{2}\bm{D}^{*}\Psi_{1})
+ \mathrm{c.c.}
\end{align}
in Eq.~\eqref{eqn:fe0}.
For simplicity we consider the case without the vector potential $\bm{A}$, and we put $\bm{D}=-\mathrm{i}\nabla$.
We also set $\bm{d}=\mathrm{i}\bm{d}_{I}$ because the real part $\mathrm{Re}\qty[\bm{d}]$ is suppressed by the particle-hole symmetry and it is actually zero according to the microscopic mean-field calculation (see Appendix.~\ref{Apdx.A}, particularly Eqs.~(\ref{Apdx_Def_vectorD}) and (\ref{d_imaginary})).
We then obtain
\begin{align}
&\bm{d}\cdot\qty(\Psi^{*}_{1}\bm{D}\Psi_{2} + \Psi^{*}_{2}\bm{D}^{*}\Psi_{1}) + \mathrm{c.c.}
\notag \\
&= \bm{d}_{\mathrm{I}}\cdot \qty( \Psi^{*}_{1}\nabla\Psi_{2} - \Psi^{*}_{2}\nabla\Psi_{1}) + \mathrm{c.c.}
\end{align}
The above term takes the form of the so-called Lifshitz invariant~\cite{LandauLifshitz}.
In the context of the Dzyaloshinskii-Moriya interaction, the vector $\bm{d}_{\mathrm{I}}$ can be interpreted as an``internal field". 
Moreover, the vector $\qty(\Psi^{*}_{1}\nabla\Psi_{2} - \Psi_{2}\nabla\Psi^{*}_{1})$ (and the term with interchanged $1$ and $2$) is similar to the form of the quantum mechanical current where the usual probability density ($\psi^* \psi$) has been replaced by an overlap of superconducting order parameters ($\Psi^{*}_{1} \Psi_{2}$, $\Psi^{*}_{2}\Psi_{1}$).
Since the overlap is determined by the phases of order parameters, this current transfers the phase.
In this sense, the vector $\bm{d}_{\mathrm{I}}$ is understood as a ``field" that drives the phase flow.
Because of the ``internal field" $\bm{d}_{\mathrm{I}}$, it is possible for the phase to couple to the external electromagnetic field to activate the Leggett mode in the linear response regime. 

\section{\label{sec:level3}Group theoretical classification of Lifshitz invariant}
This section presents the symmetry analysis of the Lifshitz invariant in multiband superconductors.
Although the Lifshitz invariant has been studied in many contexts~\cite{II:LI_DM_1st,II:LI_DM_2nd,II:LI_NCSC_1st,II:LI_NCSC_2nd,II:LI_NCSC_3rd,II:LI_com_incom_1st,II:LI_com_incom_2nd,II:LI_LiquidCrystal}, it has never been discussed in the linear response of multiband superconductors as far as we know.
The Lifshitz invariant is usually associated with the broken inversion symmetry.
However, in multiband superconductors the inversion symmetry itself is not crucial to induce the Lifshitz invariant, and the broken inversion symmetry is neither a necessary nor sufficient condition to have the Lifshitz invariant.

The basic strategy to determine whether the Lifshitz invariant is allowed or not is as follows~\cite{Sigrist_Ueda}.
We consider the representation of the term $\Psi^{*}_{i}\nabla\Psi_{j}$ ($i\neq j$). 
After calculating the direct product of the representations of the order parameters and spatial gradient and decomposing it into the direct sum of the irreducible representations, we check whether the term has a trivial irreducible representation or not. 
Since the free energy must be invariant under symmetry operations, the Lifshitz invariant is allowed to exist if the term has the trivial irreducible representation, but is not allowed if the term does not have the trivial irreducible representation.

To identify the representation of the order parameter, we need to specify the physical degrees of freedom that the order parameter has.
As an example, let us assume that the pairing symmetry is $s$-wave and the order parameter has sublattice degrees of freedom, i.e., the order parameter is defined on each lattice site in the unit cell.
These order parameters on different sites can be interchanged by symmetry operations, and may belong to a nontrivial representation, which determines whether the Lifshitz invariant can appear or not.
We first give a general procedure to derive the representation of the order parameters with sublattice degrees of freedom, which can be constructed from the \textit{induced representation} induced by a site-symmetry group (a subgroup of the crystallographic point group that fixes a certain lattice site).
The obtained representation is used to see whether the system can have the Lifshitz invariant or not, which results in a classification table of pairs of the order parameter representations for each crystallographic point group allowing the Lifshitz invariant.

Let us consider the general construction of the representation of the order parameter induced by site-symmetry groups~\cite{Serre}.
When a group $G$ and a subgroup $H$ of $G$ are given, a left coset decomposition of $H$ in $G$ is given by
\begin{align}
G = \bigcup_{\alpha}g_{\alpha}H,
\label{eq:LeftCosetDecomposition}
\end{align}
where $g_{\alpha}\in G$.
The induced representation of $G$ written by $\rho_{H;G}:= \rho_{H}\up G$ is produced by each representation $\rho_{H}$ of $H$.

We can explicitly construct a representation $\rho_{H;G}$ from the representation $\rho_{H}$.
To be precise, if the rows and/or columns of $\rho_{H}$ are labeled by $i$ and $j$, then the rows/columns of $\rho_{G}$ can be labeled by $i\alpha$ and $j\beta$.
Here, $\alpha$ and $\beta$ vary over the cosets $g_{\alpha}H$ in Eq.~\eqref{eq:LeftCosetDecomposition}.
Then we can define the representation $\rho_{H;G}$ as
\begin{align}
\qty[ \rho_{H;G}\qty(h)]_{i\alpha,j\beta} = \qty[ \tilde{\rho}_{H}\qty(g^{-1}_{\alpha}h g_{\beta})]_{ij},
\end{align}
where $h\in G$ and
\begin{align}
\qty[ \tilde{\rho}_{H}\qty(h')]_{ij} =
\begin{cases}
\qty[ \rho_{H}\qty(h')]_{ij} & \qty( \text{if } h'\in H), \\
0 & \qty( \text{else}).
\end{cases}
\end{align}
This is the general construction derived from group theory.

In our case, the group $G$ corresponds to the crystallographic point group that describes the whole system, while the group $H$ corresponds to the subgroup of $G$ that describes the site-symmetry group.
If one takes the trivial representation of $H$ (i.e., the one-dimensional representation $\qty[ \rho_{H}\qty(h)]_{ij}=1$ ($h\in H,$ $i,j=1$)), then the induced representation $\qty[\rho_{H;G}\qty(h)]_{1\alpha,1\beta}=:\qty[\rho_{H;G}\qty(h)]_{\alpha\beta}$ ($h\in G$) gives the representation of the sublattice degrees of freedom of the superconducting order parameter, where $\alpha$ and $\beta$ correspond to sublattice indices.

Here we assume that the superconducting pairing ($s$-wave, $p$-wave, etc.) and the sublattice degrees of freedom are transformed independently under symmetry operations.
That is, the representation is assumed to be the direct product of the pairing and sublattice degrees of freedom.
In addition, when we consider the representation of $\Psi^{*}_{i}\nabla\Psi_{j}$, the product of two superconducting pairings always yields the trivial irreducible representation because the two order parameters have the same pairing symmetry.

The site-symmetry group $H$ depends on each lattice site in the unit cell. 
However, if one classifies lattice sites in the unit cell by Wyckoff positions, then for each Wyckoff position the site-symmetry group is isomorphic to each other.
Thus, the site dependence of $H$ in each Wyckoff position does not affect the resulting induced representation.

When different sites belong to the same Wyckoff position, more precisely we need to consider the orbit of the group, and this allows for the correspondence with the sublattice degrees of freedom.
However, when the group orbits are different but at the same Wyckoff position, they will only appear identical in their representation.
Model (b) below is one of the examples of this case.

After obtaining the representation of the order parameter, we calculate the direct product of the representations $\Psi_{i}^{*}\nabla\Psi_{j}$, which is then decomposed into the direct sum of the irreducible representations by the reduction formula. 
The representation of $\nabla$ is solely determined from the crystallographic point group.
After checking whether the trivial representation is contained in $\Psi_{i}^{*}\nabla\Psi_{j}$, we can classify pairs of the irreducible representations of the order parameters that permit the Lifshitz invariant to show up. 

To show how our classification is obtained, let us take $D_{3h}$ as an example.
Since $\nabla$ has the same transformation property as the coordinate $\bm{r}=\qty[x, y, z]^{\mathrm{T}}$, the representation of $\nabla$, $\rho_{\nabla}$, follows from the direct sum of the representations of the basis functions $x$, $y$, and $z$.
The basis functions and direct products of the representations of crystallographic point groups are in detail given in \cite{TableGroupTheory}.
In the case of $D_{3h}$, $z$ belong to $A_{2}''$, and $x$ and $y$ belong to the representation $E'$.
Then $\rho_{\nabla}$ is given by the direct sum of these representations:
\begin{equation}
\rho_{\nabla} = A_{2}'' \oplus E'.
\end{equation}
Now we turn to the representations of order parameters.
For the pair $\qty(\Psi_{i}, \Psi_{j}) = (A_{1}', A_{2}'')$, for instance, we can evaluate the representation of $\Psi^{*}_{i}\nabla\Psi_{j}$ as
\begin{equation}
A_{1}'\otimes\qty(A_{2}''\oplus E')\otimes A_{2}'' = A_{1}'\oplus E'',
\end{equation}
which allows the Lifshitz invariant because it has the trivial irreducible representation $A_{1}'$.
For the pair $\qty(\Psi_{i}, \Psi_{j}) = (A_{2}', E'')$, on the other hand, we can calculate as
\begin{equation}
A_{2}'\otimes \qty(A_{2}''\oplus E')\otimes E'' = A_{1}''\oplus A_{2}''\oplus E' \oplus E'', 
\end{equation}
which does not allow the Lifshitz invariant since it does not have $A_{1}'$.

Here we give Table~\ref{tab:classification1}/\ref{tab:classification2}, which lists all the possible pairs of representations of the order parameters $(\Psi_{i}, \Psi_{j})$ for each crystallographic point group without/with inversion symmetry that allows the existence of the Lifshitz invariant $\Psi^{*}_{i}\nabla\Psi_{j}$.
A similar classification for $D_{2h}$ point group has recently been reported in Ref.~\cite{II:LI_Yanase_2nd}.
In Table~\ref{tab:classification1} and \ref{tab:classification2}, we notice that there are many possible combinations of the order parameter representations that allow the existence of the Lifshitz invariant, both in systems with and without inversion symmetry.
In the presence of inversion symmetry, the allowed representations are always combinations of gerade and ungerade, since $\nabla$ is parity odd.

We note that the results in Table~\ref{tab:classification1} and \ref{tab:classification2} are universal, and do not depend on what kind of physical degrees of freedom the representation of the order parameters corresponds to.
They are not even limited to superconductors but can be applied to any systems having multiple order parameters.
In the present paper, we primarily consider the case of multiband superconductors having multiple degrees of freedom.
If we assume that the order parameters have orbital degrees of freedom which also transform independently under symmetry operations, the argument can easily be extended to include orbital degrees of freedom.
When the order parameters have multiple degrees of freedom in addition to the band indices like $\Psi_{i,a}$, where $a$ is the additional degrees of freedom, the vector $\bm{d}$ can in principle be a tensor.
Even in this case the results of the classification remains the same and the argument does not change.
Although we later study $s$-wave superconductors to see the Leggett modes in a linear response in simple and concrete models, the results in the table are directly applicable to any superconducting gap symmetries since the order parameter is assumed to be represented as a direct product of the pairing and sublattice components, and the product of the two identical pairing symmetries (e.g., $p$-wave and $p$-wave) always yields the trivial irreducible representation.
\begin{table*}[t]
\centering
\caption{Pairs of representations of the order parameters $(\Psi_{i}, \Psi_{j})$ for each crystallographic point group without inversion symmetry that allow the existence of the Lifshitz invariant $\Psi^{*}_{i}\nabla\Psi_{j}-\Psi_{j}\nabla\Psi^{*}_{i}$.}
\begin{tabular}{ccc}
\hline \hline
Point group\quad\quad & Representation of $\nabla$\quad\quad &
\begin{tabular}{c}
Allowed representation of the order parameter pairs $\qty(\Psi_{i}, \Psi_{j})$ ($i\neq j$)
\end{tabular}\\ \hline
$C_{1}$ & $A$ & $(A,A)$ \\
$C_{2}$ & $A\oplus B$ & $(A,A)$, $(A,B)$, $(B,B)$  \\
$C_{1h}$ & $A'\oplus A''$ & $(A',A')$, $(A',A'')$, $(A'',A'')$  \\
$C_{3}$ & $A\oplus E$ & $(A,A)$, $(A,E)$, $(E,E)$  \\
$C_{4}$ & $A\oplus E$ & $(A,A)$, $(A,E)$, $(B,B)$, $(B,E)$, $(E,E)$  \\
$S_{4}$ & $B\oplus E$ & $(A,E)$, $(A,B)$, $(B,E)$, $(E,E)$  \\
$D_{2}$ & $B_{1}\oplus B_{2} \oplus B_{3}$ & $(A,B_{1})$, $(A,B_{2})$, $(A,B_{3})$, $(B_{1}, B_{2})$, $(B_{1}, B_{3})$, $(B_{2}, B_{3})$  \\
$C_{2v}$ & $A_{1}\oplus B_{1} \oplus B_{2}$ & $(A_{1},A_{1})$, $(A_{1},B_{1})$, $(A_{1},B_{2})$, $(A_{2}, A_{2})$, $(A_{2}, B_{1})$, $(A_{2}, B_{2})$, \\
 & &  $(B_{1}, B_{1})$, $(B_{2}, B_{2})$  \\
$C_{6}$ &  $A\oplus E_{1}$ & $(A,A)$ , $(A, E_{1})$, $(B,B)$, $(B, E_{2})$, $(E_{1}, E_{1})$, $(E_{1}, E_{2})$, \\
& &  $(E_{2}, E_{2})$ \\
$C_{3h}$ &  $A''\oplus E'$ & $(A', A'')$, $(A',E')$, $(A'', E'')$, $(E', E')$, $(E', E'')$, $(E'', E'')$ \\
$D_{3}$ &  $A_{2}\oplus E$ & $(A_{1},A_{2})$ , $(A_{1}, E)$, $(A_{2},E)$, $(E, E)$ \\
$C_{3v}$ &  $A_{1}\oplus E$ & $(A_{1},A_{1})$ , $(A_{1}, E)$, $(A_{2},E)$, $(E, E)$ \\
$D_{4}$ & $A_{2}\oplus E$ & $(A_{1}, A_{2})$, $(A_{1}, E)$, $(A_{2}, E)$, $(B_{1}, B_{2})$, $(B_{1}, E)$, $(B_{2}, E)$, \\
 & & $(E,E)$ \\
$C_{4v}$ & $A_{1}\oplus E$ & $(A_{1}, A_{1})$, $(A_{1}, E)$, $(A_{2}, A_{2})$, $(A_{2}, E)$, $(B_{1}, B_{1})$, $(B_{1}, E)$, \\
 & & $(B_{2}, B_{2})$, $(B_{2}, E)$, $(E, E)$ \\
$D_{2d}$ & $B_{2}\oplus E$ & $(A_{1}, B_{2})$, $(A_{1}, E)$, $(A_{2}, B_{1})$, $(A_{2}, E)$, $(B_{1}, E)$, $(B_{2}, E)$, \\
& & $(E, E)$ \\
$T$ & $T$ & $(A,T)$, $(E,T)$, $(T,T)$ \\
$D_{6}$ & $A_{2}\oplus E_{1}$ & $(A_{1}, A_{2})$, $(A_{1}, E_{1})$, $(A_{2}, E_{1})$, $(B_{1}, B_{2})$, $(B_{1}, E_{2})$, $(B_{2}, E_{2})$, \\
 & & $(E_{1}, E_{1})$, $(E_{1}, E_{2})$, $(E_{2}, E_{2})$ \\
$C_{6v}$ & $A_{1}\oplus E_{1}$ & $(A_{1}, A_{1})$, $(A_{1}, E_{1})$, $(A_{2}, A_{2})$, $(A_{2}, E_{1})$, $(B_{1}, B_{1})$, $(B_{1}, E_{2})$, \\
 & & $(B_{2}, B_{2})$, $(B_{2}, E_{2})$, $(E_{1}, E_{1})$, $(E_{1}, E_{2})$, $(E_{2}, E_{2})$ \\
$D_{3h}$ & $A''_{2}\oplus E'$ & $(A'_{1}, A''_{2})$, $(A'_{1}, E')$, $(A''_{1}, A'_{2})$, $(A''_{1}, E'')$, $(A'_{2}, E')$, $(A''_{2}, E'')$, \\
 & & $(E', E')$, $(E', E'')$, $(E'', E'')$ \\
$O$ & $T_{1}$ & $(A_{1}, T_{1})$, $(A_{2}, T_{2})$, $(E, T_{1})$, $(E, T_{2})$, $(T_{1}, T_{1})$, $(T_{1}, T_{2})$, \\
 & & $(T_{2}, T_{2})$ \\
$T_{d}$ & $T_{2}$ & $(A_{1}, T_{2})$, $(A_{2}, T_{1})$, $(E, T_{1})$, $(E, T_{2})$, $(T_{1}, T_{1})$, $(T_{1}, T_{2})$, \\
 & & $(T_{2}, T_{2})$ \\ \hline\hline
\end{tabular}
\label{tab:classification1}
\end{table*}

\begin{table*}[t]
\centering
\caption{Pairs of representations of the order parameters $(\Psi_{i}, \Psi_{j})$ for each crystallographic point group with inversion symmetry that allow the existence of the Lifshitz invariant $\Psi^{*}_{i}\nabla\Psi_{j}-\Psi_{j}\nabla\Psi^{*}_{i}$.}
\begin{tabular}{ccc}
\hline \hline
Point group\quad \quad & Representation of $\nabla$\quad \quad &
\begin{tabular}{c}
Allowed representation of the order parameter pairs $\qty(\Psi_{i}, \Psi_{j})$ ($i\neq j$)
\end{tabular}\\ \hline
$C_{i}$ & $A_{u}$ & $(A_{g},A_{u})$ \\
$C_{2h}$ & $A_{u}\oplus B_{u}$ & $(A_{g},A_{u})$, $(A_{g},B_{u})$, $(B_{g},A_{u})$, $(B_{g},B_{u})$  \\
$C_{3i}=S_{6}$ & $A_{u}\oplus E_{u}$ & $(A_{g},A_{u})$, $(A_{g},E_{u})$, $(E_{g},A_{u})$, $(E_{g}, E_{u})$ \\
$D_{2h}$ &  $B_{1u}\oplus B_{2u}\oplus B_{3u}$ & $(A_{g},B_{1u})$ , $(A_{g}, B_{2u})$, $(A_{g},B_{3u})$, $(B_{1g}, A_{u})$, $(B_{1g}, B_{2u})$, $(B_{1g}, B_{3u})$ \\
& & $(B_{2g}, A_{u})$, $(B_{2g}, B_{1u})$, $(B_{2g}, B_{3u})$, $(B_{3g}, A_{u})$, $(B_{3g}, B_{1u})$, $(B_{3g}, B_{2u})$ \\
$C_{4h}$ & $A_{u}\oplus E_{u}$ & $(A_{g}, A_{u})$, $(A_{g}, E_{u})$, $(B_{g}, B_{u})$, $(B_{g}, E_{u})$, $(E_{g}, A_{u})$, $(E_{g}, B_{u})$, \\
& & $(E_{g}, E_{u})$ \\
$C_{6h}$ & $A_{u}\oplus E_{1u}$ & $(A_{g}, A_{u})$, $(A_{g}, E_{1u})$, $(B_{g}, B_{u})$, $(B_{g}, E_{2u})$, $(E_{1g}, A_{u})$, $(E_{1g}, E_{1u})$, \\
& & $(E_{1g}, E_{2u})$, $(E_{2g}, A_{u})$, $(E_{2g}, E_{1u})$, $(E_{2g}, E_{2u})$ \\
$D_{3d}$ & $A_{2u}\oplus E_{u}$ & $(A_{1g}, A_{2u})$, $(A_{1g}, E_{u})$, $(A_{2g}, A_{1u})$, $(A_{2g}, E_{u})$, $(E_{g}, A_{1u})$, $(E_{g}, A_{2u})$, \\
& & $(E_{g}, E_{u})$ \\
$D_{4h}$ & $A_{2u}\oplus E_{u}$ & $(A_{1g}, A_{2u})$, $(A_{1g}, E_{u})$, $(A_{2g}, A_{1u})$, $(A_{2g}, E_{u})$, $(B_{1g}, B_{2u})$, $(B_{1g}, E_{u})$, \\
 & & $(B_{2g}, B_{1u})$, $(B_{2g}, E_{u})$, $(E_{g}, A_{1u})$, $(E_{g}, A_{2u})$, $(E_{g}, B_{1u})$, $(E_{g}, B_{2u})$, \\
 & & $(E_{g}, E_{u})$ \\
$D_{6h}$ & $A_{2u}\oplus E_{1u}$ & $(A_{1g}, A_{2u})$, $(A_{1g}, E_{1u})$, $(A_{2g}, A_{1u})$, $(A_{2g}, E_{1u})$, $(B_{1g}, B_{2u})$, $(B_{1g}, E_{2u})$, \\
 & & $(B_{2g}, B_{1u})$, $(B_{2g}, E_{2u})$, $(E_{1g}, A_{1u})$, $(E_{1g}, A_{2u})$, $(E_{1g}, E_{1u})$, $(E_{1g}, E_{2u})$, \\
 & & $(E_{2g}, B_{1u})$, $(E_{2g}, B_{2u})$, $(E_{2g}, E_{1u})$, $(E_{2g}, E_{2u})$ \\
$T_{h}$ & $T_{u}$ & $(A_{g}, T_{u})$, $(E_{g}, T_{u})$, $(T_{g}, A_{u})$, $(T_{g}, E_{u})$, $(T_{g}, T_{u})$ \\
$O_{h}$ & $T_{1u}$ & $(A_{1g}, T_{1u})$, $(A_{2g}, T_{2u})$, $(E_{g}, T_{1u})$, $(E_{g}, T_{2u})$, $(T_{1g}, A_{1u})$, $(T_{1g}, E_{u})$, \\
 & & $(T_{1g}, T_{1u})$, $(T_{1g}, T_{2u})$, $(T_{2g}, A_{2u})$, $(T_{2g}, E_{u})$, $(T_{2g}, T_{1u})$, $(T_{2g}, T_{2u})$ \\ \hline\hline
\end{tabular}
\label{tab:classification2}
\end{table*}

\section{\label{sec:level4}Application of the group theory to several models}
\begin{figure*}[t]
\centering
\includegraphics[width=172mm]{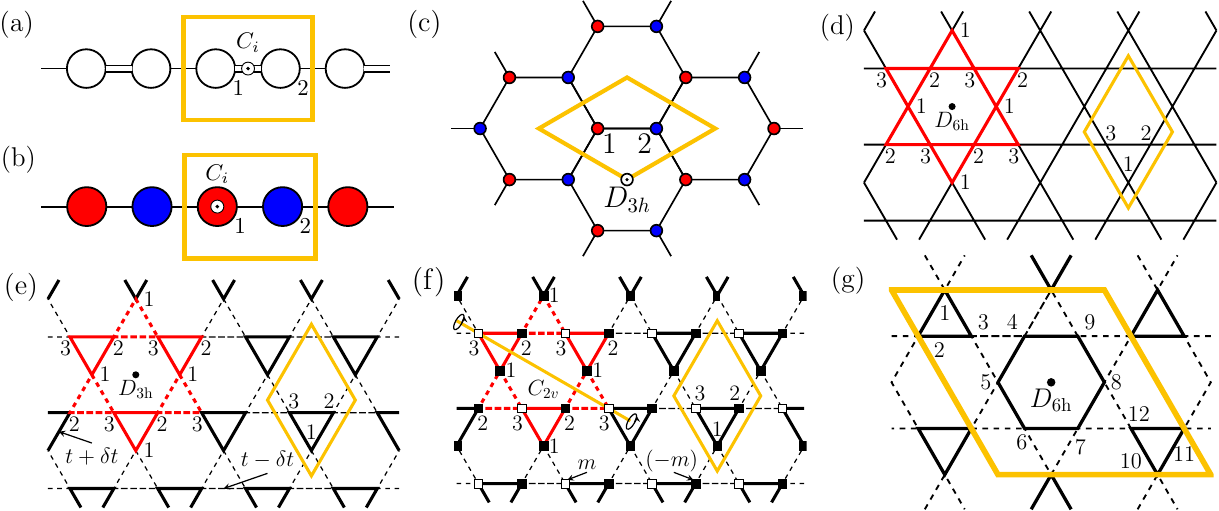}
\caption{Several target models to be checked to have the Lifshitz invariant or not. 
We simply assume the $s$-wave superconductivity with a single orbital in these models.
Unit cells of these models are enclosed by yellow lines and the subscripts of the lattice points label them. 
(a) The Rice-Mele model without on-site potential (SSH model). 
The single and double lines connecting sites correspond to hoppings $t\neq t'$.
The point group of the system is $C_{i}$. 
(b) The Rice-Mele model with the same hopping strength. 
Two sites are distinguished by on-site potentials (red and blue circles). 
The point group of the system is $C_{i}$. 
(c) Honeycomb model with two kinds of on-site potentials. 
Two sites are distinguished by on-site potentials (red and blue circles).
The point group of the system is $D_{3h}$.
(d) The Kagome lattice model with only one hopping strength. 
The black point represents the center of the symmetry operations.
The red lines make it easier to see the star of David shape, which changes in other models (e) and (f).
The point group of the system is $D_{6h}$. 
(e) Kagome lattice model after trimerization. 
The solid lines show the hopping strength ($t+\delta t$) and the dashed lines show ($t-\delta t$). 
The red solid and dashed lines show the star of David shape, which helps us see the transformation properties of order parameters under symmetry operations.
The point group of the system is $D_{3h}$. 
(f) Trimerized Kagome lattice model with on-site potentials. 
White (black) squares represent the on-site potential $m>0$ ($-m<0$).
The yellow line on the red star of David shows the axis containing the center of symmetry operation.
The point group of the system is $C_{2v}$. 
(g) Kagome lattice model with possible CDW pattern. The unit cell has 12 lattice points, constructing the tri-hexagonal pattern (3$Q$-pattern) with inner (outer) lattice points moving outward (inward).
This model incorporates the effect of modified configuration by taking two kinds of hopping strength. 
The point group of this system is $D_{6h}$.}
\label{fig:models}
\end{figure*}
We apply the general group theoretical classification obtained in the previous section to several models to check the validity of our approach. 
In the following, we assume $s$-wave superconductivity with a single atomic orbital on each lattice site in each model. 
For the Lifshitz invariant to appear, it is not necessary to break the inversion symmetry in multiband superconductors as seen in the previous section and we will see concrete examples below. 
We start by analyzing the Rice-Mele model~\cite{Rice-Mele} for a two-band superconductor previously considered in Ref.~\cite{Leggett_Kamatani} to reproduce the previous result that the Lifshitz invariant appeared to confirm the consistency. 
We then consider several different models with on-site pairing interactions: A honeycomb model with on-site potentials as an example of the system without inversion symmetry and a family of Kagome models, where we discuss lowering of symmetry by introducing different hopping parameters. 
We additionally consider a Kagome lattice model with a charge-density wave pattern, as found to occur, e.g., in $\mathrm{CsV}_{3}\mathrm{Sb}_{5}$~\cite{Kagome_Possible_CDW,Kagome_Possible_CDW_3Q}.
These models are summarized in Fig.~\ref{fig:models}.

We can cover all the cases with/without the inversion symmetry and with/without the Lifshitz invariant in these models.
We take (a) and (b) to confirm the consistency with the previous research by \cite{Leggett_Kamatani}.
(c) is the example of the system without inversion symmetry and the Lifshitz invariant.
(d) is the case with inversion symmetry but without the Lifshitz invariant.
(e) is the case without inversion symmetry but with the Lifshitz invariant,
(f) is a variant of (e), and (g) is an example of the system with inversion symmetry and the Lifshitz invariant.
The Table.~\ref{tab:properties} below summarizes all the cases.\\

\begin{table}[t]
\centering
\caption{List of the point groups (PGs), site-symmetry groups (SSGs), and the presence or absence of inversion symmetry and the Lifshitz invariant for the models shown in Fig.~\ref{fig:models}}
\begin{tabular}{c|c|c|c|c}
\hline \hline
\text{Model} & \text{PG} & \text{SSG} & \text{Inversion symmetry} & \text{Lifshitz invariant} \\ \hline
(\text{a}) & \textit{$C_{i}$} & \textit{$C_{1}$} & \checkmark & \checkmark \\ \hline
(\text{b}) & \textit{$C_{i}$} & \textit{$C_{i}$} & \checkmark &  \\ \hline
(\text{c}) & \textit{$D_{3h}$} & \textit{$D_{3h}$} & & \\ \hline
(\text{d}) & \textit{$D_{6h}$} & \textit{$D_{2h}$} & \checkmark & \\ \hline
(\text{e}) & \textit{$D_{3h}$} & \textit{$C_{2v}$} & & \checkmark \\ \hline
(\text{f}) & \textit{$C_{2v}$} & \textit{$C_{2v}$}, \textit{$C_{1h}$} & & \checkmark \\ \hline
(\text{g}) & \textit{$D_{6h}$} & \textit{$C_{2v}$} & \checkmark & \checkmark \\ \hline \hline
\end{tabular}
\label{tab:properties}
\end{table}

Before we delve into the specific models, let us review the useful method.
We practically do not have to construct the induced representation $\rho_{H;G}$ explicitly.
Instead, for each symmetry operation of $G$ we can evaluate the character $\chi_{\rho_{H;G}}$ of the induced representation directly by considering how the lattice sites in the unit cell are interchanged by the symmetry operation.
The resulting character table uniquely determines the decomposition of the induced representation into a set of irreducible representations by the reduction formula.

\textit{The Rice-Mele model without on-site potentials} (Su-Schrieffer-Heeger (SSH) model~\cite{SSH_model}). 
We shall first consider the Rice-Mele model without on-site potential, which has been studied in the context of collective modes in Ref.~\cite{Leggett_Kamatani}. 
It is a model with two sites in the unit cell with one orbital per site and attractive on-site pairing.
We treat the on-site interaction on the mean-field level by introducing two order parameters $\Psi_1,\Psi_2$.
The model is depicted in Fig.~\ref{fig:models}(a) where the two-site unit cell is shown by the yellow rhombus. 
Single and double lines connecting sites correspond to hoppings $t\ne t'$. 

The system has a bond-centered inversion symmetry, resulting in the point group  $C_{i}=\{E, i\}$. 
The two sites in the unit cell are exchanged under inversion, meaning that the subgroup is $C_{1}$. Therefore, the two-dimensional reducible representation $\rho$, under which the pair of order parameters $(\Psi_1,\Psi_2)$ transform,
has the following characters:
\begin{align}
\begin{array}{c|cc}
 & E & i \\ 
 \hline
\chi_{\rho_{C_{1};C_{i}}} & 2 & 0  \\
\end{array}
\end{align}
It is then clear that the decomposition of $\rho_{C_{1};C_{i}}$ into
irreducible representations of $C_i$ is 
$\rho_{C_{1};C_{i}} = A_{g} \oplus A_{u}$. 
The two order parameters can be decomposed into a symmetric $A_g$ component and an anti-symmetric $A_u$ component according to
\begin{align}
\mqty[\Psi_{1} \\ \Psi_{2}] = \Psi_{A_{g}}\mqty[1 \\ 1] + \Psi_{A_{u}}\mqty[1 \\ -1].
\end{align}
Next, we assess the transformation properties of the Lifshitz invariant $\Psi^{*}_{i} \nabla_\alpha \Psi_j$. The derivative transforms like a vector, i.e., under the representation $3 A_u$ of $C_i$. Thus, the Lifshitz invariant transforms as
\begin{align}
\rho_{C_{1};C_{i}} \otimes 3 A_u \otimes \rho_{C_{1};C_{i}}
&=
\qty( A_{g} \oplus A_{u}) \otimes 3 A_{u} \otimes \qty(A_{g} \oplus A_{u}) \nonumber\\
&= 6A_{g}\oplus 6A_{u},
\label{eqn.ProcedureEX.}
\end{align}

The Lifshitz invariant includes six invariant components that transform under the
trivial representation $A_g$ and is thus allowed in the free energy. These correspond to 
$A_g \nabla_\alpha A_u$ and $A_u \nabla_\alpha A_g$.
Terms of the form $A_g \nabla_\alpha A_g$, $A_u \nabla_\alpha A_u$ must vanish due to the symmetry.
These results are summarized in the second row of Table~\ref{tab:classification2}.

In summary, the symmetry analysis shows that the Leggett mode 
can appear in the linear response for the Rice-Mele model without on-site potential, consistent with that in Ref.~\cite{Leggett_Kamatani}. 

\textit{The Rice-Mele model with a uniform hopping.} 
Next, we consider the Rice-Mele model with uniform hopping (i.e., $t=t'$). 
We also introduce an on-site potential that is different for the two sublattices, as indicated by blue and red colors in Fig.~\ref{fig:models}(b). 
The point group of the system is again $C_{i}$, but the inversion is now site-centered and does not exchange sublattices in contrast to the previous case, meaning that the subgroup is $C_{i}$. 
The characters of the representation $\rho_{C_{i};C_{i}}$ of the two on-site order parameters $(\Psi_1,\Psi_2)$
are
\begin{align}
    \begin{array}{c|cc}
    & E & i \\ \hline
    \chi_{\rho_{C_{i};C_{i}}} & 2 & 2 \\
    \end{array}
\end{align}
from which it follows that $\rho_{C_{i};C_{i}} = 2A_{g}$. 
In this representation, the term $\Psi^{*}_{i}\nabla\Psi_{j}$ no longer contains an invariant,
since it transforms as
$2A_g \otimes 3A_u \otimes 2A_g = 12 A_u$. 
Hence the Lifshitz invariant cannot appear in this model and the Leggett mode does not appear in the linear response regime. This result is consistent with Ref.~\cite{Leggett_Kamatani}. 

The previous models both have $C_i$ symmetry, yet only the first case exhibits a linear collective mode response. 
This example illustrates that both the point group and the specific representation of the order parameters
determines the presence of the Lifshitz invariant, according to the classification of Table~\ref{tab:classification1} and Table~\ref{tab:classification2}.

For simplicity, we denote $\rho_{H;G}$ as $\rho$ and $\chi_{\rho{H;G}}$ as $\chi$ from now.

\textit{Honeycomb lattice.} We shall next
study a Honeycomb model with broken inversion symmetry, illustrated in Fig.~\ref{fig:models}(c). 
Red and blue colors indicate different on-site potentials. 
The point group of the system is $D_{3h}=\{E,2C_{3},3C'_{2},\sigma_{h},2S_{3},3\sigma_{v}\}$. 
We note that the point group lacks inversion symmetry. 
The two superconducting order parameters supported on each sublattice transform under the representation $\rho$ with characters
\begin{align}
    \begin{array}{c|cccccc}
     & E & 2C_{3} & 3C'_{2} & \sigma_{h} & 2S_{3} & 3\sigma_{v} \\ \hline
    \chi & 2 & 2 & 2 & 2 & 2 & 2 \\
    \end{array}
\end{align}
from which it follows that $\rho$ decomposes into the trivial irreducible representations of $D_{3h}$, i.e., $\rho=2A'_{1}$. 
The Lifshitz term transforms as $2A'_{1} \otimes (E' \oplus A''_2) \otimes 2A'_{1} = 4(E' \oplus A''_2)$, which does not contain any invariant and must therefore be absent in accordance with Table~\ref{tab:classification1}.
Even though the present model lacks a center of inversion, it still does not exhibit a linear optical collective mode response. 
The presence of the Lifshitz invariant is therefore not a simple consequence of broken inversion symmetry; instead all point group symmetries need to be carefully taken into account.

\textit{Kagome lattice.} We next turn to 
a model of the Kagome lattice depicted in Fig.~\ref{fig:models}(d). 
It has three sites per unit cell, supporting three superconducting order parameters. 
Nearest-neighbor sites are connected by identical hopping parameters $t$, resulting in the point group $D_{6h}$. 
The three order parameters transform according to the representation $\rho$ with characters
\begin{align}
    \begin{array}{c|cccccccccccc}
     & E & 2C_{6} & 2C_{3} & C_{2} & 3C'_{2} & 3C''_{2} & i & 2S_{3} & 2S_{6} & \sigma_{h} & 3\sigma_{d} & 3\sigma_{v} \\ \hline
    \chi & 3 & 0 & 0 & 3 & 1 & 1 & 3 & 0 & 0 & 3 & 1 & 1 \\
    \end{array} \,
\end{align}
from which it follows that $\rho$ is reduced to
\begin{align}
    \rho = A_{1g}\oplus E_{2g} \,.
\end{align}
We find that neither $A_{1g}$ nor $E_{2g}$ can support the Lifshitz term, since they are both even under inversion.

To change the symmetry, we
add a trimerized hopping pattern to our model as shown in Fig.~\ref{fig:models}(e). Here, straight and dashed lines correspond to $t\pm \delta t$, respectively. The point group of the system is reduced to $D_{3h}$. The characters of $\rho$ are now given by
\begin{align}
    \begin{array}{c|cccccc}
     & E & 2C_{3} & 3C'_{2} & \sigma_{h} & 2S_{3} & 3\sigma_{v} \\ \hline
    \chi & 3 & 0 & 1 & 3 & 0 & 1 \\
    \end{array}
\end{align}
and $\rho$ decomposes into
\begin{align}
    \rho = A'_{1}\oplus E' \,.
\end{align}
In this representation, the Lifshitz variant is symmetry-allowed according to Table~\ref{tab:classification1}, where we see that combinations $(A_1',E')$ as well as $(E',E')$ give rise to a linear optical collective mode response.

We can further reduce the symmetries
of our model by introducing sublattice-dependent on-site potentials shown by white and black squares in Fig.~\ref{fig:models}(f). The point group of the system is reduced to the subgroup $C_{2v}=\{E, C_{2}(z), \sigma_{v}\qty(xz),\sigma_{v}\qty(yz)\}$ of $D_{3h}$. The 
representation $E'$ splits into $A_1 \oplus B_1$, yielding the representation $\rho = 2A_{1}\oplus B_{1} $ of the three on-site order parameters. The Lifshitz term is still present, since lowering the point group leads to even fewer symmetry restrictions on the free energy. 

\textit{Kagome lattice with CDW.} Finally we shall examine a Kagome lattice model with a charge-density wave (CDW) in a tri-hexagonal pattern (3$Q$-pattern).
Such structures have, e.g., been experimentally suggested in $\mathrm{CsV}_{3}\mathrm{Sb}_{5}$ \cite{Kagome_Possible_CDW,Kagome_Possible_CDW_3Q}.
This material is also suggested to have a different CDW structure from the 3$Q$-pattern (so-called the star of david~\cite{Kagome_StarOfDavid,Intro:Kagome_SC_review}).
Nevertheless, both CDW patterns have the same symmetry, and the group theoretical results can be applied to either pattern.
The CDW unit cell is shown in 
Fig.~\ref{fig:models}(g) where straight and dashed lines correspond to different hoppings.
The point group of the system is $D_{6h}=\{E, 2C_{6}, 2C_{3}, C_{2}, 3C'_{2}, 3C''_{2}, i, 2S_{3}, 2S_{6}, \sigma_{h}, 3\sigma_{d}, 3\sigma_{v}\}$. The model has $12$ onsite order parameters that transform according to the representation $\rho$ with characters
\begin{align}
    \begin{array}{c|cccccccccccc}
     & E & 2C_{6} & 2C_{3} & C_{2} & 3C'_{2} & 3C''_{2} & i & 2S_{3} & 2S_{6} & \sigma_{h} & 3\sigma_{d} & 3\sigma_{v} \\ \hline
    \chi & 12 & 0 & 0 & 0 & 2 & 2 & 0 & 0 & 0 & 12 & 2 & 2 \\
    \end{array}
\end{align}
The reduction formula gives a decomposition representation
\begin{align}
    \rho = 2A_{1g}\oplus 2E_{2g} \oplus B_{1u} \oplus B_{2u} \oplus 2E_{1u} \,.
\end{align}
Importantly, $\rho$ includes both even and odd irreducible representations, which indeed leads to allowed Lifshitz terms and optical collective mode response according to Table~\ref{tab:classification2}. This symmetry analysis suggests that $\mathrm{CsV}_{3}\mathrm{Sb}_{5}$ 
might be an interesting experimental platform for the study of superconducting collective modes in the linear response.

\section{\label{sec:level5}Microscopic calculation of collective modes and linear optical conductivity}
In this section, we perform microscopic calculations of collective modes and the linear optical conductivity, since the group-theoretical classification just gives the necessary condition for the Lifshitz invariant and we should examine how the contribution of the Lifshitz invariant quantitatively appears in the optical responses. 
We compute the optical conductivity $\sigma\qty(\omega)$ within the effective action approach in imaginary time, focusing on the fluctuations of the order parameters.
We use the Kagome lattice model as a concrete example. 
In the following, we abbreviate the $\bm{k}$-dependence of functions whenever appropriate.

\subsection{\label{sec:level5-1}Model}
\begin{figure}[h]
\centering
\includegraphics[width=86mm]{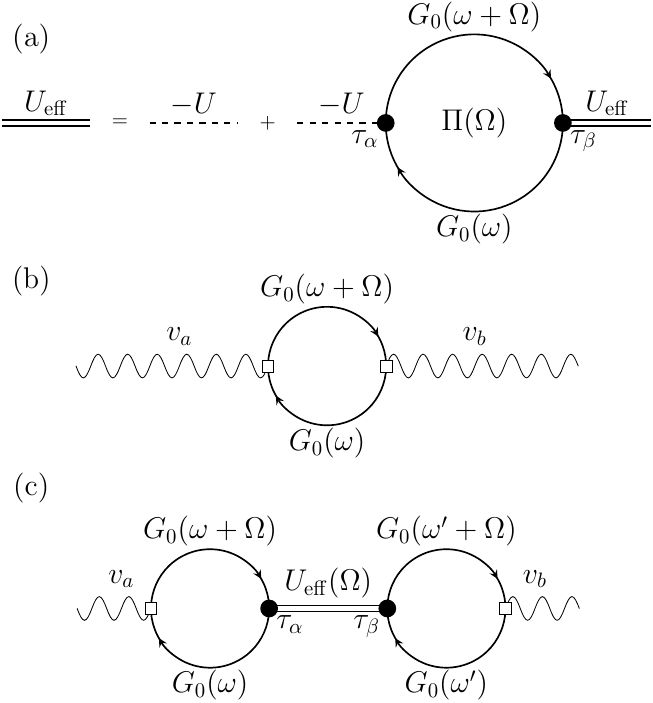}
\caption{Diagrammatic representations of (a) the effective interaction $U_{\mathrm{eff}}\qty(\omega)$ within the random phase approximation, and (b,c) the optical conductivities in superconductors. 
The diagram (b) ((c)) corresponds to the quasi-particle (collective mode) excitation.
Here $-U$ is the bare interaction.
When the spatial dimension is more than one, the optical conductivities depend on two directions $a$ and $b$.}
\label{Diagrams}
\end{figure}
Our starting point is the Hamiltonian $\mathcal{H} = \mathcal{H}_{0} + \mathcal{H}_{\mathrm{int}}$ that describes a multiband superconductor with a singlet on-site pairing, where
\begin{align}
\mathcal{H}_{0} = \sum_{\bm{k}\alpha\alpha'\sigma}\xi_{\alpha\alpha'}\qty(\bm{k})c^{\dagger}_{\bm{k}\alpha\sigma}c_{\bm{k}\alpha'\sigma}, 
\end{align}
is the kinetic part, and
\begin{align}
\mathcal{H}_{\mathrm{int}} = -U\sum_{\bm{k}\bm{k}'\alpha}c^{\dagger}_{\bm{k}\alpha\up}c^{\dagger}_{-\bm{k}\alpha\down}c_{-\bm{k}'\alpha\down}c_{\bm{k}'\alpha\up} \,.
\label{eq:hint}
\end{align}
is the interaction part. 
We have set the volume size of the system to one. 
Here $c^{\dagger}$ and $c$ are the creation and annihilation operators, $\alpha=1,2,\ldots n$ is the band index, $\xi_{\alpha,\alpha'}(\bm{k})$ represents the matrix elements of the kinetic term, and $\sigma=\up,\down$ denotes spin. To include the effect of an external electromagnetic field by a vector potential $\bm{A}$, we replace $\bm{k}$ by $\bm{k}-e\bm{A}$ (Peierls substitution) and expand the kinetic part as
\begin{align}
\xi_{\alpha\alpha'}\qty(\bm{k}- e\bm{A}) = \xi_{\alpha\alpha'}\qty(\bm{k}) - e\nabla_{\bm{k}}\xi_{\alpha\alpha'}\qty(\bm{k})\cdot\bm{A} + O\qty(\bm{A}^{2})\,, 
\end{align}
where $e$ is the electric charge.
Since we are interested in the linear optical conductivity, we only need to expand $\xi_{\alpha\alpha'}$ to the first order in the vector potential $\bm{A}$.
Then the full Hamiltonian is given by
\begin{align}
\mathcal{H} = \mathcal{H}_{0} + \mathcal{H}_{\mathrm{int}} + \mathcal{H}_{\mathrm{EM}}\,, 
\end{align}
where
\begin{align}
\mathcal{H}_{\mathrm{EM}} = e\sum_{\bm{k}\alpha\alpha'\sigma}\nabla_{\bm{k}}\xi_{\alpha\alpha'}\qty(\bm{k})\cdot\bm{A} \,c^{\dagger}_{\bm{k}\alpha\sigma}c_{\bm{k}\alpha'\sigma}\,.
\end{align}
We use the path integral approach in imaginary time $\tau$ \cite{IV:EffectiveAction_1st,IV:EffectiveAction_2nd,IV:EffectiveAction_3rd,IV:EffectiveAction_4th}. 
The partition function of the whole system can be written as $\mathcal{Z} = \int \mathcal{D}\qty(c^{\dagger}c)e^{-S[c^{\dagger},c] }$ with the Euclidean action
\begin{align}
S\qty[ c^{\dagger}, c] = \int_{0}^{\beta}d\tau \qty( \sum_{\bm{k}\alpha\sigma}c^{\dagger}_{\bm{k}\alpha\sigma}\partial_{\tau}c_{\bm{k}\alpha\sigma} + \mathcal{H} )\,.
\end{align}
We perform the Hubbard-Stratonovich transformation to decouple the fermionic interaction [Eq.~\eqref{eq:hint}], introducing bosonic fields $\Delta_{\alpha}\qty(\tau) = \Delta_{0\alpha} + \Delta_{x\alpha}\qty(\tau) - \mathrm{i}\Delta_{y\alpha}\qty(\tau)$ that have the saddle-point contribution $\Delta_{0\alpha}$ and fluctuating real and imaginary parts, $\Delta_{x\alpha}\qty(\tau), \Delta_{y\alpha}\qty(\tau)$. 

After performing the path integration over the fermionic degrees of freedom, we divide the action into the mean-field part and the fluctuation part, where we take the vector potential $\bm{A}$ up to the first order.
Integrating out the fluctuations $\Delta_{\mu\alpha}$ ($\mu=x,y$) and after analytic continuation of the Matsubara frequency $\mathrm{i}\Omega\to\omega + \mathrm{i}0^{+}$, we obtain the fluctuation part of the effective action $S_{\mathrm{FL}}$ as
\begin{align}
&S_{\mathrm{FL}} = \sum_{a,b}\frac{e^{2}}{2}\int\frac{d\omega}{2\pi}A_{a}\qty(\omega)A_{b}\qty(-\omega)\Phi_{ab}\qty(\omega) \notag \\
&+ \sum_{a,b}\frac{e^{2}}{4}\int\frac{d\omega}{2\pi}A_{a}\qty(\omega)A_{b}\qty(-\omega)Q^{\mathrm{T}}_{a}\qty(\omega)U_{\mathrm{eff}}\qty(\omega)Q_{b}\qty(-\omega),
\end{align}
with $a,b=x,y$ in the two-dimensional case.
Here we have introduced the current-current correlation function $\Phi_{ab}\qty(\mathrm{i}\Omega)$:
\begin{align}
\Phi_{ab}\qty(\mathrm{i}\Omega) &= \frac{1}{\beta}\sum_{n}\int\frac{d^{d}\bm{k}}{\qty(2\pi)^{d}}\mathrm{Tr}\qty[ v_{a}G_{0}\qty(\mathrm{i}\omega_{n}+\mathrm{i}\Omega)v_{b}G_{0}\qty(\mathrm{i}\omega_{n}) ] \notag \\
&= \int \frac{d^{d}\bm{k}}{\qty(2\pi)^{d}}\sum_{j,l}\frac{f_{jl} v_{a, jl}v_{b, lj}}{\mathrm{i}\Omega - E_{lj}},
\end{align}
with the velocity operator
\begin{align}
v_{a}\qty(\bm{k}) = \mqty[ \partial_{a}\xi\qty(\bm{k}) & O \\ O & -\partial_{a}\xi^{\mathrm{T}}\qty(-\bm{k}) ]
\end{align}
with $\partial_{a} = \partial/\partial k_{a}$ and Green's function 
\begin{align}
G_{0}\qty(\mathrm{i}\omega) = \qty(\mathrm{i}\omega - H_{\mathrm{BdG}}\qty(\bm{k}) )^{-1}.
\end{align}
$H_{\mathrm{BdG}}\qty(\bm{k})$ is the Bogoliubov-de Gennes Hamiltonian
\begin{align}
H_{\mathrm{BdG}}\qty(\bm{k}) = \mqty[ \xi\qty(\bm{k}) & \Delta_{0} \\ \Delta_{0}^{\dagger} & -\xi^{\mathrm{T}}\qty(-\bm{k}) ], 
\end{align}
where $\Delta_0$ is defined by the saddle-point equation
\begin{align}
\Delta_{0\alpha} &= -U\frac{1}{\beta}\sum_{n}\int\frac{d^{d}\bm{k}}{\qty(2\pi)^{d}}\mathrm{Tr}\qty[ \frac{1}{2}\qty(\tau_{x\alpha}-\mathrm{i}\tau_{y\alpha})G_{0}\qty(\mathrm{i}\omega_{n}) ] \notag \\
&= -U\frac{1}{\beta}\sum_{n}\int\frac{d^{d}\bm{k}}{\qty(2\pi)^{d}}\sum_{j}\frac{1}{2}\frac{\tau_{x\alpha, jj} - \mathrm{i}\tau_{y\alpha, jj}}{\mathrm{i}\omega_{n} - E_{j}} \notag \\
&= -U\int \frac{d^{d}\bm{k}}{\qty(2\pi)^{d}}\sum_{j}\frac{1}{2}\frac{\tau_{x\alpha, jj} - \mathrm{i}\tau_{y\alpha, jj}}{e^{\beta E_{j}} + 1}.
\end{align}
To simplify the notation, we have introduced the band representation $\tau_{\mu\alpha, jl}:= \mel{\varphi_{j}}{\tau_{\mu\alpha}}{\varphi_{l}}$ and $v_{jl}:= \mel{\varphi_{j}}{v}{\varphi_{l}}$ with $j,l=1,2,\cdots 2n$, where $\tau_{x\alpha}$ and $\tau_{y\alpha}$ are the generalized Pauli matrices
\begin{align}
\tau_{x\alpha}=\mqty[ O & A_{\alpha} \\ A_{\alpha} & O], \quad \tau_{y\alpha}=\mqty[ O & -\mathrm{i}A_{\alpha} \\ \mathrm{i}A_{\alpha} & O ], 
\end{align}
where $\qty[A_{\alpha}]_{\gamma\gamma'}=\delta_{\alpha\gamma}\delta_{\gamma\gamma'}$, $\ket{\varphi_{l}}$ is the $l$-th eigenvector of $H_{\mathrm{BdG}}\qty(\bm{k})$ with an eigenvalue $E_{l}$ and $E_{lj}:= E_{l} - E_{j}$.
We have also introduced the polarization bubble $\Pi\qty(\mathrm{i}\Omega)$,
\begin{align}
&\qty[\Pi\qty(\mathrm{i}\Omega)]_{\mu\alpha,\mu'\alpha'} \notag \\
&= \frac{1}{2}\frac{1}{\beta}\sum_{n}\int\frac{d^{d}\bm{k}}{\qty(2\pi)^{d}}\mathrm{Tr}\qty[ \tau_{\mu\alpha}G_{0}\qty(\mathrm{i}\omega_{n}+\mathrm{i}\Omega)\tau_{\mu'\alpha'}G_{0}\qty(\mathrm{i}\omega_{n}) ],\notag \\
&= \frac{1}{2}\int \frac{d^{d}\bm{k}}{\qty(2\pi)^{d}}\sum_{j,l}\frac{f_{jl}\tau_{\mu\alpha, jl}\tau_{\mu'\alpha', lj}}{\mathrm{i}\Omega - E_{lj}},
\end{align}
the effective interaction $U_{\mathrm{eff}}$ within the random phase approximation (RPA),
\begin{align}
U_{\mathrm{eff}} = -U + \qty(-U)\Pi\qty(-U) + \cdots = -\qty(1+U\Pi)^{-1}U, 
\label{U_eff}
\end{align}
and the vector $Q_{a}\qty(\mathrm{i}\Omega)$,
\begin{align}
&\qty[Q_{a}\qty(\mathrm{i}\Omega)]_{\mu\alpha} \notag \\
& = \frac{1}{\beta}\sum_{n}\int\frac{d^{d}\bm{k}}{\qty(2\pi)^{d}}\mathrm{Tr}\qty[ v_{a}G_{0}\qty(\mathrm{i}\omega_{n}+\mathrm{i}\Omega)\tau_{\mu\alpha}G_{0}\qty(\mathrm{i}\omega_{n}) ] \notag \\
&= \int\frac{d^{d}\bm{k}}{\qty(2\pi)^{d}}\sum_{j,l}\frac{f_{jl}v_{a,jl}\tau_{\mu\alpha,lj}}{\mathrm{i}\Omega - E_{lj}}.
\end{align}
Here we introduce $f_{jl}:= f\qty(E_{j})-f\qty(E_{l})$, where $f$ is the Fermi distribution function at zero temperature, i.e., $f=1$ for occupied bands and $f=0$ for unoccupied bands. We can get the real frequency forms of $U_{\mathrm{eff}}$ and $Q_{a}$ by analytic continuation $\mathrm{i}\Omega\to\omega+\mathrm{i}0^{+}$. The diagram for the effective interaction $U_{\mathrm{eff}}\qty(\omega)$ is depicted in Fig.~\ref{Diagrams}(a). We take the functional derivative of $S_{\mathrm{FL}}$ with respect to $A_{b}\qty(-\omega)$ to obtain the linear optical conductivity $\sigma_{ab}\qty(\omega)$ via the current density $j_{b}\qty(\omega)=E_{a}\qty(\omega)\sigma_{ab}\qty(\omega)$ ($E_{a}\qty(\omega)$ is the electric field):
\begin{align}
j_{b}\qty(\omega) &= -\frac{\delta S_{\mathrm{FL}} }{\delta A_{b}\qty(-\omega) } = \sum_{a}E_{a}\qty(\omega)\sigma_{ab}\qty(\omega), 
\end{align}
where 
\begin{align}
\sigma_{ab}\qty(\omega) &= \qty[\sigma_{1}\qty(\omega)]_{ab} + \qty[\sigma_{2}\qty(\omega)]_{ab}, \\ 
\qty[\sigma_{1}\qty(\omega)]_{ab} &= \frac{\mathrm{i}e^{2}}{\omega}\Phi_{ab}\qty(\omega), \\
\qty[\sigma_{2}\qty(\omega)]_{ab} &= \frac{\mathrm{i}e^{2}}{2\omega}Q^{\mathrm{T}}_{a}\qty(\omega)U_{\mathrm{eff}}\qty(\omega)Q_{b}\qty(-\omega), 
\label{eq:optical_conductivity}
\end{align}
and we used $E_{a}\qty(\omega)=\mathrm{i}\omega A_{a}\qty(\omega)$. Here $\sigma_{1}\qty(\omega)$ is responsible for the quasi-particle response, and $\sigma_{2}\qty(\omega)$ for the collective mode one. The diagrams for the linear optical conductivities are shown in Fig.~\ref{Diagrams}(b) and (c). The collective mode response comes from the poles of $U_{\mathrm{eff}}\qty(\omega)$, which satisfy $1+U\Pi\qty(\omega)=0$. For this $\omega$, as pointed out in \cite{Leggett_Kamatani}, $Q_{a}\qty(\omega)$ is off-resonant, or $Q_{a}\qty(\omega)$ does not have a singularity. 

Note that this formalism does not rigorously decompose the fluctuation of the order parameters into amplitude and phase. 
Nevertheless, one can see that the collective modes come from phase fluctuation (the Leggett mode). To see this, we shall write the bosonic field $\Delta_{\alpha}$ as $\qty(\Delta_{0\alpha} + \delta\Delta_{\alpha})e^{\mathrm{i}\theta_{\alpha}}$. Assume that $\delta\Delta_{\alpha}$ and $\theta_{\alpha}$ are small. Then we put
\begin{align}
\qty(\Delta_{0\alpha} + \delta\Delta_{\alpha})e^{\mathrm{i}\theta_{\alpha}} &\approx \qty(\Delta_{0\alpha} + \delta\Delta_{\alpha})\qty( 1 + \mathrm{i}\theta_{\alpha} - \frac{1}{2}\theta_{\alpha}^{2} ) \notag \\
&\approx \Delta_{0\alpha} - \qty( \frac{1}{2}\theta_{\alpha}^{2} - \mathrm{i}\theta_{\alpha})\Delta_{0\alpha}.
\label{FluctuationApprox.}
\end{align}
The first approximation is valid as long as we use the RPA.
In the second approximation we neglect the amplitude fluctuation $\delta\Delta_{\alpha}$.
This approximation is not always valid in general, but we can check that the amplitude fluctuation is small by calculating only the $\tau_{x\alpha}$ part and neglecting the $\tau_{y\alpha}$ in the linear optical conductivity.
In Appendix.~\ref{Apdx.C}, the linear optical conductivity is decomposed into two parts ($\tau_{x}$ and $\tau_{y}$ channels) and we can confirm that the amplitude fluctuation is small.

Moreover the Higgs mode is forbidden in the linear response because the real part of the constant vector $\bm{d}$ responsible for the Higgs mode linear response is zero (see Appendix.~\ref{Apdx.A}). 
We thus focus on the phase fluctuation.
In principle, it is possible to completely decompose the fluctuation into amplitude and phase as is done in \cite{Intro:Leggett_Nonlinear1} for a two-band superconductor.
However, the calculation is complicated and we do not go into details here. 

\subsection{\label{sec:level5-2}Kagome lattice superconductor}
\begin{figure*}[t]
\centering
\includegraphics[width=172mm]{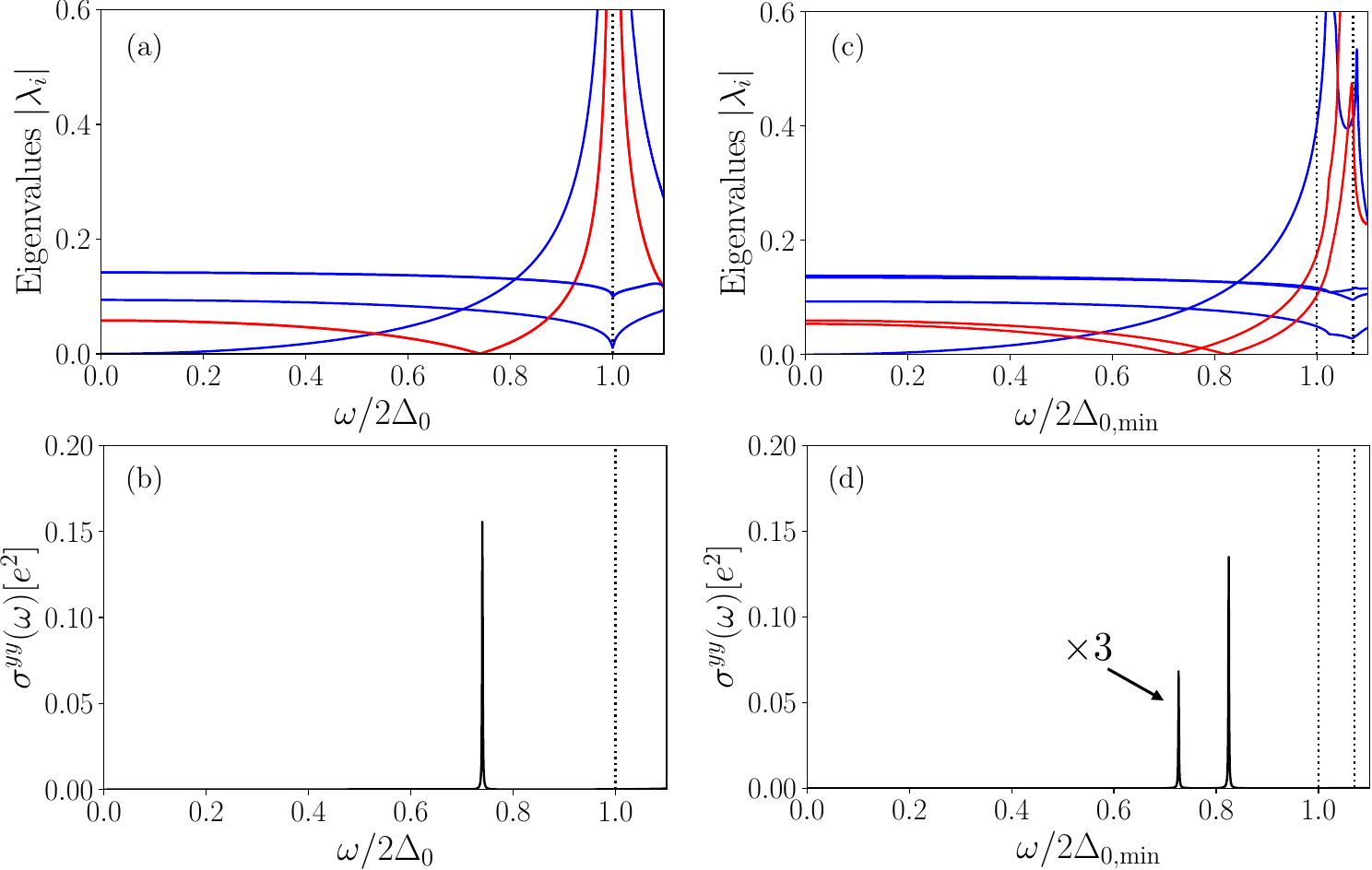}
\caption{Linear optical responses in the Kagome lattice model as a three-band superconductor. 
The parameter values are set to $t=-0.5$, $\delta t=-0.01$, $\mu=0$, $m=0$, and $U=6$. 
(a) Absolute values $\left|\lambda_{i} \right|$ of the eigenvalues $\lambda_{i}$ of $(1+U\Pi\qty(\omega))$ of the model in Fig.~\ref{fig:models}(e).
The doubly degenerate red line corresponds to the Leggett mode.
The matrix $(1+U\Pi\qty(\omega))$ appears in the denominator of the effective interaction $U_{\mathrm{eff}}$ in (\ref{U_eff}).
The zeros of the spectrum correspond to the divergence of the $U_{\mathrm{eff}}$, and hence lead to the signal of the collective modes at a specific energy $\omega$.
(b) The linear optical conductivity of $yy$ component $\sigma^{yy}\qty(\omega)$ with the vertical dotted line describing the gap value $2\Delta_{0}$. 
The calculation is performed for the model in Fig.~\ref{fig:models}(e). 
The ``pole-like" structure at $\omega = 2\Delta_{0}$ is not a pole but a cusp and does not lead to the divergence of $U_{\mathrm{eff}}$,  suggesting the absence of Higgs mode contribution as we expect from the consideration in Sec.~\ref{sec:level2}.
The peak appears at the pole of the effective interaction, indicating that the peak is coming from collective mode.
(c) Absolute values $\left|\lambda_{i} \right|$ of the eigenvalues $\lambda_{i}$ of $(1+U\Pi\qty(\omega))$ of the model in Fig.~\ref{fig:models}(f) that has the on-site potential $m=0.3$. 
Since the gap values are different in the model, the frequency is normalized by twice the minimum gap value $2\Delta_{0,\mathrm{min}}$.
The red lines correspond to the Leggett modes. 
(d) The linear optical conductivity of $yy$ component $\sigma^{yy}\qty(\omega)$ with the vertical dotted line describing gap values. 
The peaks appear at the poles of the effective interaction, indicating that the peaks are coming from collective modes.
Since the collective mode contribution in the linear response is largely controlled by the phase fluctuation, the peaks are suggested to be the Leggett modes.
The peak in (b) splits in two in (d).
In both (b) and (d), the responses are dominated by collective modes.}
\label{fig:Results}
\end{figure*}
As a concrete model for a multiband superconductor showing the linear Leggett mode, we take a Kagome lattice model with two kinds of nearest-neighbor hoppings $(t\pm\delta t)$ and an attractive on-site interaction $U$ shown in Fig.~\ref{fig:models}(e) and (f).
The bold lines show the nearest-neighbor hopping with strength ($t+\delta t$), while the dashed ones with ($t-\delta t$).
The unit cell has three lattice points as represented by the yellow diamond.
We set the lattice constant $a=1$.
The kinetic parts $\xi_{\alpha\alpha'}\qty(\bm{k})$ in the momentum space are given as follows: 
\begin{align}
\xi_{11} &= \xi_{33} = -\mu - m, \quad \xi_{22} = -\mu + m, \notag \\
\xi_{12} &= 2t\cos{\qty(\frac{1}{2}k_{x}-\frac{\sqrt{3}}{2}k_{y} )} - 2\qty(\delta t)\mathrm{i}\sin{\qty(\frac{1}{2}k_{x} - \frac{\sqrt{3}}{2}k_{y}) }, \notag \\
\xi_{13} &= 2t\cos{\qty(\frac{1}{2}k_{x}+\frac{\sqrt{3}}{2}k_{y} )} + 2\qty(\delta t)\mathrm{i}\sin{\qty(\frac{1}{2}k_{x} + \frac{\sqrt{3}}{2}k_{y}) }, \notag \\
\xi_{23} &= 2t\cos{\qty(k_{x})} + 2\qty(\delta t)\mathrm{i}\sin{\qty(k_{x})}, \notag \\
\xi_{21} &= \xi_{12}^{*}, \quad \xi_{31} = \xi_{13}^{*}, \quad \xi_{32} = \xi_{23}^{*},
\end{align}
where we denote the chemical potential by $\mu$ and the on-site potential by $m$.
Refer to Appendix.~\ref{Apdx.D} for the details of derivation. In this model, there is no on-site potential and $m$ is set to be zero. In the model of Fig.~\ref{fig:models}(f), there are two kinds of on-site potentials, one of which is the white square with the potential $m$ and the other one is the black square with the potential $(-m)$. We calculate the poles of the effective interaction $U_{\mathrm{eff}}$ and the linear optical conductivities to see the linear Leggett mode in these models in the next section.

\subsection{\label{sec:level5-3}Results}
Here we shall see the linear response of superconductors with the Lifshitz invariant using the trimerized Kagome model depicted in Fig.~\ref{fig:models}(e).
The classification in Sec.~\ref{sec:level4} has already confirmed from the point of lattice symmetry that these models are allowed to have the Lifshitz invariant and we should analyze how the Lifshitz invariant and the Leggett modes qualitatively contribute to the linear optical response.
We set $t=-0.5$, $\delta t=-0.01$, $\mu=0$, $m=0$ and $U=6$. 
We first see the characteristic frequencies of the Leggett mode, which correspond to the poles of the effective interaction; $1+U\Pi\qty(\omega)=0$. The absolute values of the eigenvalues $\lambda_{i}$ of the inverse matrix of the effective interaction $U_{\mathrm{eff}}^{-1}\qty(\omega)$ are shown in Fig.~\ref{fig:Results}(a). 
Fig.~\ref{fig:Results}(b) represents the linear optical conductivity of $yy$ component $\sigma^{yy}\qty(\omega)$ with the vertical dotted line describing the gap value $2\Delta_{0}$. 
Since there is no on-site potential, all the gap values on each lattice site are the same.
The contributions from the quasi-particles are negligible and the collective mode is dominant.
We would expect six eigenvalues because the matrix $U_{\mathrm{eff}}\qty(\omega)$ is $6\times6$. 
However, only four of them appear in Fig.~\ref{fig:Results}(a). 
The reason is that the two out of six eigenvalues are degenerate.
The red line corresponding to the Leggett mode is doubly degenerate.
The blue line with the biggest value at $\omega=0$ is also doubly degenerate.
The pole $\omega=0$ coincides with the Nambu-Goldstone mode, not contributing to the optical conductivity.
The ``pole-like" structure at $\omega=2\Delta_{0}$ is not an actual pole but a cusp indicating the suppressed contribution of the Higgs mode, which is confirmed in the optical conductivity $\sigma^{yy}(\omega)$ in Fig.~\ref{fig:Results}(b). 
The pole of the effective interaction at $\omega\approx0.75\cdot 2\Delta_{0}$ appears as a peak in the optical conductivity, which comes from the Leggett mode.
As we show in Appendix.~\ref{Apdx.C}, if we neglect the $\tau_{y}$ channel contribution in the calculation of the optical conductivity, the peak at $\hbar\omega\approx 0.75\cdot2\Delta_{0}$ disappears.
On the other hand, if we neglect the $\tau_{x}$ channel, the peak remains with the same peak height.
From these results, we can confirm that the peak in the optical conductivity signals the Leggett mode. 

Next, we consider the trimerized Kagome model with on-site potentials depicted in Fig.~\ref{fig:models}(f) to see the effect of on-site potentials.
The definitions of bold and dashed lines are the same as in Fig.~\ref{fig:models}(e). 
The white squares display the on-site potential $m$, while the black ones show $(-m)$ with $m=0.3$.
The absolute values of the eigenvalues of the inverse matrix of the effective interaction with on-site potentials are shown in Fig.~\ref{fig:Results}(c). 
Fig.~\ref{fig:Results}(d) represents the linear optical conductivity of $yy$ component $\sigma^{yy}\qty(\omega)$ with the vertical dotted lines describing the gap values. 
The collective mode response is dominant as in Fig.~\ref{fig:Results}(b). 
Because of the on-site potentials, one of the three gap values is different from the others, which is minimal. 
We call it $\Delta_{0,\mathrm{min}}$ and normalize the energy $\omega$ by $2\Delta_{0,\mathrm{min}}$. 
The degeneracies of the eigenvalues are also resolved, and all six eigenvalues become non-degenerate.
The eigenvalue that is responsible for the Leggett mode in Fig.~\ref{fig:Results}(b) splits in Fig.~\ref{fig:Results}(d) and we can see two peaks below $2\Delta_{0,\mathrm{min}}$, one of which is multiplied by a factor of three for visibility. 
The physical interpretation of the splitting based on the free energy argument is as follows.
The three order parameters live in the identical Mexican-hat potentials without the on-site potentials because all sites are equivalent and the couplings between the two of them are also the same. 
When we add the on-site potentials, however, one of the three Mexican-hat potentials becomes different from the others in our setting, and the couplings between the order parameters in identical potentials and different potentials are not equal, letting the Leggett mode peak split. 
The splitting originates from the lowering of the symmetry of the system, resolving the degeneracy of the Leggett mode mass. For the cases of $\sigma^{xx}\qty(\omega)$ and $\sigma^{xy}\qty(\omega)$ with the same parameter values, $\sigma^{yy}\qty(\omega)$ for $\delta t=0$, and $\sigma^{yy}\qty(\omega)$ with $\mu=-2t$ (filling the flat band of the Kagome lattice), we refer to Appendix.~\ref{Apdx.E}.

\section{\label{sec:level6}Discussions}
We have studied the Lifshitz invariant in multiband superconductors and its effect on optical conductivities. 
We first used the macroscopic GL theory to see the linear coupling between the phase of the order parameter and the external field, 
interpreted the term to be the Lifshitz invariant,
and classified all the combinations of irreducible representations of order parameters in crystallographic point groups that allow the Lifshitz invariant to appear by the conditions for the free energy to be invariant under symmetry operations.
The Lifshitz invariant in multiband superconductors has been shown to be interpreted as a coupling between the ``internal field" and the ``current" of the overlaps of the order parameters, which is controlled by the lattice geometry. Because of the ``internal field", it was possible for the phase of the order parameter to linearly connect to the external field. 

We also showed that the wide range of multiband superconductors can have the Lifshitz invariant according to the group theory.
The reason that there has not been experimental detection of the Leggett mode in the linear response so far may be that the constant vector $\bm{d}$ in the GL theory would be practically small in many systems.
Another possible explanation is that impurities in the real materials would suppress the Leggett mode in the linear response, whose effect has been neglected in our clean limit model.
In previous papers~\cite{Intro:Higgs_Leggett2,Intro:Higgs_Leggett3}, the signals of the Leggett mode in a nonlinear response is relatively suppressed by the effect of nonmagnetic impurities as compared to the Higgs-mode and quasiparticle contributions.
It is thus interesting to study the impurity effects in the optical conductivity in the presence of the Lifshitz invariant, which we leave as a future problem.
About the high harmonic generations, the signals of the Leggett mode were reported to be quite smaller than those of  the Higgs mode, and they were hardly affected by nonmagnetic impurities.
The similar results are expected in the linear response regime, though the impurity can disturb the coherence between the two phases contributing to the Leggett mode.
Thus we leave this issue for future research. 

The condition for the Lifshitz invariant to appear is whether the system follows the nontrivial representation due to the sublattice geometry.
As we saw in Sec.~\ref{sec:level4}, the inversion symmetry itself is neither a necessary nor sufficient condition, and both of the cases with and without the inversion symmetry can have the Lifshitz invariant, which is different from the previous studies where the Lifshitz invariant had commonly been related to the broken inversion symmetry~\cite{II:LI_DM_1st,II:LI_DM_2nd,II:LI_LiquidCrystal,II:LI_NCSC_1st,II:LI_NCSC_2nd,II:LI_NCSC_3rd}.
The multiband nature, or the sublattice geometry plays a quite important role for the system to have the Lifshitz invariant.
Since the Lifshitz invariant has the first-order spatial derivative (of the linear $\bm{q}$ term in momentum space), we may expect an instability toward non-uniform spatial modulation of the order parameter with finite $\bm{q}$.
As already stated in Sec.~\ref{sec:level2-1}, however, the instability occurs only when $\left|\bm{d}\right|$ is large enough.
The detailed analysis about the effect of the vector $\bm{d}$ to the ground state is given in Appendix.~\ref{Apdx.B}. 

We would like to comment on open issues about the treatment of the order parameters in the group theoretical argument. 
We implicitly assumed that the order parameters are defined on each lattice point.
This assumption seems to be well justified because the order parameters reflect the symmetry of the system even though the size of Cooper pairs is much larger than the lattice constant.
Nevertheless, this situation may not be valid when we consider the retardation effect of phonons seriously.
We also assumed that sublattice and other degrees of freedom form a direct product in the order parameter representation.
If the system has a symmetry that intertwines these degrees of freedom (which cannot be represented by the direct product), there will be other interesting situations that are not studied in the present work.

We additionally constructed the microscopic three-band superconducting models based on the Kagome lattice to see the linear Leggett mode in the optical conductivity.
The degeneracy of the Leggett mode was resolved by adding the on-site potentials and reducing the symmetry.

Finally, we list possible experimental observations of the linear Leggett mode.
One possible candidate is $\mathrm{CsV}_{3}\mathrm{Sb}_{5}$~\cite{Kagome_StarOfDavid,Intro:Kagome_SC_review,Kagome_Possible_CDW,Kagome_Possible_CDW_3Q}. 
This material is reported to have a CDW phase above the superconducting transition temperature $T_{c}$, and the phase is thought to coexist with superconductivity below $T_{c}$. 
The superconducting pairing symmetry is predicted to be $s$-wave~\cite{Kagome_s_wave} and anisotropic~\cite{Kagome_Anisotropic_s_wave}.
The CDW would be responsible for the lattice modulation, causing the two different hopping strengths as in our model of Fig.~\ref{fig:models}(e) and (f). 
It has also been reported that the material does not break the time-reversal symmetry in the CDW phase for a high-quality sample~\cite{Kagome_Preserved_TRS}, which is in accordance with our model preserving the time-reversal symmetry. 
The experiment of the optical Kerr effect has also concluded that it is highly unlikely that the material breaks the time-reversal symmetry~\cite{Kagome_NoTRSB}.
Although our model for the numerical calculation does not completely reproduce the CDW pattern or its modulation, the model could indicate the possible experimental confirmation of the Leggett mode in the linear optical conductivity since our model and the actual material (simplified as in Fig.~\ref{fig:models}(g)) can have the Lifshitz invariant and both would show the similar property arising from the Lifshitz invariant. 
Hence, by measuring the linear optical conductivity in the superconducting phase of $\mathrm{CsV}_{3}\mathrm{Sb}_{5}$, there would be at least one peak coming from the Leggett mode and we could obtain information about the phase difference between order parameters.

\section{\label{sec:level7} Acknowledgement}
We are indebted to E. K\"{o}nig, A. Schnyder, N. Heinsdorf, S. Klein, T. Ishii, K. Takasan, T. Morimoto, S. Kitamura, M. Sigrist, R. Shimano, H. Watanabe and S. Onari for helpful comments and discussions. We thank the Max Planck-UBC-UTokyo Center for Quantum Materials for valuable collaborations and financial support. R.N. acknowledges the hospitality of the Max Planck Institute for Solid State Research. N.T. acknowledges support by JSPS KAKENHI (Grant No. JP20K03811) and JST FOREST (Grant No. JPMJFR2131).

\onecolumngrid
\appendix
\section{Microscopic calculation for the free energy and the linear optical conductivities}
\label{Apdx.A}
In this appendix, we show the microscopic derivation of the formalism of linear optical conductivity. 
We also microscopically derive the GL free energy and prove that the constant vector $\bm{d}$ is purely imaginary, which suppresses the Higgs mode in the linear response regime. 
Besides, we confirm that $\bm{d}$ vanishes when the normal part (kinetic part) of the BdG Hamiltonian $\xi\qty(\bm{k})$ is real, or the differences of the hopping strengths are zero.\\

We get started from the Hamiltonian without the electromagnetic field below.
\begin{align}
\mathcal{H} &= \mathcal{H}_{0} + \mathcal{H}_{\mathrm{int}}, \\
\mathcal{H}_{0} &= \sum_{\bm{k}\sigma\alpha\alpha'}\xi_{\alpha\alpha'}\qty(\bm{k})c^{\dagger}_{\bm{k}\alpha\sigma}c_{\bm{k}\alpha'\sigma}, \\
\mathcal{H}_{\mathrm{int}} &= -U\sum_{\bm{k}\bm{k}'\alpha}c^{\dagger}_{\bm{k}\alpha\up}c^{\dagger}_{-\bm{k}\alpha\down}c_{-\bm{k}'\alpha\down}c_{\bm{k}'\alpha\up}.
\end{align}
Here we have already assumed that the two-body interaction occurs in the Cooper pairing, $\qty(\bm{k}, \up)$ and $\qty(-\bm{k}, \down)$. To take into account the electromagnetic field, we use the \textit{velocity gauge}:
\begin{align}
\mathcal{H}\qty(\bm{k}) \to \mathcal{H}\qty(\bm{k}-e\bm{A}).
\end{align}
We shall expand the Hamiltonian for small $\bm{A}$. The kinetic term generates the electromagnetic term $\mathcal{H}_{\mathrm{EM}}$, and the interaction term remains unchanged.
\begin{align}
\mathcal{H}_{\mathrm{EM}} = -e\sum_{\bm{k}\sigma\alpha\alpha'}\sum_{a}\qty[ \partial_{a}\xi_{\alpha\alpha'}\qty(\bm{k}) ]A_{a}c^{\dagger}_{\bm{k}\alpha\sigma}c_{\bm{k}\alpha'\sigma} + O\qty(\bm{A}^{2}),
\end{align}
where $\partial_{a} := \partial/\partial k_{a}$. Here, we focus on the linear response. This means that we should focus on the linear $\bm{A}$ term. From now we neglect the $O(\bm{A}^{2})$ term.\\

Now let us use the path integral formulation to get an effective action $S$. We first consider the finite temperature case. The partition function $\mathcal{Z}$ is written as
\begin{align}
\mathcal{Z} &= \int \mathcal{D}\qty(c^{\dagger},c)e^{-S\qty[ c^{\dagger}, c]}, \quad \text{and} \quad S\qty[ c^{\dagger}, c] = \int_{0}^{\beta}d\tau \qty( \sum_{\bm{k}\sigma\alpha}c^{\dagger}_{\bm{k}\alpha\sigma}\partial_{\tau}c_{\bm{k}\alpha\sigma} + \mathcal{H}).
\end{align}
For simplicity, we let the volume size of the whole system be one. Hubbard-Stratonovich transformation puts the interaction part of the Hamiltonian into the coupling between the auxiliary boson field $\Delta$ and $\Delta^{*}$, and fermion (electron here).
\begin{align}
&\exp\qty( -U\sum_{\alpha}\sum_{\bm{k}\bm{k}'} c^{\dagger}_{\bm{k}\alpha\up}c^{\dagger}_{-\bm{k}\alpha\down}c_{-\bm{k}'\alpha\down}c_{\bm{k}'\alpha\up} ) \notag \\
&=\int \mathcal{D}\qty( \Delta^{*}, \Delta) \exp\qty( \sum_{\alpha}\qty[ \frac{\left| \Delta_{\alpha}\right|^{2}}{U} - \sum_{\bm{k}}\qty( \Delta^{*}_{\alpha}c_{\bm{k}\alpha\up}c_{-\bm{k}\alpha\down} + \Delta_{\alpha}c^{\dagger}_{-\bm{k}\alpha\down}c^{\dagger}_{\bm{k}\alpha\up} ) ] ).
\end{align}
We introduce the Nambu basis $\Psi^{\dagger}_{\bm{k}}$ and $\Psi_{\bm{k}}$ to express the action in a concise way:
\begin{align}
\Psi^{\dagger}_{\bm{k}} = \qty[ c^{\dagger}_{\bm{k},1\up} \cdots\quad c^{\dagger}_{\bm{k}, n\up}\quad c_{-\bm{k},1\down} \cdots\quad c_{-\bm{k}, n\down} ].
\end{align}
Note that we implicitly assumed that the creation/annihilation operators depend on the imaginary time $\tau$. Then the action $S\qty[ c^{\dagger}, c, \Delta^{*}, \Delta]$ can be written as
\begin{align}
S\qty[ c^{\dagger}, c, \Delta^{*}, \Delta] &= \int_{0}^{\beta}d\tau \qty( \sum_{\bm{k}\sigma\alpha}c^{\dagger}_{\bm{k}\alpha\sigma}\partial_{\tau}c_{\bm{k}\alpha\sigma} + \sum_{\bm{k}\sigma\alpha\alpha'}\xi_{\alpha\alpha'}\qty(\bm{k})c^{\dagger}_{\bm{k}\alpha\sigma}c_{\bm{k}\alpha'\sigma} - U\sum_{\alpha}\sum_{\bm{k}\bm{k}'}c^{\dagger}_{\bm{k}\alpha\up}c^{\dagger}_{-\bm{k}\alpha\down}c_{-\bm{k}'\alpha\down}c_{\bm{k}'\alpha\up} ) \notag \\
& \quad +\int_{0}^{\beta}d\tau \qty(-e)\sum_{a}\sum_{\bm{k}\sigma\alpha\alpha'}\qty( \partial_{a}\xi_{\alpha\alpha'}\qty(\bm{k}) )A_{a}c^{\dagger}_{\bm{k}\alpha\sigma}c_{\bm{k}\alpha'\sigma}\notag \\
&= \int_{0}^{\beta}d\tau\sum_{\alpha}\Delta^{*}_{\alpha}\frac{1}{U}\Delta_{\alpha} \notag \\
& \quad + \sum_{\bm{k}\alpha}\int_{0}^{\beta}d\tau \qty( \sum_{\sigma\alpha'}c^{\dagger}_{\bm{k}\alpha\sigma}\qty( \delta_{\alpha\alpha'}\partial_{\tau} + \xi_{\alpha\alpha'}\qty(\bm{k}) )c_{\bm{k}\alpha'\sigma} - \Delta^{*}_{\alpha}c_{\bm{k}\alpha\up}c_{-\bm{k}\alpha\down} - \Delta_{\alpha}c^{\dagger}_{-\bm{k}\alpha\down}c^{\dagger}_{\bm{k}\alpha\up}) \notag \\
&\quad + \int_{0}^{\beta}d\tau\qty(-e)\sum_{a}\sum_{\bm{k}\sigma\alpha\alpha'}\qty( \partial_{a}\xi_{\alpha\alpha'}\qty(\bm{k}) )A_{a}c^{\dagger}_{\bm{k}\alpha\sigma}c_{\bm{k}\alpha'\sigma} \notag \\
&= \int_{0}^{\beta}d\tau\sum_{\alpha}\Delta^{*}_{\alpha}\qty(\tau)\frac{1}{U}\Delta_{\alpha}\qty(\tau) - \int_{0}^{\beta}d\tau \sum_{\bm{k}}\Psi^{\dagger}_{\bm{k}}\qty(\tau)G^{-1}\qty(\bm{k},\tau)\Psi_{\bm{k}}\qty(\tau),
\end{align}
where
\begin{align}
G^{-1}\qty(\bm{k}, \tau) &= \mqty[ -I_{n}\partial_{\tau} - \xi\qty(\bm{k}) & -\Delta\qty(\tau) \\ -\Delta^{*}\qty(\tau) & -I_{n}\partial_{\tau} + \xi^{\mathrm{T}}\qty(-\bm{k}) ] + \sum_{a}e A_{a}\qty(\tau)\mqty[ \partial_{a}\xi\qty(\bm{k}) & O \\ O & -\partial_{a}\xi^{\mathrm{T}}\qty(-\bm{k}) ], \notag \\
&= \mqty[ -I_{n}\partial_{\tau} - \xi\qty(\bm{k}) & -\Delta_{0} \\ -\Delta^{*}_{0} & -I_{n}\partial_{\tau} + \xi^{\mathrm{T}}\qty(-\bm{k}) ] + \mqty[ O & -\delta\Delta\qty(\tau) \\ -\delta\Delta^{*}\qty(\tau) & O ] + \sum_{a}e A_{a}\qty(\tau)\mqty[ \partial_{a}\xi\qty(\bm{k}) & O \\ O & \partial_{a}\xi\qty(\bm{k}) ],
\end{align}
is the inverse Green function with $I_{n}$ being the $n\times n$ unit matrix.
For simplicity we abbreviate $I_{n}$ from now.
We used the time-reversal symmetry in the second equality, $\qty[\Delta\qty(\tau)]_{\alpha\alpha'}=\Delta_{\alpha}\qty(\tau)\delta_{\alpha\alpha'}$, $\Delta_{\alpha}\qty(\tau)=\Delta_{0,\alpha} + \delta\Delta_{\alpha}\qty(\tau)$, and $\Delta_{0, \alpha}$ is the gap value at the saddle point (corresponding to the mean-field value).
Here we used the following relationship:
\begin{align}
c^{\dagger}_{\bm{k}\alpha\sigma}\qty(\tau)\partial_{\tau}c_{\bm{k}\alpha\sigma}\qty(\tau) &= \frac{1}{\beta}\sum_{m,n}\qty[ c^{\dagger}_{\bm{k}\alpha\sigma}\qty(\omega_{n}) e^{\mathrm{i}\omega_{n}\tau} ]\partial_{\tau}\qty[ c_{\bm{k}\alpha\sigma}\qty(\omega_{m})e^{-\mathrm{i}\omega_{m}\tau}] \notag \\
&= \frac{1}{\beta}\sum_{m,n}\qty( -\mathrm{i}\omega_{m})c^{\dagger}_{\bm{k}\alpha\sigma}\qty(\omega_{n})c_{\bm{k}\alpha\sigma}\qty(\omega_{m})e^{\mathrm{i}\qty(\omega_{n}-\omega_{m})\tau} \notag \\
&= \frac{1}{\beta}\sum_{m,n}\qty( -\mathrm{i}\omega_{m}) \qty[ -c_{\bm{k}\alpha\sigma}\qty(\omega_{m})c^{\dagger}_{\bm{k}\alpha\sigma}\qty(\omega_{n}) + \delta_{mn} ]e^{\mathrm{i}\qty(\omega_{n}-\omega_{m})\tau} \notag \\
&= -\qty[\partial_{\tau}c_{\bm{k}\alpha\sigma}\qty(\tau)]c^{\dagger}_{\bm{k}\alpha\sigma}\qty(\tau),
\end{align}
and
\begin{align}
\int_{0}^{\beta}d\tau \qty( -\qty[\partial_{\tau}c_{\bm{k}\alpha\sigma}\qty(\tau)]c^{\dagger}_{\bm{k}\alpha\sigma}\qty(\tau) ) &= \qty[ -c_{\bm{k}\alpha\sigma}\qty(\tau)c^{\dagger}_{\bm{k}\alpha\sigma}\qty(\tau) ]_{0}^{\beta} + \int_{0}^{\beta}d\tau c_{\bm{k}\alpha\sigma}\qty(\tau)\partial_{\tau}c^{\dagger}_{\bm{k}\alpha\sigma}\qty(\tau) \notag \\
&= \int_{0}^{\beta}d\tau c_{\bm{k}\alpha\sigma}\qty(\tau)\partial_{\tau}c^{\dagger}_{\bm{k}\alpha\sigma}\qty(\tau),
\end{align}
where we used the periodicity of the operators.
Then we move to the Fourier space by
\begin{align}
\Psi_{\bm{k}}\qty(\tau) = \frac{1}{\beta}\sum_{n}\Psi_{\bm{k}}\qty(\omega_{n})e^{-\mathrm{i}\omega_{n}\tau}.
\end{align}
This puts the imaginary time derivative $\partial_{\tau}$ into $-\mathrm{i}\omega_{n}$, and we obtain the action expressed in Fourier space.
\begin{align}
&S[c^{\dagger}, c, \Delta^{*}, \Delta] \notag \\
&\quad = \beta\sum_{\alpha}\frac{\left|\Delta_{0,\alpha}\right|^{2}}{U} + \sum_{\alpha}\sum_{n}\delta\Delta^{*}_{\alpha}\qty(\mathrm{i}\omega_{n})\frac{1}{U}\delta\Delta_{\alpha}\qty(\mathrm{i}\omega_{n})- \frac{1}{\beta^{2}}\sum_{\bm{k}}\sum_{m,n}\Psi^{\dagger}\qty(\mathrm{i}\omega_{m}) G^{-1}\qty(\mathrm{i}\omega_{m},\mathrm{i}\omega_{n}; \bm{k}) \Psi\qty(\mathrm{i}\omega_{n}), 
\end{align}
where
\begin{align}
G^{-1}\qty(\mathrm{i}\omega_{m},\mathrm{i}\omega_{n}; \bm{k}) &\quad= \mqty[ \mathrm{i}\omega_{n} - \xi\qty(\bm{k}) & -\Delta \\ -\Delta^{*} & \mathrm{i}\omega_{n} + \xi^{\mathrm{T}}\qty(-\bm{k}) ]\beta\delta_{mn} + \sum_{a}e A_{a}\qty(\mathrm{i}\omega_{m} - \mathrm{i}\omega_{n})\mqty[ \partial_{a}\xi\qty(\bm{k}) & O \\ O & \partial_{a}\xi\qty(\bm{k}) ]\beta \notag \\
&\quad= \mqty[ \mathrm{i}\omega_{n} - \xi\qty(\bm{k}) & -\Delta_{0} \\ -\Delta^{*}_{0} & \mathrm{i}\omega_{n} + \xi^{\mathrm{T}}\qty(-\bm{k}) ]\beta\delta_{mn} \notag \\
&\quad \quad+ \mqty[ O & -\delta\Delta\qty(\mathrm{i}\omega_{m} - \mathrm{i}\omega_{n}) \\ -\delta\Delta^{*}\qty(\mathrm{i}\omega_{m} - \mathrm{i}\omega_{n}) & O ]\beta + \sum_{a}e A_{a}\qty(\mathrm{i}\omega_{m} - \mathrm{i}\omega_{n})\mqty[ \partial_{a}\xi\qty(\bm{k}) & O \\ O & \partial_{a}\xi\qty(\bm{k}) ]\beta,
\end{align}
is inverse Green's function in Fourier space.
Here, we define $\delta\Delta\qty(\mathrm{i}\omega_{m}-\mathrm{i}\omega_{n})$ and $A_{a}\qty(\mathrm{i}\omega_{m}-\mathrm{i}\omega_{n})$ by
\begin{align}
\delta\Delta\qty(\mathrm{i}\omega_{m}-\mathrm{i}\omega_{n}) = \frac{1}{\beta}\int_{0}^{\beta}d\tau \delta\Delta\qty(\tau)e^{\mathrm{i}\qty(\omega_{m}-\omega_{n})\tau} \quad \text{and} \quad A_{a}\qty(\mathrm{i}\omega_{m}-\mathrm{i}\omega_{n}) = \frac{1}{\beta}\int_{0}^{\beta}d\tau A_{a}(\tau)e^{\mathrm{i}\qty(\omega_{m}-\omega_{n})\tau},
\end{align}
so that both $\delta\Delta$ and $A_{a}$ have the same dimension before and after the transformation.
We omit the $\delta\Delta$ terms in the first order because they offset with unimportant terms (the $L=1$ term in the expression below). 
Performing the fermionic path integral yields the action $S\qty[ \Delta^{*}, \Delta]$:
\begin{align}
S\qty[ \Delta^{*}, \Delta] = \beta\sum_{\alpha}\frac{\left|\Delta_{0,\alpha}\right|^{2}}{U} + \sum_{\alpha}\sum_{m}\delta\Delta_{\alpha}^{*}\qty(\mathrm{i}\omega_{m})\frac{1}{U}\delta\Delta_{\alpha}\qty(\mathrm{i}\omega_{m}) - \frac{1}{\beta^{2}}\sum_{\bm{k}}\sum_{n,m}\mathrm{Tr}\ln{\qty[ -G^{-1}\qty(\mathrm{i}\omega_{m},\mathrm{i}\omega_{n}; \bm{k}) ]}.
\end{align}
Choosing the reference state, and corresponding Green's function $G_{0}$ and the self-energy $\Sigma$ to be $G^{-1} = G^{-1}_{0} - \Sigma$, we rewrite the trace term as
\begin{align}
\mathrm{Tr}\ln{\qty[ -G^{-1} ] } &= \mathrm{Tr}\ln{ \qty[ -\qty(G^{-1}_{0} - \Sigma) ] } \notag \\
&= \mathrm{Tr}\ln{ \qty[ -G^{-1}_{0}\qty( 1 - G_{0}\Sigma) ] } \notag \\
&= \mathrm{Tr}\ln{ \qty[ -G^{-1}_{0} ]}  + \mathrm{Tr}\ln{ \qty[ 1- G_{0}\Sigma ] } \notag \\
&= \mathrm{Tr}\ln{ \qty[ -G^{-1}_{0} ]}  - \mathrm{Tr}\qty[ \sum_{L=1}^{\infty} \frac{\qty(G_{0}\Sigma)^{L}}{L} ].
\end{align}
There are two important ways to choose the reference state. One is normal and the other is the superconducting ground state.\\

\textit{The Ginzburg-Landau effective action}. We first derive the Ginzburg-Landau effective action $S_{\mathrm{eff}}$ in the equilibrium superconducting state. We first decompose Green's function into two parts; reference state Green's function and the self-energy. We neglect the electromagnetic parts to obtain the action in an equilibrium state.
\begin{align}
G^{-1}\qty(\mathrm{i}\omega_{m}, \mathrm{i}\omega_{n}; \bm{k}) &= \mqty[ \mathrm{i}\omega_{n} - \xi\qty(\bm{k}) & -\Delta \\ -\Delta^{*} & \mathrm{i}\omega_{n} + \xi^{\mathrm{T}}\qty(-\bm{k}) ]\beta\delta_{mn} \notag \\
&= \mqty[ \mathrm{i}\omega_{n} - \xi\qty(\bm{k}) & O \\ O & \mathrm{i}\omega_{n} + \xi^{\mathrm{T}}\qty(-\bm{k}) ]\beta\delta_{mn} - \mqty[ O & \Delta \\ \Delta^{*} & O ]\beta\delta_{mn} \notag \\
&= G^{-1}_{0}\qty(\mathrm{i}\omega_{m}, \mathrm{i}\omega_{n}; \bm{k}) - \Sigma.
\end{align}
We are interested in the linear $\bm{d}$ term, or the term with $\Delta^{*}$ and $\Delta$. Hence we focus on the term $L=2$:
\begin{align}
\mathrm{Tr}\qty[\frac{1}{2}G_{0}\qty(\mathrm{i}\omega_{m}, \mathrm{i}\omega_{n}; \bm{k}+\bm{q})\Sigma G_{0}\qty(\mathrm{i}\omega_{m}, \mathrm{i}\omega_{n}; \bm{k})\Sigma].
\end{align}
We shall consider the simpler form of Green's function. By defining the normal state Green function $g\qty(\mathrm{i}\omega_{n}; \bm{k}):= \qty( \mathrm{i}\omega_{n} - \xi\qty(\bm{k}) )^{-1}$, we can write the reference state Green function in the form of
\begin{align}
G_{0}\qty(\mathrm{i}\omega_{m}, \mathrm{i}\omega_{n}; \bm{k}) = \mqty[ g\qty(\mathrm{i}\omega_{n}; \bm{k}) & O \\ O & - g^{\mathrm{T}}\qty(-\mathrm{i}\omega_{n}; -\bm{k} ) ]\frac{1}{\beta}\delta_{mn}.
\end{align}
Then we put
\begin{align}
&G_{0}\qty(\mathrm{i}\omega_{m}, \mathrm{i}\omega_{n}; \bm{k}+\bm{q}) \Sigma G_{0}\qty(\mathrm{i}\omega_{m}, \mathrm{i}\omega_{n}; \bm{k}) \Sigma \notag \\
&= \mqty[ -g\qty(\mathrm{i}\omega_{n}; \bm{k}+\bm{q})\Delta g^{\mathrm{T}}\qty(-\mathrm{i}\omega_{n}; -\bm{k})\Delta^{*} & O \\ O & -g^{\mathrm{T}}\qty(-\mathrm{i}\omega_{n}; -\bm{k}-\bm{q})\Delta^{*}g\qty(\mathrm{i}\omega_{n}; \bm{k})\Delta ]\delta_{mn},
\end{align}
yielding
\begin{align}
&-\mathrm{Tr}\qty[ \frac{1}{2}G_{0}\qty(\mathrm{i}\omega_{m}, \mathrm{i}\omega_{n}; \bm{k}+\bm{q})\Sigma G_{0}\qty(\mathrm{i}\omega_{m}, \mathrm{i}\omega_{n}; \bm{k})\Sigma] \notag \\
&= \frac{1}{2}\sum_{\bm{k}}\sum_{\bm{q}}\sum_{n}\mathrm{Tr}\mqty[ g\qty(\mathrm{i}\omega_{n}; \bm{k}+\bm{q})\Delta g^{\mathrm{T}}\qty(-\mathrm{i}\omega_{n}; -\bm{k})\Delta^{*} & O \\ O & g^{\mathrm{T}}\qty(-\mathrm{i}\omega_{n}; -\bm{k}-\bm{q})\Delta^{*}g\qty(\mathrm{i}\omega_{n}; \bm{k})\Delta ] \notag \\
&= \frac{1}{2}\sum_{\bm{k}}\sum_{\bm{q}}\sum_{n}\sum_{\alpha\alpha'}\qty[ \Delta^{*}_{\alpha}g_{\alpha\alpha'}\qty(-\mathrm{i}\omega_{n}; -\bm{k})g_{\alpha\alpha'}\qty(\mathrm{i}\omega_{n}; \bm{k}+\bm{q})\Delta_{\alpha'} + \Delta_{\alpha}g_{\alpha'\alpha}\qty(\mathrm{i}\omega_{n}; \bm{k})g_{\alpha'\alpha}\qty(-\mathrm{i}\omega_{n}; -\bm{k}-\bm{q})\Delta^{*}_{\alpha'} ].
\end{align}
At last, we get the effective action $S_{\mathrm{eff}}$:
\begin{align}
S_{\mathrm{eff}} = \beta\sum_{\alpha}\frac{\left|\Delta_{\alpha}\right|^{2}}{U} - \beta\sum_{\alpha,\alpha'}\sum_{\bm{q}}\qty[ \Delta_{\alpha}^{*}F^{(1)}_{\alpha\alpha'}\qty(\bm{q})\Delta_{\alpha'} + \Delta_{\alpha}F^{(2)}_{\alpha\alpha'}\qty(\bm{q})\Delta^{*}_{\alpha'} ], 
\label{eqn:EffectiveAction}
\end{align}
with
\begin{align}
F^{(1)}_{\alpha\alpha'}\qty(\bm{q}) &= \frac{1}{2\beta}\sum_{n}\sum_{\bm{k}}g_{\alpha\alpha'}\qty(-\mathrm{i}\omega_{n}; -\bm{k})g_{\alpha\alpha'}\qty(\mathrm{i}\omega_{n}; \bm{k}+\bm{q}), \\
F^{(2)}_{\alpha\alpha'}\qty(\bm{q}) &= \frac{1}{2\beta}\sum_{n}\sum_{\bm{k}} g_{\alpha'\alpha}\qty(\mathrm{i}\omega_{n}; \bm{k})g_{\alpha'\alpha}\qty(-\mathrm{i}\omega_{n}; -\bm{k}-\bm{q}). 
\end{align}
We use the property of the normal state Green function $g\qty(\mathrm{i}\omega; \bm{k})$ to derive the Lifshitz invariant term in the free energy in ~\eqref{eqn:fe0}. Since the kinetic part $\xi\qty(\bm{k})$ is Hermite, the component of the transposed normal state Green function $g_{\alpha'\alpha}\qty(\mathrm{i}\omega_{n}; \bm{k})$ is written as
\begin{align}
g_{\alpha'\alpha}\qty(\mathrm{i}\omega_{n}; \bm{k}) &= \qty[\qty( \mathrm{i}\omega_{n} - \xi^{*}\qty(\bm{k}) )^{-1}]_{\alpha\alpha'} \notag \\
&= \qty[ \qty( -\mathrm{i}\omega_{n} - \xi\qty(\bm{k}) )^{-1}]^{*}_{\alpha\alpha'} \notag \\
&= g^{*}_{\alpha\alpha'}\qty(-\mathrm{i}\omega_{n}; \bm{k}).
\label{g:property1}
\end{align}
Using this property and defining the vector $\bm{d}_{\alpha\alpha'}$ as
\begin{align}
\bm{d}_{\alpha\alpha'} := \frac{1}{2\beta}\sum_{n}\sum_{\bm{k}}g_{\alpha\alpha'}\qty(-\mathrm{i}\omega_{n}; -\bm{k})\nabla_{\bm{k}}g_{\alpha\alpha'}\qty(\mathrm{i}\omega_{n}; \bm{k}),
\label{Apdx_Def_vectorD}
\end{align}
we expand the second term of ~\eqref{eqn:EffectiveAction} with respect to $\bm{k}$.
\begin{align}
F^{\qty(1)}_{\alpha\alpha'}\qty(\bm{q}) &= \frac{1}{2\beta}\sum_{n}\sum_{\bm{k}}g_{\alpha\alpha'}\qty(-\mathrm{i}\omega_{n}; -\bm{k})\qty[ g_{\alpha\alpha'}\qty(\mathrm{i}\omega_{n}; \bm{k}) + \qty(\nabla_{\bm{k}}g_{\alpha\alpha'}\qty(\mathrm{i}\omega_{n}; \bm{k}))\cdot\bm{q} + O\qty(\bm{q}^{2})], \\
F^{\qty(2)}_{\alpha\alpha'}\qty(\bm{q}) &= \frac{1}{2\beta}\sum_{n}\sum_{\bm{k}}g_{\alpha'\alpha}\qty(\mathrm{i}\omega_{n}; \bm{k})\qty[ g_{\alpha'\alpha}\qty(-\mathrm{i}\omega_{n}; -\bm{k}) + \qty(\nabla_{-\bm{k}}g_{\alpha'\alpha}\qty(-\mathrm{i}\omega_{n}; -\bm{k}))\cdot\qty(-\bm{q}) + O\qty(\bm{q}^{2})] \notag \\
&= \frac{1}{2\beta}\sum_{n}\sum_{\bm{k}}g_{\alpha'\alpha}\qty(-\mathrm{i}\omega_{n}; -\bm{k})\qty[ g^{*}_{\alpha\alpha'}\qty(\mathrm{i}\omega_{n}; \bm{k}) - \qty( \nabla_{\bm{k}}g^{*}_{\alpha\alpha'}\qty(\mathrm{i}\omega_{n}; \bm{k}) ) \cdot\bm{q} + O\qty(\bm{q}^{2})]. 
\end{align}
We pick up the linear $\bm{q}$ terms to see the Lifshitz invariant terms, paying attention to the fact that $\bm{q}$ is treated as a canonical momentum in coordinate space. Here we restrict ourselves to the two-band case $(\alpha,\alpha'=1,2)$. In this case the vectors we have to consider are $\bm{d} = \bm{d}_{12}$ and $\bm{d}^{*} = \bm{d}_{21}$. Taking this into account, $F^{\qty(1)}_{\alpha\alpha'}$ yields
\begin{align}
&\Delta_{1}^{*}\frac{1}{2\beta}\sum_{n}\sum_{\bm{k}}g_{12}\qty(-\mathrm{i}\omega_{n};-\bm{k})\qty[\nabla_{\bm{k}}g_{12}\qty(\mathrm{i}\omega_{n};\bm{k})]\cdot\bm{q}\Delta_{2} + \Delta^{*}_{2}\frac{1}{2\beta}\sum_{n}\sum_{\bm{k}}g_{21}\qty(-\mathrm{i}\omega_{n};-\bm{k})\qty[\nabla_{\bm{k}}g_{21}\qty(\mathrm{i}\omega_{n}; \bm{k})]\cdot\bm{q}\Delta_{1} \notag \\
&= \Delta_{1}^{*}\frac{1}{2\beta}\sum_{n}\sum_{\bm{k}}g_{12}\qty(-\mathrm{i}\omega_{n};-\bm{k})\qty[\nabla_{\bm{k}}g_{12}\qty(\mathrm{i}\omega_{n};\bm{k})]\cdot\bm{q}\Delta_{2} + \Delta_{2}^{*}\frac{1}{2\beta}\sum_{n}\sum_{\bm{k}}g_{12}^{*}\qty(-\mathrm{i}\omega_{n};-\bm{k})\qty[\nabla_{\bm{k}}g^{*}_{12}\qty(\mathrm{i}\omega_{n};\bm{k})]\cdot\bm{q}\Delta_{1} \notag \\
&= \Delta^{*}_{1}\bm{d}\cdot\bm{q}\Delta_{2} + \Delta^{*}_{2}\bm{d}^{*}\cdot\bm{q}\Delta_{1},
\end{align}
where $\mathrm{i}\omega_{n}$ is replaced by $-\mathrm{i}\omega_{n}$ of the second term in the first equality. $F^{\qty(2)}_{\alpha\alpha'}$ similarly puts
\begin{align}
\Delta_{1}\bm{d}^{*}\cdot\qty(-\bm{q})\Delta^{*}_{2} + \Delta_{2}\bm{d}\cdot\qty(-\bm{q})\Delta^{*}_{1}.
\end{align}
We move on to the coordinate space ($\bm{q}\to\bm{D}$ and $-\bm{q}\to\bm{D}^{*}$) and obtain
\begin{align}
\bm{d}\cdot\qty(\Delta^{*}_{1}\bm{D}\Delta_{2}) + \bm{d}^{*}\cdot\qty(\Delta^{*}_{2}\bm{D}\Delta_{1}) + \bm{d}^{*}\cdot\qty(\Delta_{1}\bm{D}^{*}\Delta^{*}_{2}) + \bm{d}\cdot\qty(\Delta_{2}\bm{D}^{*}\Delta^{*}_{1}),
\end{align}
which is exactly the Lifshitz invariant term in the free energy except for the coefficient.\\

We can prove that the vector $\bm{d}_{\alpha\alpha'}$ is purely imaginary in usual lattice models. Fourier transformation gives the wavenumber dependence $\bm{k}$ via the plane wave $e^{\mathrm{i}\bm{k}\cdot\bm{r}} = \cos{\qty(\bm{k}\cdot\bm{r})} + \mathrm{i}\sin{\qty(\bm{k}\cdot\bm{r})}$ and the real (imaginary) components of the kinetic part of the BdG Hamiltonian $\xi\qty(\bm{k})$ is even (odd) function of $\bm{k}$. Hence $\xi\qty(\bm{k})$ has the preperty $\xi\qty(-\bm{k})=\xi^{\mathrm{T}}\qty(\bm{k})$. With this relation we can say that
\begin{align}
g_{\alpha'\alpha}\qty(\mathrm{i}\omega_{n}; \bm{k}) &= \qty[ \qty(\mathrm{i}\omega_{n} - \xi\qty(\bm{k}) )^{-1} ]_{\alpha'\alpha} \notag \\
&= \qty[ \qty(\mathrm{i}\omega_{n} - \xi\qty(-\bm{k}) )^{-1}]_{\alpha\alpha'} \notag \\
&= g_{\alpha\alpha'}\qty(\mathrm{i}\omega_{n}; -\bm{k}),
\end{align}
and combining this relation with (\ref{g:property1}) we get
\begin{align}
g_{\alpha\alpha'}\qty(\mathrm{i}\omega_{n};\bm{k}) = g^{*}_{\alpha\alpha'}\qty(-\mathrm{i}\omega_{n}; -\bm{k}).
\end{align}
We thus prove that $\bm{d}_{\alpha\alpha'}$ is purely imaginary:
\begin{align}
\bm{d}_{\alpha\alpha'} &= \frac{1}{2\beta}\sum_{n}\sum_{\bm{k}}g_{\alpha\alpha'}\qty(-\mathrm{i}\omega_{n};-\bm{k})\nabla_{\bm{k}}g_{\alpha\alpha'}\qty(\mathrm{i}\omega_{n};\bm{k}) \notag \\
&= \frac{1}{2\beta}\sum_{n}\sum_{\bm{k}}g^{*}_{\alpha\alpha'}\qty(\mathrm{i}\omega_{n}; \bm{k})\nabla_{\bm{k}}g^{*}_{\alpha\alpha'}\qty(-\mathrm{i}\omega_{n};-\bm{k}) \notag \\
&= \frac{1}{2\beta}\sum_{n}\sum_{\bm{k}}g^{*}_{\alpha\alpha'}\qty(-\mathrm{i}\omega_{n}; -\bm{k})\nabla_{\bm{k}}g^{*}_{\alpha\alpha'}\qty(\mathrm{i}\omega_{n};-\bm{k}) \notag \\
&= -\frac{1}{2\beta}\sum_{n}\sum_{\bm{k}}g^{*}_{\alpha\alpha'}\qty(-\mathrm{i}\omega_{n}; \bm{k})\nabla_{-\bm{k}}g^{*}_{\alpha\alpha'}\qty(\mathrm{i}\omega_{n};-\bm{k}) \notag \\
&= -\frac{1}{2\beta}\sum_{n}\sum_{\bm{k}}g^{*}_{\alpha\alpha'}\qty(-\mathrm{i}\omega_{n}; -\bm{k})\nabla_{\bm{k}}g^{*}_{\alpha\alpha'}\qty(-\mathrm{i}\omega_{n};\bm{k}) \notag \\
& = -\bm{d}^{*}_{\alpha\alpha'},
\label{d_imaginary}
\end{align}
and the Higgs mode does not contribute to the linear response.

We can also prove that the vector $\bm{d}_{\alpha\alpha'}$ is produced by the imaginary part of the kinetic term $\xi\qty(\bm{k})$. When the system has no hopping difference, $\xi\qty(\bm{k})$ is real and satisfy the condition $\xi\qty(\bm{k}) = \xi^{*}\qty(\bm{k}) = \xi\qty(-\bm{k})$. Because of this property it follows that
\begin{align}
g_{\alpha\alpha'}\qty(\mathrm{i}\omega_{n}; \bm{k}) = g^{*}_{\alpha\alpha'}\qty(-\mathrm{i}\omega_{n}; \bm{k}).
\end{align}
By the similar procedure of (\ref{d_imaginary}) we can obtain
\begin{align}
\bm{d}_{\alpha\alpha'} = \bm{d}^{*}_{\alpha\alpha'},
\end{align}
which shows that $\bm{d}_{\alpha\alpha'}$ is real and becomes zero because it is purely imaginary. Therefore the Leggett mode in the linear response is induced by the imaginary part of the kinetic component $\xi\qty(\bm{k})$, or the hopping difference in the system.\\

\textit{Gap equation}. 
In this case, we are interested in the gap equation for the saddle point (the mean-field) value and the free energy, or the effective action in equilibrium. Thus we choose the reference state as the superconducting ground state and neglect the electromagnetic terms and the fluctuations of the order parameters. This puts inverse Green's function to be
\begin{align}
G^{-1}\qty(\mathrm{i}\omega_{m}, \mathrm{i}\omega_{n}; \bm{k}) &= \mqty[ \mathrm{i}\omega_{n} - \xi\qty(\bm{k}) & -\Delta_{0} \\ -\Delta^{*}_{0} & \mathrm{i}\omega_{n} + \xi^{\mathrm{T}}\qty(-\bm{k}) ]\beta\delta_{mn} \notag \\
&= G^{-1}_{0}\qty(\mathrm{i}\omega_{m}, \mathrm{i}\omega_{n}; \bm{k}).
\end{align}
The effective action is given by
\begin{align}
S_{\mathrm{eff}} = \beta\sum_{\alpha}\frac{\left|\Delta_{0,\alpha}\right|^{2}}{U} - \frac{1}{\beta^{2}}\sum_{\bm{k}}\sum_{n,m}\mathrm{Tr}\ln{\qty[ -G^{-1}_{0}\qty(\mathrm{i}\omega_{m}, \mathrm{i}\omega_{n}; \bm{k}) ] },
\end{align}
neglecting fluctuations.
The gap equation for $\Delta_{0,\alpha}$ is derived by the minimization of the effective action concerning $\Delta^{*}_{0,\alpha}$;
\begin{align}
\frac{\delta S_{\mathrm{eff}}}{\delta \Delta^{*}_{0,\alpha}} = \beta\frac{\Delta_{0,\alpha}}{U} - \frac{1}{\beta}\sum_{\bm{k}}\sum_{n,m}\frac{\delta}{\delta \Delta^{*}_{0,\alpha} }\mathrm{Tr}\ln{\qty[ -G^{-1}_{0}\qty(\mathrm{i}\omega_{m}, \mathrm{i}\omega_{n}; \bm{k}) ] }.
\end{align}
The functional derivative of the trace term proceeds as
\begin{align}
\frac{\delta}{\delta \Delta^{*}_{0,\alpha}}\mathrm{Tr}\ln{ \qty[ -G^{-1}_{0}\qty(\mathrm{i}\omega_{m}, \mathrm{i}\omega_{n}; \bm{k}) ] } &= \mathrm{Tr}\qty[ \qty(-G^{-1}_{0}\qty(\mathrm{i}\omega_{m}, \mathrm{i}\omega_{n}; \bm{k}) )^{-1} \frac{\delta}{\delta \Delta^{*}_{0,\alpha}}\qty( -G^{-1}_{0}\qty(\mathrm{i}\omega_{m}, \mathrm{i}\omega_{n}; \bm{k}) ) ] \notag \\
&= \mathrm{Tr}\qty[ G_{0}\qty(\mathrm{i}\omega_{m}, \mathrm{i}\omega_{n}; \bm{k})\frac{\delta}{\delta \Delta^{*}_{0,\alpha}}G^{-1}_{0}\qty(\mathrm{i}\omega_{m}, \mathrm{i}\omega_{n}; \bm{k}) ], 
\end{align}
where the Green function is defined without $\beta$.
By definition, the functional derivative of the inverse Green function becomes
\begin{align}
\frac{\delta}{\delta \Delta^{*}_{0,\alpha}}G^{-1}_{0}\qty(\mathrm{i}\omega_{m}, \mathrm{i}\omega_{n}; \bm{k}) &= \mqty[ O & O \\ -A_{\alpha} & O ]\beta\delta_{mn} \notag \\
&= -\frac{1}{2}\qty(\tau_{x,\alpha} - \mathrm{i}\tau_{y,\alpha})\beta\delta_{mn}, 
\end{align}
where $\qty[ A_{\alpha} ]_{\gamma\gamma'} = \delta_{\alpha\gamma}\delta_{\gamma\gamma'}$ and 
\begin{align}
\tau_{x,\alpha} = \mqty[ O & A_{\alpha} \\ A_{\alpha} & O ] \quad \text{and} \quad \tau_{y,\alpha} = \mqty[ O & -\mathrm{i}A_{\alpha} \\ \mathrm{i}A_{\alpha} & O ],
\end{align}
are the generalized Pauli matrices.
Thus we put
\begin{align}
\frac{\delta}{\delta \Delta_{0,\alpha}}\mathrm{Tr}\ln{ \qty[ -G^{-1}\qty( \mathrm{i}\omega_{m}, \mathrm{i}\omega_{n}; \bm{k}) ] } = -\mathrm{Tr}\qty[ G_{0}\qty(\mathrm{i}\omega_{m}, \mathrm{i}\omega_{n}; \bm{k})\frac{1}{2}\qty(\tau_{x,\alpha} - \mathrm{i}\tau_{y,\alpha}) ]\beta\delta_{mn}.
\end{align}
We finally reach the gap equation for $\Delta_{0,\alpha}$:
\begin{align}
&\beta\frac{\Delta_{0,\alpha}}{U} + \frac{1}{\beta}\sum_{n}\sum_{\bm{k}}\mathrm{Tr}\qty[ \frac{1}{2}\qty(\tau_{x,\alpha} - \mathrm{i}\tau_{y,\alpha})G_{0}\qty(\mathrm{i}\omega_{n}, \mathrm{i}\omega_{n}; \bm{k}) ]\beta = 0 \notag \\
&\Leftrightarrow \Delta_{0,\alpha} = -U\frac{1}{\beta}\sum_{n}\sum_{\bm{k}}\mathrm{Tr}\qty[ \frac{1}{2}\qty(\tau_{x,\alpha} - \mathrm{i}\tau_{y,\alpha})G_{0}\qty(\mathrm{i}\omega_{n}, \mathrm{i}\omega_{n}; \bm{k}) ].
\end{align}
Note that the summation over the frequency index $m$ and the Kronecker delta $\delta_{mn}$ gives $\beta$.

\textit{Linear optical conductivities}. We are interested in the optical conductivities of quasiparticles and collective modes here. Hence we choose the reference state to be the superconducting ground state and put
\begin{align}
G_{0}^{-1}\qty(\mathrm{i}\omega_{m}, \mathrm{i}\omega_{n}; \bm{k}) &= \mqty[ \mathrm{i}\omega_{n} - \xi\qty(\bm{k}) & -\Delta_{0} \\ -\Delta^{*}_{0} & \mathrm{i}\omega_{n} + \xi^{\mathrm{T}}\qty(-\bm{k}) ]\beta\delta_{mn}, \\
\Sigma\qty(\mathrm{i}\omega_{m}, \mathrm{i}\omega_{n}; \bm{k}) &= \mqty[ O & \delta\Delta\qty(\mathrm{i}\omega_{m} - \mathrm{i}\omega_{n}) \\ \delta\Delta^{*}\qty(\mathrm{i}\omega_{m} - \mathrm{i}\omega_{n}) & O ]\beta - \sum_{a}e A_{a}\qty(\mathrm{i}\omega_{m} - \mathrm{i}\omega_{n})\mqty[ \partial_{a}\xi\qty(\bm{k}) & O \\ O & \partial_{a}\xi\qty(\bm{k}) ]\beta.
\end{align}
We expand the action $S\qty[\Delta^{*}, \Delta]$ at the Gaussian level ($L=2$) and use RPA to obtain the effective action. Since we direct our attention to linear optical conductivities, we just keep the vector potential $\bm{A}$ in the first order. Besides, we decompose the fluctuation $\delta\Delta_{\alpha}$ into $\Delta_{x,\alpha} - \mathrm{i}\Delta_{y,\alpha}$ for calculation (we omit $\delta$ after the decomposition for simplicity).
Note that we should include the effect of the non-trace term
\begin{align}
&\sum_{\alpha}\sum_{m}\delta\Delta_{\alpha}^{*}\qty(\mathrm{i}\omega_{m})\frac{1}{U}\delta\Delta_{\alpha}\qty(\mathrm{i}\omega_{m}) \notag \\
&= \sum_{\alpha}\sum_{m}\qty[ \Delta_{x,\alpha}\qty(-\mathrm{i}\omega_{m}) - \mathrm{i}\Delta_{y,\alpha}\qty(-\mathrm{i}\omega_{m}) ]\frac{1}{U} \qty[ \Delta_{x,\alpha}\qty(\mathrm{i}\omega_{m}) + \mathrm{i}\Delta_{y,\alpha}\qty(\mathrm{i}\omega_{m}) ] \notag \\
&= \sum_{\alpha}\sum_{m}\frac{1}{U} \qty[ \Delta_{x,\alpha}\qty(-\mathrm{i}\omega_{m})\Delta_{x,\alpha}\qty(\mathrm{i}\omega_{m}) + \Delta_{y,\alpha}\qty(-\mathrm{i}\omega_{m})\Delta_{y,\alpha}\qty(\mathrm{i}\omega_{m})] \notag \\
&\quad + \sum_{\alpha}\sum_{m}\frac{1}{U} \mathrm{i}\qty[ \Delta_{x,\alpha}\qty(-\mathrm{i}\omega_{m})\Delta_{y,\alpha}\qty(\mathrm{i}\omega_{m}) - \Delta_{y,\alpha}\qty(-\mathrm{i}\omega_{m})\Delta_{x,\alpha}\qty(\mathrm{i}\omega_{m}) ] \notag \\
&= \sum_{\alpha}\sum_{m}\frac{1}{U} \qty[ \Delta_{x,\alpha}\qty(-\mathrm{i}\omega_{m})\Delta_{x,\alpha}\qty(\mathrm{i}\omega_{m}) + \Delta_{y,\alpha}\qty(-\mathrm{i}\omega_{m})\Delta_{y,\alpha}\qty(\mathrm{i}\omega_{m})] \notag \\
&= \sum_{\mu,\alpha}\sum_{m} \Delta_{\mu\alpha}\qty(-\mathrm{i}\omega_{m})\frac{1}{U}\Delta_{\mu\alpha}\qty(\mathrm{i}\omega_{m}),
\end{align}
where $\mu$ specifies the index of the Pauli matrix and the imaginary part vanished by the summation over $\qty(\mathrm{i}\omega_{m})$.

Now we proceed to the trace term ($L=2$). Before calculation, we define the velocity operator $v_{a}\qty(\bm{k})$:
\begin{align}
v_{a}\qty(\bm{k}) := \mqty[ \partial_{a}\xi\qty(\bm{k}) & O \\ O & \partial_{a}\xi\qty(\bm{k}) ].
\end{align}
Then the self-energy $\Sigma\qty(\mathrm{i}\omega_{m}, \mathrm{i}\omega_{n}; \bm{k})$ is
\begin{align}
\Sigma\qty(\mathrm{i}\omega_{m}, \mathrm{i}\omega_{n}; \bm{k}) = \beta\sum_{\mu,\alpha}\Delta_{\mu\alpha}\qty(\mathrm{i}\omega_{m} - \mathrm{i}\omega_{n})\tau_{\mu\alpha} - \beta e\sum_{a}A_{a}\qty(\mathrm{i}\omega_{m}-\mathrm{i}\omega_{n})v_{a}\qty(\bm{k}).
\end{align}
The trace term is written as
\begin{align}
&\sum_{\bm{k}}\sum_{n,m}\mathrm{Tr}\qty[ \frac{1}{2}G_{0}\Sigma G_{0}\Sigma ] \notag \\
&= \frac{1}{2}\frac{1}{\beta}\sum_{\bm{k}}\sum_{\mu\mu',\alpha\alpha'}\sum_{n,m}\mathrm{Tr}\qty[ G_{0}\qty(\mathrm{i}\omega_{m}+\mathrm{i}\omega_{n}, \bm{k})\Delta_{\mu\alpha}\qty(-\mathrm{i}\omega_{m})\tau_{\mu\alpha}G_{0}\qty(\mathrm{i}\omega_{n},\bm{k})\Delta_{\mu'\alpha'}\qty(\mathrm{i}\omega_{n})\tau_{\mu'\alpha'} ] \notag \\
&\quad +\frac{1}{2}\frac{1}{\beta}\sum_{\bm{k}}\sum_{\mu\alpha}\sum_{n,m}\sum_{a}\mathrm{Tr}\qty[ G_{0}\qty(\mathrm{i}\omega_{m} + \mathrm{i}\omega_{n}; \bm{k})\Delta_{\mu\alpha}\qty(-\mathrm{i}\omega_{m})\tau_{\mu\alpha}G_{0}\qty(\mathrm{i}\omega_{n}; \bm{k})\qty(-e)A_{a}\qty(\mathrm{i}\omega_{n})v_{a}\qty(\bm{k})  ] \notag \\
&\quad + \frac{1}{2}\frac{1}{\beta}\sum_{\bm{k}}\sum_{\mu'\alpha'}\sum_{n,m}\sum_{a}\mathrm{Tr}\qty[ G_{0}\qty(\mathrm{i}\omega_{m} + \mathrm{i}\omega_{n}; \bm{k}) \qty(-e) A_{a}\qty( - \mathrm{i}\omega_{m})v_{a}\qty(\bm{k})G_{0}\qty( \mathrm{i}\omega_{n}; \bm{k})\Delta_{\mu'\alpha'}\qty(\mathrm{i}\omega_{n})\tau_{\mu'\alpha'} ] \notag \\
&\quad + \frac{1}{2}\frac{1}{\beta}\sum_{\bm{k}}\sum_{n,m}\sum_{a,b}\mathrm{Tr}\qty[ G_{0}\qty(\mathrm{i}\omega_{m}+\mathrm{i}\omega_{n}; \bm{k})\qty(-e) A_{a}\qty( - \mathrm{i}\omega_{m})v_{a}\qty(\bm{k})G_{0}\qty(\mathrm{i}\omega_{n},\bm{k})\qty(-e)A_{b}\qty(\mathrm{i}\omega_{n})v_{b}\qty(\bm{k}) ],
\end{align}
where we define the Green function $G_{0}$ without $\beta$.
We treat each term separately.
We can alter the first term as
\begin{align}
&\frac{1}{2}\frac{1}{\beta}\sum_{\bm{k}}\sum_{\mu\mu',\alpha\alpha'}\sum_{n,m}\mathrm{Tr}\qty[ G_{0}\qty(\mathrm{i}\omega_{m}+\mathrm{i}\omega_{n}, \bm{k})\Delta_{\mu\alpha}\qty(-\mathrm{i}\omega_{m})\tau_{\mu\alpha}G_{0}\qty(\mathrm{i}\omega_{n})\Delta_{\mu'\alpha'}\qty(\mathrm{i}\omega_{m})\tau_{\mu'\alpha'} ] \notag \\
&= \sum_{\mu\mu',\alpha\alpha'}\sum_{m}\Delta_{\mu\alpha}\qty(-\mathrm{i}\omega_{m}) \frac{1}{2\beta}\sum_{n}\sum_{\bm{k}}\mathrm{Tr}\qty[ G_{0}\qty( \mathrm{i}\omega_{m}+\mathrm{i}\omega_{n}; \bm{k})\tau_{\mu\alpha}G_{0}\qty( \mathrm{i}\omega_{n}; \bm{k})\tau_{\mu'\alpha'} ]\Delta_{\mu'\alpha'}\qty(\mathrm{i}\omega_{m}) \notag \\
&= \sum_{\mu\mu', \alpha\alpha'}\sum_{m}\Delta_{\mu\alpha}\qty(-\mathrm{i}\omega_{m})\qty[ \Pi\qty(-\mathrm{i}\omega_{m}) ]_{\mu\alpha, \mu'\alpha'}\Delta_{\mu'\alpha'}\qty(\mathrm{i}\omega_{m}),
\end{align}
where
\begin{align}
\qty[ \Pi\qty(\mathrm{i}\Omega) ]_{\mu\alpha, \mu'\alpha'} &= \frac{1}{2\beta}\sum_{n}\sum_{\bm{k}}\mathrm{Tr}\qty[ \tau_{\mu\alpha}G_{0}\qty(\mathrm{i}\omega_{n}+\mathrm{i}\Omega; \bm{k})\tau_{\mu'\alpha'}G_{0}\qty(\mathrm{i}\omega_{n}; \bm{k}) ] \notag \\
&= \frac{1}{2\beta}\sum_{n}\sum_{\bm{k}}\mathrm{Tr}\qty[ G_{0}\qty(\mathrm{i}\omega_{n}; \bm{k})\tau_{\mu\alpha}G_{0}\qty(\mathrm{i}\omega_{n}+\mathrm{i}\Omega; \bm{k})\tau_{\mu'\alpha'} ] \notag \\
&= \frac{1}{2\beta}\sum_{n}\sum_{\bm{k}}\mathrm{Tr}\qty[ G_{0}\qty(\mathrm{i}\omega_{n}-\mathrm{i}\Omega; \bm{k})\tau_{\mu\alpha}G_{0}\qty(\mathrm{i}\omega_{n}; \bm{k})\tau_{\mu'\alpha'} ],
\end{align}
is the polarization bubble.
The second term can be rewritten as
\begin{align}
&\frac{1}{2}\frac{1}{\beta}\sum_{\bm{k}}\sum_{\mu\alpha}\sum_{n,m}\sum_{a}\mathrm{Tr}\qty[ G_{0}\qty(\mathrm{i}\omega_{m} + \mathrm{i}\omega_{n}; \bm{k})\Delta_{\mu\alpha}\qty(-\mathrm{i}\omega_{m})\tau_{\mu\alpha}G_{0}\qty(\mathrm{i}\omega_{n}; \bm{k})\qty(-e)A_{a}\qty(\mathrm{i}\omega_{n})v_{a}\qty(\bm{k})  ] \notag \\
& = \frac{1}{2}\qty(-e)\sum_{\mu\alpha}\sum_{m}\sum_{a}\Delta_{\mu\alpha}\qty(-\mathrm{i}\omega_{m})A_{a}\qty(\mathrm{i}\omega_{m}) \qty[Q_{a}\qty(\mathrm{i}\omega_{m})]_{\mu\alpha},
\end{align}
where
\begin{align}
\qty[ Q_{a}\qty(\mathrm{i}\Omega) ]_{\mu\alpha} &= \frac{1}{\beta}\sum_{n}\sum_{\bm{k}}\mathrm{Tr}\qty[ v_{a}\qty(\bm{k})G_{0}\qty(\mathrm{i}\omega_{n} + \mathrm{i}\Omega; \bm{k})\tau_{\mu\alpha}G_{0}\qty(\mathrm{i}\omega_{n}; \bm{k}) ] \notag \\
&= \frac{1}{\beta}\sum_{n}\sum_{\bm{k}}\mathrm{Tr}\qty[ G_{0}\qty(\mathrm{i}\omega_{n} + \mathrm{i}\Omega; \bm{k})\tau_{\mu\alpha}G_{0}\qty(\mathrm{i}\omega_{n}; \bm{k})v_{a}\qty(\bm{k}) ].
\end{align}
The third term is similar to the second one:
\begin{align}
&\frac{1}{2}\frac{1}{\beta}\sum_{\bm{k}}\sum_{\mu'\alpha'}\sum_{n,m}\sum_{a}\mathrm{Tr}\qty[ G_{0}\qty(\mathrm{i}\omega_{m} + \mathrm{i}\omega_{n}; \bm{k}) \qty(-e) A_{a}\qty( - \mathrm{i}\omega_{m})v_{a}\qty(\bm{k})G_{0}\qty( \mathrm{i}\omega_{n}; \bm{k})\Delta_{\mu'\alpha'}\qty(\mathrm{i}\omega_{n})\tau_{\mu'\alpha'} ] \notag \\
&= \frac{1}{2}\qty(-e)\sum_{\mu'\alpha'}\sum_{m}\sum_{a}\Delta_{\mu'\alpha'}\qty(\mathrm{i}\omega_{m})A_{a}\qty(-\mathrm{i}\omega_{m}) \qty[Q_{a}\qty(-\mathrm{i}\omega_{m}) ]_{\mu'\alpha'}.
\end{align}
The last one is converted as follows.
\begin{align}
&\frac{1}{2}\frac{1}{\beta}\sum_{\bm{k}}\sum_{n,m}\sum_{a,b}\mathrm{Tr}\qty[ G_{0}\qty(\mathrm{i}\omega_{m}+\mathrm{i}\omega_{n}; \bm{k})\qty(-e) A_{a}\qty( - \mathrm{i}\omega_{m})v_{a}G_{0}\qty(\mathrm{i}\omega_{n},\bm{k})\qty(\bm{k})\qty(-e)A_{b}\qty(\mathrm{i}\omega_{n})v_{b}\qty(\bm{k}) ] \notag \\
&= \frac{1}{2}\qty(-e)^{2}\sum_{m}\sum_{a,b} A_{a}\qty(-\mathrm{i}\omega_{m})A_{b}\qty(\mathrm{i}\omega_{m}) \Phi_{ab}\qty(-\mathrm{i}\omega_{m} ),
\end{align}
where $\Phi_{ab}\qty(\mathrm{i}\Omega)$:
\begin{align}
\Phi_{ab}\qty(\mathrm{i}\Omega) &= \frac{1}{\beta}\sum_{n}\sum_{\bm{k}}\mathrm{Tr}\qty[ v_{a}\qty(\bm{k})G_{0}\qty(\mathrm{i}\omega_{n}+\mathrm{i}\Omega; \bm{k})v_{b}\qty(\bm{k})G_{0}\qty(\mathrm{i}\omega_{n}; \bm{k}) ] \notag \\
&= \frac{1}{\beta}\sum_{n}\sum_{\bm{k}}\mathrm{Tr}\qty[ G_{0}\qty(\mathrm{i}\omega_{n}; \bm{k})v_{a}\qty(\bm{k})G_{0}\qty(\mathrm{i}\omega_{n}+\mathrm{i}\Omega; \bm{k})v_{b}\qty(\bm{k}) ] \notag \\
&= \frac{1}{\beta}\sum_{n}\sum_{\bm{k}}\mathrm{Tr}\qty[ G_{0}\qty(\mathrm{i}\omega_{n} - \mathrm{i}\Omega; \bm{k})v_{a}\qty(\bm{k})G_{0}\qty(\mathrm{i}\omega_{n}; \bm{k})v_{b}\qty(\bm{k}) ],
\end{align}
is the current-current correlation function.
Summarizing all the terms puts the effective action $S_{\mathrm{eff}}\qty[ \Delta^{*}, \Delta]$ into
\begin{align}
S_{\mathrm{eff}}\qty[ \Delta^{*}, \Delta] = S_{\mathrm{MF}} + S_{\mathrm{FL}}\qty[\Delta^{*}, \Delta],
\end{align}
with
\begin{align}
S_{\mathrm{MF}} = \beta\sum_{\alpha}\frac{\left|\Delta_{\alpha}\right|^{2}}{U} - \sum_{n}\sum_{\bm{k}}\mathrm{Tr}\ln{\qty[ -G^{-1}_{0}\qty(\mathrm{i}\omega_{n}; \bm{k} ) ]},
\end{align}
and
\begin{align}
S_{\mathrm{FL}}\qty[\Delta^{*}, \Delta] &= \sum_{\mu,\alpha}\sum_{m} \Delta_{\mu\alpha}\qty(-\mathrm{i}\omega_{m})\frac{1}{U}\Delta_{\mu'\alpha'}\qty(\mathrm{i}\omega_{m})\delta_{\mu\mu'}\delta_{\alpha\alpha'} \notag \\
&\quad + \sum_{\mu\mu', \alpha\alpha'}\sum_{m}\Delta_{\mu\alpha}\qty(-\mathrm{i}\omega_{m})\qty[ \Pi\qty(-\mathrm{i}\omega_{m}) ]_{\mu\alpha,\mu'\alpha'}\Delta_{\mu'\alpha'}\qty(\mathrm{i}\omega_{m}) \notag \\
&\quad + \frac{1}{2}\qty( - e)\sum_{\mu\alpha}\sum_{m}\sum_{a}\Delta_{\mu\alpha}\qty(-\mathrm{i}\omega_{m})A_{a}\qty(\mathrm{i}\omega_{m})\qty[Q_{a}\qty(\mathrm{i}\omega_{m})]_{\mu\alpha} \notag \\
&\quad + \frac{1}{2}\qty( - e)\sum_{\mu'\alpha'}\sum_{m}\sum_{a}\Delta_{\mu'\alpha'}\qty(\mathrm{i}\omega_{m})A_{a}\qty(-\mathrm{i}\omega_{m})\qty[Q_{a}\qty(-\mathrm{i}\omega_{m})]_{\mu'\alpha'} \notag \\
&\quad + \frac{1}{2}\qty( - e )^{2}\sum_{m}\sum_{a,b} A_{a}\qty(-\mathrm{i}\omega_{m})A_{b}\qty(\mathrm{i}\omega_{m}) \Phi_{ab}\qty(-\mathrm{i}\omega_{m} ).
\end{align}
Keep in mind that the partition function $\mathcal{Z}$ has the form of
\begin{align}
\mathcal{Z} = \int \mathcal{D}\qty( \Delta^{*}, \Delta )e^{-S_{\mathrm{MF}} - S_{\mathrm{FL}}\qty[ \Delta^{*}, \Delta] }.
\end{align}
We use the identity:
\begin{align}
&\int \mathcal{D}\qty(\phi^{\mathrm{T}}, \phi)\exp{-\frac{1}{2}\sum_{m}\qty[ \phi^{\mathrm{T}}\qty(-\mathrm{i}\omega_{m}) M\qty(\mathrm{i}\omega_{m})\phi\qty(\mathrm{i}\omega_{m}) + \phi^{\mathrm{T}}\qty(-\mathrm{i}\omega_{m})b\qty(\mathrm{i}\omega_{m}) + b^{\mathrm{T}}\qty(-\mathrm{i}\omega_{m})\phi\qty(\mathrm{i}\omega_{m}) ]} \notag \\
&=\exp{ \frac{1}{2}\sum_{m}b^{\mathrm{T}}\qty(-\mathrm{i}\omega_{m})M^{-1}\qty(\mathrm{i}\omega_{m})b\qty(\mathrm{i}\omega_{m}) },
\end{align}
to carry out the bosonic path integral.
We define $M$ as
\begin{align}
\qty[ M\qty(\mathrm{i}\omega_{m})]_{\mu\alpha, \mu'\alpha'} &= -\qty( \frac{1}{U}\delta_{\mu\mu'}\delta_{\alpha\alpha'} + \qty[ \Pi\qty(-\mathrm{i}\omega_{m}) ]_{\mu\alpha, \mu'\alpha'}) \notag \\
&= -\frac{1}{U}\qty( \delta_{\mu\mu'}\delta_{\alpha\alpha'} + U\qty[ \Pi\qty(-\mathrm{i}\omega_{m}) ]_{\mu\alpha, \mu'\alpha'}), \notag \\
&= \qty[ U_{\mathrm{eff}}^{-1}\qty( -\mathrm{i}\omega_{m})]_{\mu\alpha, \mu'\alpha'},
\end{align}
where $U_{\mathrm{eff}}$ is an effective interaction within the random phase approximation (RPA), and we obtain the effective action $S_{\mathrm{eff}}$:
\begin{align}
S_{\mathrm{eff}} &= S_{\mathrm{MF}} + S_{\mathrm{FL}}, \\
S_{\mathrm{FL}} &= \frac{e^{2}}{2}\sum_{m}\sum_{a,b} A_{a}\qty(-\mathrm{i}\omega_{m})A_{b}\qty(\mathrm{i}\omega_{m}) \Phi_{ab}\qty(-\mathrm{i}\omega_{m} ) \notag \\
&\quad + \frac{e^{2}}{4}\sum_{m}\sum_{a,b}A_{a}\qty(-\mathrm{i}\omega_{m})Q^{\mathrm{T}}_{a}\qty(-\mathrm{i}\omega_{m})U_{\mathrm{eff}}\qty(-\mathrm{i}\omega_{m}) A_{b}\qty(\mathrm{i}\omega_{m})Q_{b}\qty(\mathrm{i}\omega_{m}).
\end{align}
We move to the zero temperature case for simplicity, making a replacement
\begin{equation}
\frac{1}{\beta}\sum_{m}F\qty(\mathrm{i}\omega_{m}) \to \int\frac{d\omega}{2\pi}F\qty(\omega).
\end{equation}
Then we have (with $\omega\to -\omega$)
\begin{align}
S_{\mathrm{FL}} = \beta\sum_{a,b}\frac{e^{2}}{2}\int \frac{d\omega}{2\pi}A_{a}\qty(\omega)A_{b}\qty(-\omega)\Phi_{ab}\qty(\omega) + \beta\sum_{a,b}\frac{e^{2}}{4}\int\frac{d\omega}{2\pi}A_{a}\qty(\omega)A_{b}\qty(-\omega)Q^{\mathrm{T}}_{a}\qty(\omega)U_{\mathrm{eff}}\qty(\omega)Q_{b}\qty(-\omega).
\end{align}
To compute the current $j_{a}\qty(\omega)$ (we focus on the two-dimensional case $a=x,y$), we take a functional derivative of $S$ about $A_{a}\qty(-\omega)$. Here we exploit the properties $\Phi_{ab}(\mathrm{i}\Omega) = \Phi_{ba}\qty(-\mathrm{i}\Omega)$, $\Pi\qty(\mathrm{i}\Omega) = \Pi^{\mathrm{T}}\qty(-\mathrm{i}\Omega)$, $U_{\mathrm{eff}}^{-1}\qty(\mathrm{i}\Omega) = \qty[ U_{\mathrm{eff}}^{-1}\qty(-\mathrm{i}\Omega) ]^{\mathrm{T}}$ and $Q^{\mathrm{T}}_{a}\qty(\omega)U_{\mathrm{eff}}\qty(\omega)Q_{b}\qty(-\omega) = Q^{\mathrm{T}}_{b}\qty(-\omega)U_{\mathrm{eff}}\qty(-\omega)Q_{a}\qty(\omega)$:

\begin{align}
\Phi_{ba}\qty(-\mathrm{i}\Omega) &= \frac{1}{\beta}\sum_{n}\sum_{\bm{k}}\mathrm{Tr}\qty[G_{0}\qty(\mathrm{i}\omega_{n} + \mathrm{i}\Omega; \bm{k})v_{b}\qty(\bm{k})G_{0}\qty(\mathrm{i}\omega_{n}; \bm{k})v_{a}\qty(\bm{k}) ] \notag \\
&= \frac{1}{\beta}\sum_{n}\sum_{\bm{k}}\mathrm{Tr}\qty[v_{a}\qty(\bm{k}) G_{0}\qty(\mathrm{i}\omega_{n} + \mathrm{i}\Omega; \bm{k})v_{b}\qty(\bm{k})G_{0}\qty(\mathrm{i}\omega_{n}; \bm{k})] \notag \\
&= \Phi_{ab}\qty(\mathrm{i}\Omega),
\end{align}

\begin{align}
\qty[ \Pi\qty(-\mathrm{i}\Omega) ]_{\mu'\alpha', \mu\alpha} &= \frac{1}{2\beta}\sum_{n}\sum_{\bm{k}}\mathrm{Tr}\qty[ G_{0}\qty(\mathrm{i}\omega_{n}+\mathrm{i}\Omega; \bm{k})\tau_{\mu'\alpha'}G_{0}\qty(\mathrm{i}\omega_{n}; \bm{k})\tau_{\mu\alpha} ] \notag \\
&= \frac{1}{2\beta}\sum_{n}\sum_{\bm{k}}\mathrm{Tr}\qty[ \tau_{\mu\alpha}G_{0}\qty(\mathrm{i}\omega_{n}+\mathrm{i}\Omega; \bm{k})\tau_{\mu'\alpha'}G_{0}\qty(\mathrm{i}\omega_{n}; \bm{k}) ] \notag \\
&= \qty[ \Pi\qty(\mathrm{i}\Omega) ]_{\mu\alpha, \mu'\alpha'},
\end{align}

\begin{align}
\qty[ U_{\mathrm{eff}}^{-1}\qty(-\mathrm{i}\Omega) ]_{\mu'\alpha', \mu\alpha} &= -\frac{1}{U}\qty( \delta_{\mu'\mu}\delta_{\alpha\alpha'} + U\qty[ \Pi\qty( \mathrm{i}\Omega) ]_{\mu'\alpha', \mu\alpha} ) \notag \\
&= -\frac{1}{U}\qty( \delta_{\mu\mu'}\delta_{\alpha\alpha'} + U\qty[ \Pi\qty( -\mathrm{i}\Omega) ]_{\mu\alpha, \mu'\alpha'} ) \notag \\
&= \qty[ U_{\mathrm{eff}}^{-1}\qty(\mathrm{i}\Omega) ]_{\mu\alpha, \mu'\alpha'},
\end{align}
and finally
\begin{align}
Q^{\mathrm{T}}_{a}\qty(-\omega)U_{\mathrm{eff}}\qty(-\omega)Q_{b}\qty(\omega) &= Q^{\mathrm{T}}_{b}\qty(\omega)U^{\mathrm{T}}_{\mathrm{eff}}\qty(-\omega)Q_{a}\qty(-\omega) \notag \\
&= Q^{\mathrm{T}}_{b}\qty(\omega)U_{\mathrm{eff}}\qty(\omega)Q_{a}\qty(-\omega).
\end{align}
With these preparations, we can calculate the first-order current for each direction $j_{b}\qty(\omega)$ by taking the functional derivative of the effective action $S_{\mathrm{eff}}$ about $A_{b}\qty(-\omega)$. For simplicity, we assume that the frequency $\omega$ for the functional derivative is positive (the same argument holds for the negative $\omega$). Note that the integrals in $S_{\mathrm{FL}}$ are written as
\begin{align}
\sum_{a,b}\int_{-\infty}^{\infty}\frac{d\omega}{2\pi}A_{a}\qty(\omega)A_{b}\qty(-\omega)\Phi_{ab}\qty(\omega) &= \sum_{a,b}\qty( \int_{0}^{\infty} + \int_{-\infty}^{0})\frac{d\omega}{2\pi}A_{a}\qty(\omega)A_{b}\qty(-\omega)\Phi_{ab}\qty(\omega) \notag \\
&= \sum_{a,b}\qty[ \int_{0}^{\infty}\frac{d\omega}{2\pi}A_{a}\qty(\omega)A_{b}\qty(-\omega)\Phi_{ab}\qty(\omega) + \int_{0}^{\infty}\frac{d\omega}{2\pi}A_{a}\qty(-\omega)A_{b}\qty(\omega)\Phi_{ab}\qty(-\omega) ] \notag \\
&= \sum_{a,b}\qty[ \int_{0}^{\infty}\frac{d\omega}{2\pi}A_{a}\qty(\omega)A_{b}\qty(-\omega)\Phi_{ab}\qty(\omega) + \int_{0}^{\infty}\frac{d\omega}{2\pi}A_{a}\qty(-\omega)A_{b}\qty(\omega)\Phi_{ba}\qty(\omega) ] \notag \\
&= 2\sum_{a,b}\int_{0}^{\infty}\frac{d\omega}{2\pi}A_{a}\qty(\omega)A_{b}\qty(-\omega)\Phi_{ab}\qty(\omega).
\end{align}
and similarly
\begin{align}
&\sum_{a,b}\int_{-\infty}^{\infty}\frac{d\omega}{2\pi}A_{a}\qty(\omega)A_{b}\qty(-\omega)Q^{\mathrm{T}}_{a}\qty(\omega)U_{\mathrm{eff}}\qty(\omega)Q_{b}\qty(-\omega) \notag \\
&= 2\sum_{a,b}\int_{0}^{\infty}\frac{d\omega}{2\pi} A_{a}\qty(\omega)A_{b}\qty(-\omega)Q^{\mathrm{T}}_{a}\qty(\omega)U_{\mathrm{eff}}\qty(\omega)Q_{b}\qty(-\omega),
\end{align}
Now we can take the functional derivative with peace of mind. We can compute the first-order current $j_{b}\qty(\omega)$ as
\begin{align}
j_{b}\qty(\omega) &= -\frac{\delta S_{\mathrm{eff}}}{\beta\delta A_{b}\qty(-\omega) } \notag \\
&= - \sum_{a}A_{a}\qty(\omega) \qty[ e^{2}\Phi_{ab}\qty(\omega) + \frac{e^{2}}{2}Q^{\mathrm{T}}_{a}\qty(\omega)U_{\mathrm{eff}}\qty(\omega)Q_{b}\qty(-\omega) ].
\end{align}
Since the vector potential $A_{a}(\omega)$ is connected to the electric field for $a$-direction by $E_{a}\qty(\omega) = \mathrm{i}\omega A_{a}\qty(\omega)$, we arrive at the expression
\begin{align}
j_{b}\qty(\omega) = \sum_{a} E_{a}\qty(\omega) \qty[ \frac{\mathrm{i}e^{2}}{\omega}\Phi_{ab}\qty(\omega) + \frac{\mathrm{i}e^{2}}{2\omega}Q^{\mathrm{T}}_{a}\qty(\omega)U_{\mathrm{eff}}\qty(\omega)Q_{b}\qty(-\omega) ].
\end{align}
Using the definition of the optical (or electrical) conductivity in Fourier space $j_{b}\qty(\omega) = \sum_{a}E_{a}\qty(\omega)\sigma_{ab}\qty(\omega)$, we put the optical conductivity $\sigma_{ab}\qty(\omega)$ as
\begin{align}
\sigma_{ab}\qty(\omega) = \qty[\sigma_{1}\qty(\omega)]_{ab} + \qty[\sigma_{2}\qty(\omega) ]_{ab},
\end{align}
where
\begin{align}
\qty[ \sigma_{1}\qty(\omega) ]_{ab} &= \frac{\mathrm{i}e^{2}}{\omega}\Phi_{ab}\qty(\omega),
\end{align}
is the optical conductivity of quasi-particle, and
\begin{align}
\qty[ \sigma_{2}\qty(\omega) ]_{ab} &= \frac{\mathrm{i}e^{2}}{2\omega}Q^{\mathrm{T}}_{a}\qty(\omega)U_{\mathrm{eff}}\qty(\omega)Q_{b}\qty(-\omega).
\end{align}
is the optical conductivity of the collective mode.

\section{Stability of the uniform solution in the Ginzburg-Landau theory with the Lifshitz invariant}
\label{Apdx.B}
In this appendix, we show that the spatially uniform solution has a lower free energy than a spatially modulated one for a two-band superconductor in the Ginzburg-Landau theory with the Lifshitz invariant.
Throughout this appendix, we ignore the vector potential $\bm{A}$.
The free energy density $\mathcal{F}$ is written as
\begin{align}
\mathcal{F} &= \sum_{i=1,2}\qty[ a_{i}\left| \psi_{i}\right|^{2} + \frac{b_{i}}{2}\left| \psi_{i}\right|^{4} + \frac{1}{2m^{*}_{i}}\left|\bm{D}\psi_{i}\right|^{2} ] \notag \\
&+ \qty[ \epsilon\psi^{*}_{1}\psi_{2} + \eta\qty(\bm{D}^{*}\psi^{*}_{1})\cdot\qty(\bm{D}\psi_{2}) + \mathrm{c.c.} ] \notag \\
&+ \qty[ \bm{d}\cdot\qty( \psi^{*}_{1}\bm{D}\psi_{2} + \psi^{*}_{2}\bm{D}^{*}\psi_{1}) + \mathrm{c.c.} ]. 
\label{Apdx.B_original}
\end{align}
We substitute
\begin{equation}
\bm{D} = -\mathrm{i}\nabla, \quad \bm{d}=\mathrm{i}\bm{d}_{I}, \quad \psi_{1} = \psi_{1,0}e^{\mathrm{i}\theta_{1,0}}e^{\mathrm{i}\bm{q}_{1}\cdot\bm{r}}, \quad \psi_{2} = \psi_{2,0}e^{\mathrm{i}\theta_{2,0}}e^{\mathrm{i}\bm{q}_{2}\cdot\bm{r}}, \quad \qty(\psi_{1,0}, \psi_{2,0}\in \mathbb{R}),
\end{equation}
into Eq.~(\ref{Apdx.B_original}) and obtain the form below.
\begin{align}
\mathcal{F} &= \sum_{i=1,2}\qty[ a_{i}\left|\psi_{i,0}\right|^{2} + \frac{b_{i}}{2}\left|\psi_{i,0}\right|^{4} + \frac{1}{2m^{*}_{i}}\left|\bm{q}_{i}\psi_{i,0}\right|^{2} ] \notag \\
&+ 2\epsilon\psi_{1,0}\psi_{2,0}\cos{\qty[\theta_{1,0} - \theta_{2,0} + \qty(\bm{q}_{1}-\bm{q}_{2})\cdot\bm{r} ] } \notag \\
&+ 2\eta\psi_{1,0}\psi_{2,0}\qty(\bm{q}_{1}\cdot\bm{q}_{2})\cos{\qty[\theta_{1,0} - \theta_{2,0} + \qty(\bm{q}_{1}-\bm{q}_{2})\cdot\bm{r} ] } \notag \\
&+ 2\psi_{1,0}\psi_{2,0}\qty[ \bm{d}_{I}\cdot\qty(\bm{q}_{1}+\bm{q}_{2})]\sin{ \qty[ \theta_{1,0}-\theta_{2,0} + \qty(\bm{q}_{1} - \bm{q}_{2})\cdot\bm{r} ] }. 
\end{align}
When $\bm{q}_{1}\neq\bm{q}_{2}$, the oscillating terms vanish by integration, and the system comes back to the two-band superconductor without interband couplings.
The free energy must be stable even in this case, meaning that $m^{*}_{i}>0$.
This case is not our focus, so we put $\bm{q}_{1}=\bm{q}_{2} = \bm{q}$ with $m^{*}_{i}>0$.
Setting $\phi = \theta_{1,0}-\theta_{2,0}$, we get
\begin{align}
\mathcal{F} &= \sum_{i=1,2}\qty[ a_{i}\left|\psi_{i,0}\right|^{2} + \frac{b_{i}}{2}\left|\psi_{i,0}\right|^{4} + \frac{1}{2m^{*}_{i}}\left|\bm{q}_{i}\psi_{i,0}\right|^{2} ] \notag \\
&+ 2\psi_{1,0}\psi_{2,0}\qty(\epsilon + \eta q^{2})\cos{\phi} + 4\psi_{1,0}\psi_{2,0}\qty(\bm{d}_{I}\cdot\bm{q})\sin{\phi}.
\end{align}
Note that the condition $m^{*}_{i}>0$ should be satisfied as well.
To minimize the free energy, we take a functional derivative in terms of $\phi$:
\begin{equation}
\frac{\delta\mathcal{F}}{\delta\phi} = -2\psi_{1,0}\psi_{2,0}\qty(\epsilon + \eta q^{2})\sin{\phi} + 4\psi_{1,0}\psi_{2,0}\qty(\bm{d}_{I}\cdot\bm{q})\cos{\phi} = 0. 
\end{equation}
We then obtain
\begin{equation}
\tan{\phi} = \frac{2\qty(\bm{d}_{I}\cdot\bm{q})}{\epsilon + \eta q^{2}}. 
\end{equation}
This yields
\begin{align}
\mathcal{F} &= \sum_{i=1,2}\qty[ a_{i}\psi_{i,0}^{2} + \frac{b_{i}}{2}\psi_{i,0}^{4} + \frac{1}{2m^{*}_{i}}\psi_{i,0}^{2}q^{2} ] \notag \\
&+ 2\psi_{1,0}\psi_{2,0}\cfrac{1}{\sqrt{1 + \qty[ \cfrac{2\bm{d}_{I}\cdot\bm{q}}{\epsilon + \eta q^{2}} ]^{2}} } \qty[ \epsilon + \eta q^{2} + \frac{4\qty(\bm{d}_{I}\cdot\bm{q})^{2}}{\epsilon + \eta q^{2}} ] \notag \\
&= \sum_{i=1,2}\qty[ a_{i}\psi_{i,0}^{2} + \frac{b_{i}}{2}\psi_{i,0}^{4} + \frac{1}{2m^{*}_{i}}\psi_{i,0}^{2}q^{2} ] + 2\psi_{1,0}\psi_{2,0}\qty(\epsilon + \eta q^{2})\sqrt{ 1 + \qty(\frac{2\bm{d}_{I}\cdot\bm{q}}{\epsilon + \eta q^{2}})^{2} }.
\label{general}
\end{align}
This expression is valid independent of the sign of $\qty(\epsilon + \eta q^{2})$.
For simplicity we put
\begin{equation}
a_{1} = a_{2} = a,\quad b_{1}=b_{2} = b, \quad m^{*}_{1} = m^{*}_{2} = m^{*}, \quad \psi_{1,0}=\psi_{2,0}=\psi_{0}. \notag
\end{equation}
The above condition corresponds to the state that every site is equivalent. For large $q$,
\begin{equation}
\frac{2\bm{d}_{I}\cdot\bm{q}}{\epsilon + \eta q^{2}}\to 0.
\end{equation}
Then the free energy is reduced to
\begin{equation}
\frac{\mathcal{F}}{\psi_{0}^{2}} \approx 2\qty(a + \frac{b}{2}\psi_{0}^{2} + \epsilon) + \qty(\frac{1}{m^{*}} + 2\eta)q^{2}.
\end{equation}
This must be stable, suggesting that 
\begin{equation}
\frac{1}{m^{*}} + 2\eta > 0
\end{equation}
is satisfied.
Under this condition, we now focus on the small $q$ case.
\begin{align}
\qty[ \frac{2\bm{d}_{I}\cdot\bm{q}}{\epsilon + \eta q^{2}} ]^{2} &= \frac{4}{\epsilon^{2}}\qty[ \frac{\bm{d}_{I}\cdot\bm{q}}{1 + \qty(\eta/\epsilon)q^{2}} ]^{2} \notag \\
&\approx \frac{4}{\epsilon^{2}}\qty(\bm{d}_{I}\cdot\bm{q})^{2}\qty( 1 - \frac{2\eta}{\epsilon}q^{2} ), \\
\sqrt{ 1 + \qty(\frac{2\bm{d}_{I}\cdot\bm{q}}{\epsilon + \eta q^{2}})^{2} } &\approx \sqrt{ 1 + \frac{4}{\epsilon^{2}}\qty(\bm{d}_{I}\cdot\bm{q})^{2} - \frac{8\eta}{\epsilon^{3}}\qty(\bm{d}_{I}\cdot\bm{q})^{2}q^{2} } \notag \\
&\approx 1 + \frac{2}{\epsilon^{2}}\qty(\bm{d}_{I}\cdot\bm{q})^{2}, \\
\qty(\epsilon + \eta q^{2})\sqrt{ 1 + \qty(\frac{2\bm{d}_{I}\cdot\bm{q}}{\epsilon + \eta q^{2}})^{2} } &\approx \epsilon + \eta q^{2} + \frac{2\qty(\epsilon + \eta q^{2})}{\epsilon^{2}}\qty(\bm{d}_{I}\cdot\bm{q})^{2} \notag \\
&= \epsilon + \qty(\eta + \frac{2}{\epsilon}d_{I}^{2})q^{2}.
\end{align}
Here we assume that $\bm{d}_{I}$ and $\bm{q}$ point in the same direction.
Hence for small $\bm{q}$ we have
\begin{equation}
\frac{\mathcal{F}}{\psi_{0}^{2}} \approx 2\qty(a + \frac{b}{2}\psi_{0}^{2} + \epsilon) + \qty(\frac{1}{m^{*}} + 2\eta + \frac{2d_{I}^{2}}{\epsilon})q^{2}.
\end{equation}
We should classify the cases depending on the sign of $\epsilon$.
When $\epsilon < 0$, meaning that the Josephson-like coupling between the order parameters is attractive, there are two cases as below.
\begin{align}
d_{I}^{2} > -\frac{\epsilon}{2}\qty(\frac{1}{m^{*}} + 2\eta) \quad&\to\quad q\neq 0, \notag \\
d_{I}^{2} < -\frac{\epsilon}{2}\qty(\frac{1}{m^{*}} + 2\eta) \quad&\to\quad q=0,
\end{align}
which is still valid even when we substitute the expression with $\hbar$.
Therefore, we can conclude for the system with negative $\epsilon$ that if $d_{I}^{2}$ is large enough, the order parameters are modulated spatially.
If not, there is no spatial modulation of the order parameters.
On the flip side, when $\epsilon>0$, namely that the Josephson-like coupling is repulsive, the coefficient of $q^{2}$ in the free energy is always positive.
Hence the free energy becomes the lowest at $q=0$, or $\tan{\phi}=0$.
This has two options $\phi=0$ and $\phi=\pi$. Looking at the expression (\ref{general}), the free energy is found to be the lowest for $\phi=\pi$.
This indicates that the two order parameters are in the opposite phase, which can be interpreted as the system having a spatially modulated order.\\

Usually, $d_{I}^{2}$ is small enough to satisfy the condition of $q=0$, and thus we can conclude that in this case the superconducting order parameter is not spatially modulated by the Lifshitz invariant.

\section{Decomposition of the optical conductivity into amplitude and phase fluctuation channels}
\label{Apdx.C}
\begin{figure}[h]
\centering
\includegraphics[scale=0.5]{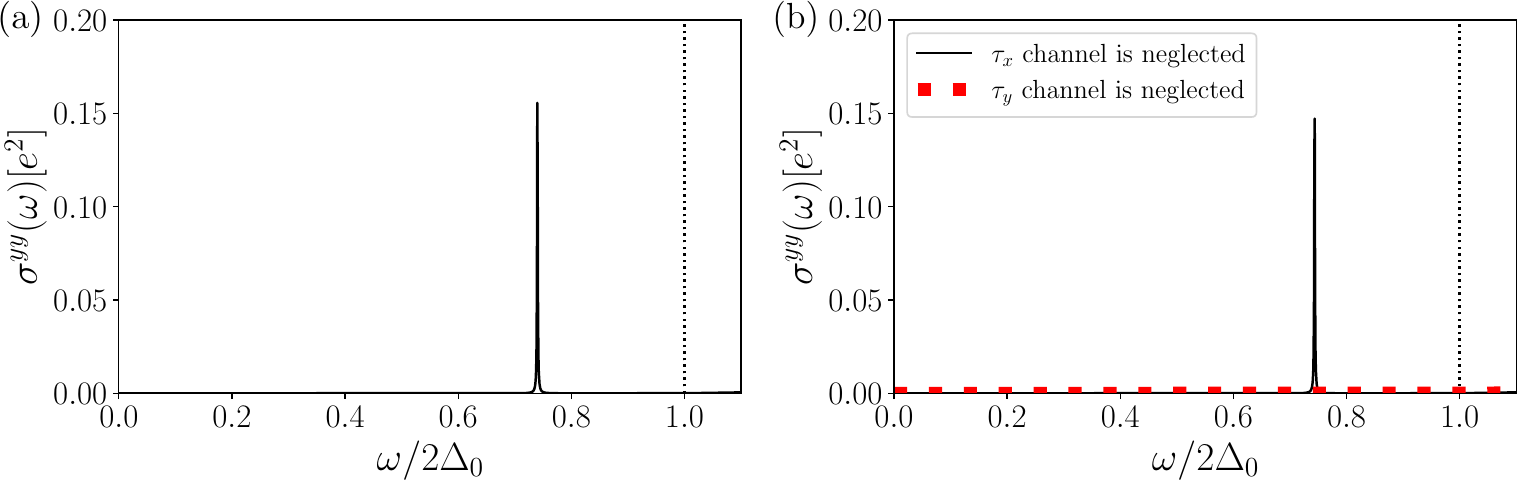}
\caption{[(a),(b)] Linear optical conductivity in the Kagome lattice model as a three-band superconductor with (a) full $\tau_{x}$ and $\tau_{y}$ channels (same figure as in Fig.~\ref{fig:Results}(b)) and (b) either one of the channels ($\tau_{x}$ or $\tau_{y}$) being neglected.
The parameter values are set to $t=-0.5$, $\delta t=-0.01$, $\mu=0$, $m=0$, and $U=6$.
(b) The red dotted line represents the contribution without the phase fluctuation ($\tau_{y}$ channel) the while the black solid line without the amplitude fluctuation ($\tau_{x}$ channel).
The red dotted line overlaps with the horizontal axis, which shows that the contribution from the amplitude fluctuation is small.}
\label{fig_Apdx.C}
\end{figure}

In this appendix, we show how to distinguish the origin of the resonance peak in the optical conductivity such as those seen in Fig.~\ref{fig:Results}.
To this end, we calculate the linear optical conductivity with neglecting $\tau_{y}$ or $\tau_{x}$ vertices in Eq.~(\ref{eq:optical_conductivity}).
We can say that the peak comes from the amplitude fluctuation when it appears with neglecting $\tau_{y}$.
On the other hand, we can say that the peak comes from the phase fluctuation when it appears with neglecting $\tau_{x\alpha}$.
The results of linear optical conductivities $\sigma^{yy}\qty(\omega)$ are given below.
The red dotted line represents the linear optical conductivity from the amplitude fluctuation ($\tau_{x\alpha}$ channel) and the black solid line from the phase fluctuation ($\tau_{y\alpha}$ channel).
In Fig.~\ref{fig_Apdx.C}, the red dotted lines overlap with the horizontal axes and the peak is absent without $\tau_{y\alpha}$ channel contribution.
Hence it is concluded that the peak in the optical conductivity comes from the Leggett mode.

\section{Derivation of the kinetic part of the Hamiltonian of Kagome lattice}
\label{Apdx.D}
\begin{figure}[h]
\centering
\includegraphics[scale=0.8]{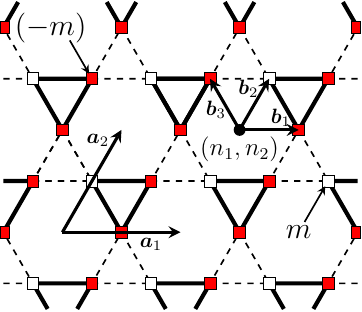}
\caption{The Kagome lattice model with two kinds of hopping strength and on-site potentials. The definitions of the bold and dashed lines are the same as in Fig.~\ref{fig:models}(a) in the main text.
White squares show the on-site potential $m$, while red squares do $(-m)$ with $m\geq0$. The vectors $\bm{a}_{1}$ and $\bm{a}_{2}$ specify the center of the hexagon $(n_{1}, n_{2})$, and $\bm{b}_{1}$, $\bm{b}_{2}$, and $\bm{b}_{3}$ specify the lattice point.}
\label{Apdx.D_Kagome_Lattice}
\end{figure}
We show the derivation of the kinetic part of the Hamiltonian for the Kagome lattice shown in Fig.~\ref{Apdx.D_Kagome_Lattice}.
The definitions of the bold and dashed lines are the same as in Fig.~\ref{fig:models}(e) in the main text.
We only take into account the nearest neighbor hopping and derive the kinetic part of the Hamiltonian. We impose an on-site potential and chemical potential on each lattice point. We call the point $1$, $2$, or $3$ depending on the vector $\bm{b}$ specifying the point. Every lattice point $\bm{r}$ is specified by $n_{1}$, $n_{2}$ and $i=1,2,3$:
\begin{align}
\bm{r} = n_{1}\bm{a}_{1} + n_{2}\bm{a}_{2} + \delta_{1i}\bm{b}_{1} + \delta_{2i}\bm{b}_{2} + \delta_{3i}\bm{b}_{3},
\end{align}
where
\begin{align}
\bm{a}_{1} = \mqty[2a \\ 0], \quad \bm{a}_{2} = \mqty[a \\ \sqrt{3}a], \quad \bm{b}_{1} = \mqty[a \\ 0], \quad \bm{b}_{2} = \mqty[ a/2 \\ \sqrt{3}a/2], \quad \bm{b}_{3} = \mqty[-a/2 \\ \sqrt{3}a/2],
\end{align}
and $a$ is the lattice constant. We take $a=1$ unit and $n_{1}$, $n_{2}$ are integers. Each lattice point has an expression
\begin{align}
\bm{r}\qty(n_{1}, n_{2}; 1) &= \mqty[ 2n_{1}+n_{2}+1 \\ \sqrt{3}n_{2} ], \\
\bm{r}\qty(n_{1}, n_{2}; 2) &= \mqty[ 2n_{1}+n_{2}+1/2 \\ \sqrt{3}n_{2}+\sqrt{3}/2 ], \\
\bm{r}\qty(n_{1}, n_{2}; 3) &= \mqty[ 2n_{1}+n_{2}-1/2 \\ \sqrt{3}n_{2}-\sqrt{3}/2 ].
\end{align}
The kinetic part of the Hamiltonian is put
\begin{align}
&\qty(t+\delta t)\sum_{n_{1}, n_{2}}\qty( c_{n_{1},n_{2};1}^{\dagger}c_{n_{1}, n_{2};2} + c_{n_{1},n_{2};2}^{\dagger}c_{n_{1}+1, n_{2};3} + c_{n_{1}+1,n_{2};3}^{\dagger}c_{n_{1}, n_{2};1} + \text{c.c.} ) \notag \\
&\quad +\qty( t-\delta t)\sum_{n_{1},n_{2}}\qty( c_{n_{1}-1,n_{2}+1;1}^{\dagger}c_{n_{1}, n_{2};2} + c_{n_{1},n_{2};2}^{\dagger}c_{n_{1}, n_{2};3} + c_{n_{1},n_{2};3}^{\dagger}c_{n_{1}-1, n_{2}+1;1} + \text{c.c.} ).
\end{align}
We can define the Fourier transform of annihilation operators as below.
\begin{align}
c_{n_{1}, n_{2}; 1\sigma} &= \frac{1}{N}\sum_{\bm{k}}c_{\bm{k},1\sigma}\exp[ \mathrm{i}\qty{ k_{x}\qty( 2n_{1} + n_{2} + 1) + k_{y}\cdot \sqrt{3}n_{2} } ], \\
c_{n_{1}, n_{2}; 2\sigma} &= \frac{1}{N}\sum_{\bm{k}}c_{\bm{k},2\sigma}\exp[ \mathrm{i}\qty{ k_{x}\qty( 2n_{1} + n_{2} + \frac{1}{2}) + k_{y}\qty( \sqrt{3}n_{2} + \frac{\sqrt{3}}{2} ) } ], \\
c_{n_{1}, n_{2}; 3\sigma} &= \frac{1}{N}\sum_{\bm{k}}c_{\bm{k},3\sigma}\exp[ \mathrm{i}\qty{ k_{x}\qty( 2n_{1} + n_{2} - \frac{1}{2}) + k_{y}\qty( \sqrt{3}n_{2} + \frac{\sqrt{3}}{2} ) } ].
\end{align}
Inserting the expressions into the original kinetic part of the Hamiltonian, we get that in the wavenumber space. Since the whole expression is lengthy, we write down the coefficients of the pairs of the operators $c^{\dagger}c$.
\begin{align}
c_{\bm{k},1}^{\dagger}c_{\bm{k},2}\quad &: \quad 2t\cos\qty(\frac{1}{2}k_{x} - \frac{\sqrt{3}}{2}k_{y}) + 2\qty(\delta t)\mathrm{i}\sin\qty( -\frac{1}{2}k_{x} + \frac{\sqrt{3}}{2}k_{y}), \\
c_{\bm{k},2}^{\dagger}c_{\bm{k},3}\quad &: \quad 2t\cos\qty(k_{x}) + 2\qty(\delta t)\mathrm{i}\sin\qty( k_{x}), \\
c_{\bm{k},3}^{\dagger}c_{\bm{k},1}\quad &: \quad 2t\cos\qty(\frac{1}{2}k_{x} + \frac{\sqrt{3}}{2}k_{y}) + 2\qty(\delta t)\mathrm{i}\sin\qty( -\frac{1}{2}k_{x} - \frac{\sqrt{3}}{2}k_{y}),  \\
c_{\bm{k},2}^{\dagger}c_{\bm{k},1}\quad &: \quad 2t\cos\qty(\frac{1}{2}k_{x} - \frac{\sqrt{3}}{2}k_{y}) + 2\qty(\delta t)\mathrm{i}\sin\qty( \frac{1}{2}k_{x} - \frac{\sqrt{3}}{2}k_{y}),  \\
c_{\bm{k},3}^{\dagger}c_{\bm{k},2}\quad &: \quad 2t\cos\qty(k_{x}) - 2\qty(\delta t)\mathrm{i}\sin\qty( k_{x}),  \\
c_{\bm{k},1}^{\dagger}c_{\bm{k},3}\quad &: \quad 2t\cos\qty(\frac{1}{2}k_{x} + \frac{\sqrt{3}}{2}k_{y}) + 2\qty(\delta t)\mathrm{i}\sin\qty( \frac{1}{2}k_{x} + \frac{\sqrt{3}}{2}k_{y}).
\end{align}

\section{Optical conductivities for the different parameter values}
\label{Apdx.E}
We show the numerical results of the linear optical conductivities for different parameter choices.
Since we see the $yy$ component of the linear optical conductivities in the main text, we first check other components $\sigma^{xx}\qty(\omega)$ and $\sigma^{xy}\qty(\omega)$. We are also interested in characteristic cases such as $\delta t= 0$ (the point group of the system is $D_{6h}$) and $\mu=-2t$ for $t<0$ (filling the flat band).
Other parameters are set to be the same in the main text.
The system of $\delta t=0$ is schematically depicted in the main text in Fig.~\ref{fig:models}(d). The results corresponding to the cases above are presented below.
\begin{figure}[h]
\centering
\includegraphics[scale=0.45]{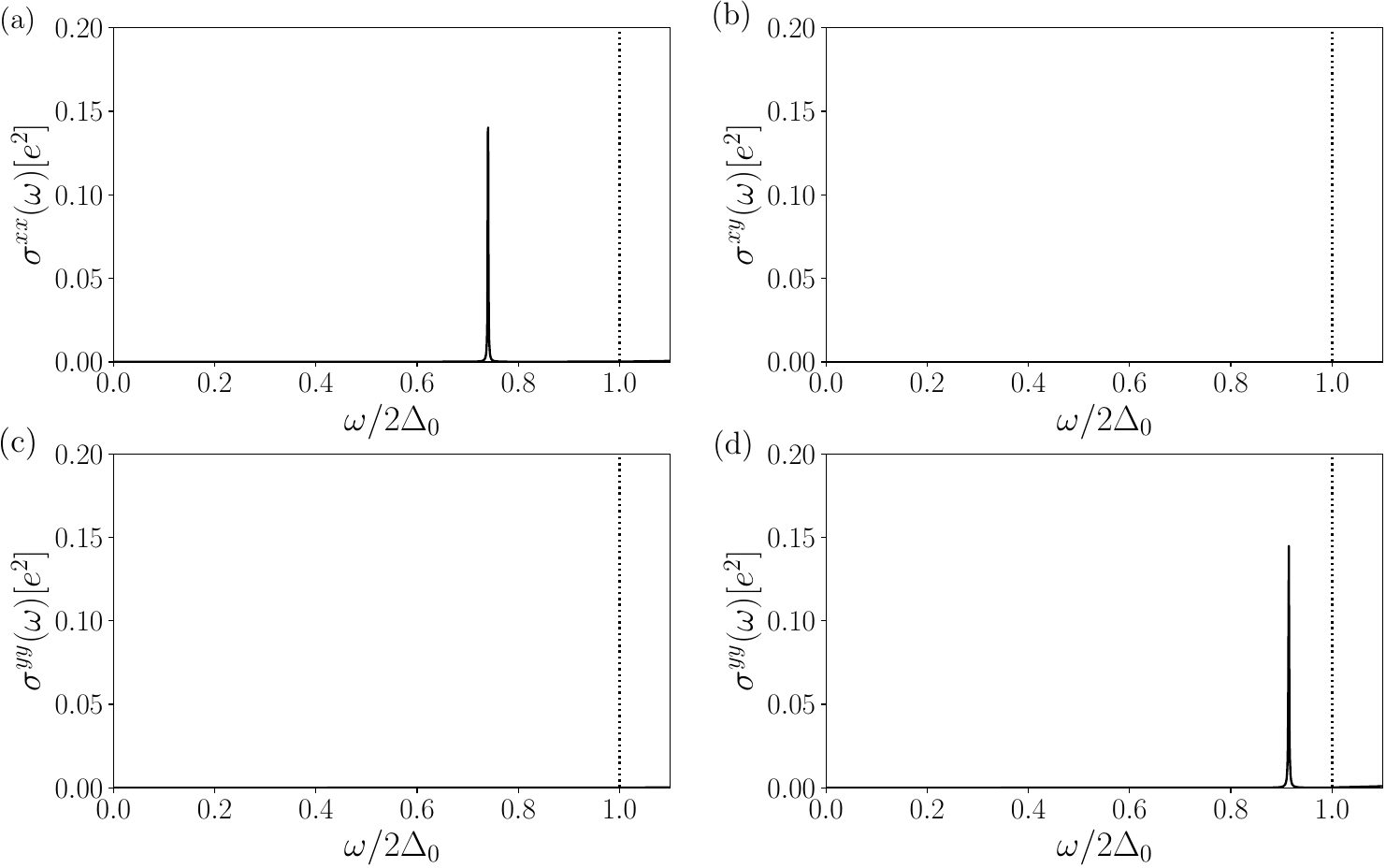}
\caption{Different cases of the linear optical conductivities in the Kagome lattice model. The parameter values are set to $t=-0.5$, $\delta t=-0.01$, $\mu=0$, $m=0$, and $U=6$. [(a),(b)] The linear optical conductivity of $xx$ (a) and $xy$ (b) component with the vertical dotted line representing the gap value. (c) The linear optical conductivity of $yy$ component $\sigma^{yy}\qty(\omega)$ with $\delta t=0$. (d) The linear optical conductivity of $yy$ component $\sigma^{yy}\qty(\omega)$ with $\mu=-2t$. Since the Kagome lattice model has a flat band at $E=-2t$ ($E$ is the energy) regardless of the value of $\delta t$, this flat band is filled in this case.}
\label{Fig7}
\end{figure}
The peaks in $\sigma^{xx}\qty(\omega)$ [Fig.~\ref{Fig7}(a)] and $\sigma^{yy}\qty(\omega)$ [Fig.~\ref{fig:Results}(b)] reflect the isotropic $s$-wave pairing symmetry of the order parameters. The fact that the peak does not appear in $\sigma^{xy}\qty(\omega)$ in Fig.~\ref{Fig7}(b) implies that the system does not break the time-reversal symmetry because the finite $\sigma^{xy}\qty(\omega)$ in the linear response regime is connected to the Hall-like response. When the system preserves the time-reversal symmetry and there is no external magnetic field, the Hall response should be zero, which is consistent with the disappearance of the peak. In Fig.~\ref{Fig7}(c), the peak does not come out, which is consistent with the group-theoretical prediction above. The result in Fig.~\ref{Fig7}(d) gives the case with a filled flat band at $E=-2t$, where $E$ is the energy. No qualitative change below the gap $2\Delta_{0}$ suggests that the flat band is not directly associated with the Leggett mode in linear response.

\twocolumngrid
\bibliography{main.bib}

\begin{thebibliography}{98}%
\makeatletter
\providecommand \@ifxundefined [1]{%
 \@ifx{#1\undefined}
}%
\providecommand \@ifnum [1]{%
 \ifnum #1\expandafter \@firstoftwo
 \else \expandafter \@secondoftwo
 \fi
}%
\providecommand \@ifx [1]{%
 \ifx #1\expandafter \@firstoftwo
 \else \expandafter \@secondoftwo
 \fi
}%
\providecommand \natexlab [1]{#1}%
\providecommand \enquote  [1]{``#1''}%
\providecommand \bibnamefont  [1]{#1}%
\providecommand \bibfnamefont [1]{#1}%
\providecommand \citenamefont [1]{#1}%
\providecommand \href@noop [0]{\@secondoftwo}%
\providecommand \href [0]{\begingroup \@sanitize@url \@href}%
\providecommand \@href[1]{\@@startlink{#1}\@@href}%
\providecommand \@@href[1]{\endgroup#1\@@endlink}%
\providecommand \@sanitize@url [0]{\catcode `\\12\catcode `\$12\catcode
  `\&12\catcode `\#12\catcode `\^12\catcode `\_12\catcode `\%12\relax}%
\providecommand \@@startlink[1]{}%
\providecommand \@@endlink[0]{}%
\providecommand \url  [0]{\begingroup\@sanitize@url \@url }%
\providecommand \@url [1]{\endgroup\@href {#1}{\urlprefix }}%
\providecommand \urlprefix  [0]{URL }%
\providecommand \Eprint [0]{\href }%
\providecommand \doibase [0]{https://doi.org/}%
\providecommand \selectlanguage [0]{\@gobble}%
\providecommand \bibinfo  [0]{\@secondoftwo}%
\providecommand \bibfield  [0]{\@secondoftwo}%
\providecommand \translation [1]{[#1]}%
\providecommand \BibitemOpen [0]{}%
\providecommand \bibitemStop [0]{}%
\providecommand \bibitemNoStop [0]{.\EOS\space}%
\providecommand \EOS [0]{\spacefactor3000\relax}%
\providecommand \BibitemShut  [1]{\csname bibitem#1\endcsname}%
\let\auto@bib@innerbib\@empty
\bibitem [{\citenamefont {Anderson}(1958)}]{Intro:1_Anderson}%
  \BibitemOpen
  \bibfield  {author} {\bibinfo {author} {\bibfnamefont {P.~W.}\ \bibnamefont
  {Anderson}},\ }\bibfield  {title} {\bibinfo {title} {{Random-Phase
  Approximation in the Theory of Superconductivity}},\ }\href
  {https://journals.aps.org/pr/abstract/10.1103/PhysRev.112.1900} {\bibfield
  {journal} {\bibinfo  {journal}
  {\href{https://journals.aps.org/pr/abstract/10.1103/PhysRev.112.1900}{Phys.
  Rev.}}\ }\textbf {\bibinfo {volume} {112}} (\bibinfo {year}
  {1958})}\BibitemShut {NoStop}%
\bibitem [{\citenamefont {Schmid}(1968)}]{Intro:2_Schmid}%
  \BibitemOpen
  \bibfield  {author} {\bibinfo {author} {\bibfnamefont {A.}~\bibnamefont
  {Schmid}},\ }\bibfield  {title} {\bibinfo {title} {{The approach to
  equilibrium in a pure superconductor the relaxation of the Cooper pair
  density}},\ }\href@noop {} {\bibfield  {journal} {\bibinfo  {journal}
  {\href{https://link.springer.com/article/10.1007/BF02422735}{Phys. Kondens.
  Mater.}}\ }\textbf {\bibinfo {volume} {8}} (\bibinfo {year}
  {1968})}\BibitemShut {NoStop}%
\bibitem [{\citenamefont {Littlewood}\ and\ \citenamefont
  {Varma}(1981)}]{Intro:3_Littlewood_Varma}%
  \BibitemOpen
  \bibfield  {author} {\bibinfo {author} {\bibfnamefont {P.~B.}\ \bibnamefont
  {Littlewood}}\ and\ \bibinfo {author} {\bibfnamefont {C.~M.}\ \bibnamefont
  {Varma}},\ }\bibfield  {title} {\bibinfo {title} {{Gauge-Invariant Theory of
  the Dynamical Interaction of Charge Density Waves and Superconductivity}},\
  }\href {https://doi.org/10.1103/PhysRevLett.47.811} {\bibfield  {journal}
  {\bibinfo  {journal} {Phys. Rev. Lett.}\ }\textbf {\bibinfo {volume} {47}},\
  \bibinfo {pages} {811} (\bibinfo {year} {1981})}\BibitemShut {NoStop}%
\bibitem [{\citenamefont {Littlewood}\ and\ \citenamefont
  {Varma}(1982)}]{Intro:4_Littlewood_Varma}%
  \BibitemOpen
  \bibfield  {author} {\bibinfo {author} {\bibfnamefont {P.~B.}\ \bibnamefont
  {Littlewood}}\ and\ \bibinfo {author} {\bibfnamefont {C.~M.}\ \bibnamefont
  {Varma}},\ }\bibfield  {title} {\bibinfo {title} {{Amplitude collective modes
  in superconductors and their coupling to charge-density waves}},\ }\href
  {https://doi.org/10.1103/PhysRevB.26.4883} {\bibfield  {journal} {\bibinfo
  {journal} {Phys. Rev. B}\ }\textbf {\bibinfo {volume} {26}},\ \bibinfo
  {pages} {4883} (\bibinfo {year} {1982})}\BibitemShut {NoStop}%
\bibitem [{\citenamefont {Pekker}\ and\ \citenamefont
  {Varma}(2015)}]{Intro:5_Pekker_Varma}%
  \BibitemOpen
  \bibfield  {author} {\bibinfo {author} {\bibfnamefont {D.}~\bibnamefont
  {Pekker}}\ and\ \bibinfo {author} {\bibfnamefont {C.}~\bibnamefont {Varma}},\
  }\bibfield  {title} {\bibinfo {title} {{Amplitude/Higgs Modes in Condensed
  Matter Physics }},\ }\href@noop {} {\bibfield  {journal} {\bibinfo  {journal}
  {\href{https://www.annualreviews.org/doi/abs/10.1146/annurev-conmatphys-031214-014350}{Annu.
  Rev. Condens. Matter Phys.}}\ }\textbf {\bibinfo {volume} {6}} (\bibinfo
  {year} {2015})}\BibitemShut {NoStop}%
\bibitem [{\citenamefont {Shimano}\ and\ \citenamefont
  {Tsuji}(2020)}]{Intro:6_Shimano_Tsuji}%
  \BibitemOpen
  \bibfield  {author} {\bibinfo {author} {\bibfnamefont {R.}~\bibnamefont
  {Shimano}}\ and\ \bibinfo {author} {\bibfnamefont {N.}~\bibnamefont
  {Tsuji}},\ }\bibfield  {title} {\bibinfo {title} {{Higgs Mode in
  Superconductors }},\ }\href@noop {} {\bibfield  {journal} {\bibinfo
  {journal}
  {\href{https://www.annualreviews.org/doi/abs/10.1146/annurev-conmatphys-031119-050813}{Annu.
  Rev. Condens. Matter Phys.}}\ }\textbf {\bibinfo {volume} {11}} (\bibinfo
  {year} {2020})}\BibitemShut {NoStop}%
\bibitem [{\citenamefont {Tsuji}\ \emph {et~al.}(2024)\citenamefont {Tsuji},
  \citenamefont {Danshita},\ and\ \citenamefont
  {Tsuchiya}}]{Intro:6_2_Tsuji-Ency}%
  \BibitemOpen
  \bibfield  {author} {\bibinfo {author} {\bibfnamefont {N.}~\bibnamefont
  {Tsuji}}, \bibinfo {author} {\bibfnamefont {I.}~\bibnamefont {Danshita}},\
  and\ \bibinfo {author} {\bibfnamefont {S.}~\bibnamefont {Tsuchiya}},\
  }\bibfield  {title} {\bibinfo {title} {{Higgs and Nambu–Goldstone modes in
  condensed matter physics}},\ }in\ \href
  {https://doi.org/https://doi.org/10.1016/B978-0-323-90800-9.00256-0} {\emph
  {\bibinfo {booktitle} {Encyclopedia of Condensed Matter Physics (Second
  Edition)}}}\ (\bibinfo  {publisher} {Academic Press},\ \bibinfo {address}
  {Oxford},\ \bibinfo {year} {2024})\ \bibinfo {edition} {{2nd}}\ ed.,\ pp.\
  \bibinfo {pages} {174--186}\BibitemShut {NoStop}%
\bibitem [{\citenamefont {Nambu}\ and\ \citenamefont
  {Jona-Lasinio}(1961)}]{Intro_NG1}%
  \BibitemOpen
  \bibfield  {author} {\bibinfo {author} {\bibfnamefont {Y.}~\bibnamefont
  {Nambu}}\ and\ \bibinfo {author} {\bibfnamefont {G.}~\bibnamefont
  {Jona-Lasinio}},\ }\bibfield  {title} {\bibinfo {title} {{Dynamical Model of
  Elementary Particles Based on an Analogy with Superconductivity. I}},\ }\href
  {https://doi.org/10.1103/PhysRev.122.345} {\bibfield  {journal} {\bibinfo
  {journal} {Phys. Rev.}\ }\textbf {\bibinfo {volume} {122}},\ \bibinfo {pages}
  {345} (\bibinfo {year} {1961})}\BibitemShut {NoStop}%
\bibitem [{\citenamefont {Goldstone}(1961)}]{Intro_NG2}%
  \BibitemOpen
  \bibfield  {author} {\bibinfo {author} {\bibfnamefont {J.}~\bibnamefont
  {Goldstone}},\ }\bibfield  {title} {\bibinfo {title} {{Field Theories with
  ``Superconductor" Solutions}},\ }\href
  {https://doi.org/https://doi.org/10.1007/BF02812722} {\bibfield  {journal}
  {\bibinfo  {journal} {Nuovo Cim.}\ }\textbf {\bibinfo {volume} {19}},\
  \bibinfo {pages} {154} (\bibinfo {year} {1961})}\BibitemShut {NoStop}%
\bibitem [{\citenamefont {Anderson}(1963)}]{Intro:Anderson-Higgs1}%
  \BibitemOpen
  \bibfield  {author} {\bibinfo {author} {\bibfnamefont {P.~W.}\ \bibnamefont
  {Anderson}},\ }\bibfield  {title} {\bibinfo {title} {{Plasmons, Gauge
  Invariance, and Mass}},\ }\href {https://doi.org/10.1103/PhysRev.130.439}
  {\bibfield  {journal} {\bibinfo  {journal} {Phys. Rev.}\ }\textbf {\bibinfo
  {volume} {130}},\ \bibinfo {pages} {439} (\bibinfo {year}
  {1963})}\BibitemShut {NoStop}%
\bibitem [{\citenamefont {Higgs}(1964)}]{Intro:Anderson-Higgs2}%
  \BibitemOpen
  \bibfield  {author} {\bibinfo {author} {\bibfnamefont {P.~W.}\ \bibnamefont
  {Higgs}},\ }\bibfield  {title} {\bibinfo {title} {{Broken Symmetries and the
  Masses of Gauge Bosons}},\ }\href
  {https://doi.org/10.1103/PhysRevLett.13.508} {\bibfield  {journal} {\bibinfo
  {journal} {Phys. Rev. Lett.}\ }\textbf {\bibinfo {volume} {13}},\ \bibinfo
  {pages} {508} (\bibinfo {year} {1964})}\BibitemShut {NoStop}%
\bibitem [{\citenamefont {Tsuji}\ and\ \citenamefont
  {Aoki}(2015)}]{Intro:HiggsMode1}%
  \BibitemOpen
  \bibfield  {author} {\bibinfo {author} {\bibfnamefont {N.}~\bibnamefont
  {Tsuji}}\ and\ \bibinfo {author} {\bibfnamefont {H.}~\bibnamefont {Aoki}},\
  }\bibfield  {title} {\bibinfo {title} {{Theory of Anderson pseudospin
  resonance with Higgs mode in superconductors}},\ }\href
  {https://doi.org/10.1103/PhysRevB.92.064508} {\bibfield  {journal} {\bibinfo
  {journal} {Phys. Rev. B}\ }\textbf {\bibinfo {volume} {92}},\ \bibinfo
  {pages} {064508} (\bibinfo {year} {2015})}\BibitemShut {NoStop}%
\bibitem [{\citenamefont {Kemper}\ \emph {et~al.}(2015)\citenamefont {Kemper},
  \citenamefont {Sentef}, \citenamefont {Moritz}, \citenamefont {Freericks},\
  and\ \citenamefont {Devereaux}}]{Intro:HiggsMode2}%
  \BibitemOpen
  \bibfield  {author} {\bibinfo {author} {\bibfnamefont {A.~F.}\ \bibnamefont
  {Kemper}}, \bibinfo {author} {\bibfnamefont {M.~A.}\ \bibnamefont {Sentef}},
  \bibinfo {author} {\bibfnamefont {B.}~\bibnamefont {Moritz}}, \bibinfo
  {author} {\bibfnamefont {J.~K.}\ \bibnamefont {Freericks}},\ and\ \bibinfo
  {author} {\bibfnamefont {T.~P.}\ \bibnamefont {Devereaux}},\ }\bibfield
  {title} {\bibinfo {title} {{Direct observation of Higgs mode oscillations in
  the pump-probe photoemission spectra of electron-phonon mediated
  superconductors}},\ }\href {https://doi.org/10.1103/PhysRevB.92.224517}
  {\bibfield  {journal} {\bibinfo  {journal} {Phys. Rev. B}\ }\textbf {\bibinfo
  {volume} {92}},\ \bibinfo {pages} {224517} (\bibinfo {year}
  {2015})}\BibitemShut {NoStop}%
\bibitem [{\citenamefont {Cea}\ \emph {et~al.}(2016)\citenamefont {Cea},
  \citenamefont {Castellani},\ and\ \citenamefont
  {Benfatto}}]{Intro:HiggsMode3}%
  \BibitemOpen
  \bibfield  {author} {\bibinfo {author} {\bibfnamefont {T.}~\bibnamefont
  {Cea}}, \bibinfo {author} {\bibfnamefont {C.}~\bibnamefont {Castellani}},\
  and\ \bibinfo {author} {\bibfnamefont {L.}~\bibnamefont {Benfatto}},\
  }\bibfield  {title} {\bibinfo {title} {{Nonlinear optical effects and
  third-harmonic generation in superconductors: Cooper pairs versus Higgs mode
  contribution}},\ }\href {https://doi.org/10.1103/PhysRevB.93.180507}
  {\bibfield  {journal} {\bibinfo  {journal} {Phys. Rev. B}\ }\textbf {\bibinfo
  {volume} {93}},\ \bibinfo {pages} {180507} (\bibinfo {year}
  {2016})}\BibitemShut {NoStop}%
\bibitem [{\citenamefont {Tsuji}\ \emph {et~al.}(2016)\citenamefont {Tsuji},
  \citenamefont {Murakami},\ and\ \citenamefont {Aoki}}]{Intro:HiggsMode4}%
  \BibitemOpen
  \bibfield  {author} {\bibinfo {author} {\bibfnamefont {N.}~\bibnamefont
  {Tsuji}}, \bibinfo {author} {\bibfnamefont {Y.}~\bibnamefont {Murakami}},\
  and\ \bibinfo {author} {\bibfnamefont {H.}~\bibnamefont {Aoki}},\ }\bibfield
  {title} {\bibinfo {title} {{Nonlinear light--Higgs coupling in
  superconductors beyond BCS: Effects of the retarded phonon-mediated
  interaction}},\ }\href {https://doi.org/10.1103/PhysRevB.94.224519}
  {\bibfield  {journal} {\bibinfo  {journal} {Phys. Rev. B}\ }\textbf {\bibinfo
  {volume} {94}},\ \bibinfo {pages} {224519} (\bibinfo {year}
  {2016})}\BibitemShut {NoStop}%
\bibitem [{\citenamefont {Jujo}(2018)}]{Intro:HiggsMode5}%
  \BibitemOpen
  \bibfield  {author} {\bibinfo {author} {\bibfnamefont {T.}~\bibnamefont
  {Jujo}},\ }\bibfield  {title} {\bibinfo {title} {{Quasiclassical Theory on
  Third-Harmonic Generation in Conventional Superconductors with Paramagnetic
  Impurities}},\ }\href {https://doi.org/10.7566/JPSJ.87.024704} {\bibfield
  {journal} {\bibinfo  {journal} {J. Phys. Soc. Jpn.}\ }\textbf {\bibinfo
  {volume} {87}},\ \bibinfo {pages} {024704} (\bibinfo {year}
  {2018})}\BibitemShut {NoStop}%
\bibitem [{\citenamefont {Silaev}(2019)}]{Intro:HiggsMode6}%
  \BibitemOpen
  \bibfield  {author} {\bibinfo {author} {\bibfnamefont {M.}~\bibnamefont
  {Silaev}},\ }\bibfield  {title} {\bibinfo {title} {{Nonlinear electromagnetic
  response and Higgs-mode excitation in BCS superconductors with impurities}},\
  }\href {https://doi.org/10.1103/PhysRevB.99.224511} {\bibfield  {journal}
  {\bibinfo  {journal} {Phys. Rev. B}\ }\textbf {\bibinfo {volume} {99}},\
  \bibinfo {pages} {224511} (\bibinfo {year} {2019})}\BibitemShut {NoStop}%
\bibitem [{\citenamefont {{Schwarz, L. and Fauseweh, B. and Tsuji, N. and
  Cheng, N. and Bittner, N. and Krull, H. and Berciu, M. and Uhrig, G. S. and
  Schnyder, A. P. and Kaiser, S. and Manske, D.}}(2020)}]{Intro:HiggsMode7}%
  \BibitemOpen
  \bibfield  {author} {\bibinfo {author} {\bibnamefont {{Schwarz, L. and
  Fauseweh, B. and Tsuji, N. and Cheng, N. and Bittner, N. and Krull, H. and
  Berciu, M. and Uhrig, G. S. and Schnyder, A. P. and Kaiser, S. and Manske,
  D.}}},\ }\bibfield  {title} {\bibinfo {title} {{Classification and
  characterization of nonequilibrium Higgs modes in unconventional
  superconductors }},\ }\href@noop {} {\bibfield  {journal} {\bibinfo
  {journal} {\href{https://www.nature.com/articles/s41467-019-13763-5}{Nat.
  Commun.}}\ }\textbf {\bibinfo {volume} {11}} (\bibinfo {year}
  {2020})}\BibitemShut {NoStop}%
\bibitem [{\citenamefont {Tsuji}\ and\ \citenamefont
  {Nomura}(2020)}]{Intro:HiggsMode8}%
  \BibitemOpen
  \bibfield  {author} {\bibinfo {author} {\bibfnamefont {N.}~\bibnamefont
  {Tsuji}}\ and\ \bibinfo {author} {\bibfnamefont {Y.}~\bibnamefont {Nomura}},\
  }\bibfield  {title} {\bibinfo {title} {{Higgs-mode resonance in third
  harmonic generation in NbN superconductors: Multiband electron-phonon
  coupling, impurity scattering, and polarization-angle dependence}},\ }\href
  {https://doi.org/10.1103/PhysRevResearch.2.043029} {\bibfield  {journal}
  {\bibinfo  {journal} {Phys. Rev. Res.}\ }\textbf {\bibinfo {volume} {2}},\
  \bibinfo {pages} {043029} (\bibinfo {year} {2020})}\BibitemShut {NoStop}%
\bibitem [{\citenamefont {Haenel}\ \emph {et~al.}(2021)\citenamefont {Haenel},
  \citenamefont {Froese}, \citenamefont {Manske},\ and\ \citenamefont
  {Schwarz}}]{Intro:HiggsMode9}%
  \BibitemOpen
  \bibfield  {author} {\bibinfo {author} {\bibfnamefont {R.}~\bibnamefont
  {Haenel}}, \bibinfo {author} {\bibfnamefont {P.}~\bibnamefont {Froese}},
  \bibinfo {author} {\bibfnamefont {D.}~\bibnamefont {Manske}},\ and\ \bibinfo
  {author} {\bibfnamefont {L.}~\bibnamefont {Schwarz}},\ }\bibfield  {title}
  {\bibinfo {title} {{Time-resolved optical conductivity and Higgs oscillations
  in two-band dirty superconductors}},\ }\href
  {https://doi.org/10.1103/PhysRevB.104.134504} {\bibfield  {journal} {\bibinfo
   {journal} {Phys. Rev. B}\ }\textbf {\bibinfo {volume} {104}},\ \bibinfo
  {pages} {134504} (\bibinfo {year} {2021})}\BibitemShut {NoStop}%
\bibitem [{\citenamefont {{M. Udina, J. Fiore, T. Cea, C. Castellani, G.
  Seibold, and L. Benfatto}}(2022)}]{Intro:HiggsMode10}%
  \BibitemOpen
  \bibfield  {author} {\bibinfo {author} {\bibnamefont {{M. Udina, J. Fiore, T.
  Cea, C. Castellani, G. Seibold, and L. Benfatto}}},\ }\bibfield  {title}
  {\bibinfo {title} {{THz non-linear optical response in cuprates: predominance
  of the BCS response over the Higgs mode }},\ }\href@noop {} {\bibfield
  {journal} {\bibinfo  {journal}
  {\href{https://pubs.rsc.org/en/content/articlelanding/2022/fd/d2fd00016d}{Faraday
  Discuss.}}\ }\textbf {\bibinfo {volume} {237}} (\bibinfo {year}
  {2022})}\BibitemShut {NoStop}%
\bibitem [{\citenamefont {Matsunaga}\ \emph {et~al.}(2013)\citenamefont
  {Matsunaga}, \citenamefont {Hamada}, \citenamefont {Makise}, \citenamefont
  {Uzawa}, \citenamefont {Terai}, \citenamefont {Wang},\ and\ \citenamefont
  {Shimano}}]{Intro:Ex_PumpProbe_HiggsMode_Matsunaga2013}%
  \BibitemOpen
  \bibfield  {author} {\bibinfo {author} {\bibfnamefont {R.}~\bibnamefont
  {Matsunaga}}, \bibinfo {author} {\bibfnamefont {Y.~I.}\ \bibnamefont
  {Hamada}}, \bibinfo {author} {\bibfnamefont {K.}~\bibnamefont {Makise}},
  \bibinfo {author} {\bibfnamefont {Y.}~\bibnamefont {Uzawa}}, \bibinfo
  {author} {\bibfnamefont {H.}~\bibnamefont {Terai}}, \bibinfo {author}
  {\bibfnamefont {Z.}~\bibnamefont {Wang}},\ and\ \bibinfo {author}
  {\bibfnamefont {R.}~\bibnamefont {Shimano}},\ }\bibfield  {title} {\bibinfo
  {title} {{Higgs Amplitude Mode in the BCS Superconductors
  ${\mathrm{Nb}}_{1\mathrm{\text{\ensuremath{-}}}x}{\mathrm{Ti}}_{x}\mathbf{N}$
  Induced by Terahertz Pulse Excitation}},\ }\href
  {https://doi.org/10.1103/PhysRevLett.111.057002} {\bibfield  {journal}
  {\bibinfo  {journal} {Phys. Rev. Lett.}\ }\textbf {\bibinfo {volume} {111}},\
  \bibinfo {pages} {057002} (\bibinfo {year} {2013})}\BibitemShut {NoStop}%
\bibitem [{\citenamefont {Matsunaga}\ \emph {et~al.}(2014)\citenamefont
  {Matsunaga}, \citenamefont {Tsuji}, \citenamefont {Fujita}, \citenamefont
  {Sugioka}, \citenamefont {Makise}, \citenamefont {Uzawa}, \citenamefont
  {Terai}, \citenamefont {Wang}, \citenamefont {Aoki},\ and\ \citenamefont
  {Shimano}}]{Intro:Ex_THG_HiggsMode1}%
  \BibitemOpen
  \bibfield  {author} {\bibinfo {author} {\bibfnamefont {R.}~\bibnamefont
  {Matsunaga}}, \bibinfo {author} {\bibfnamefont {N.}~\bibnamefont {Tsuji}},
  \bibinfo {author} {\bibfnamefont {H.}~\bibnamefont {Fujita}}, \bibinfo
  {author} {\bibfnamefont {A.}~\bibnamefont {Sugioka}}, \bibinfo {author}
  {\bibfnamefont {K.}~\bibnamefont {Makise}}, \bibinfo {author} {\bibfnamefont
  {Y.}~\bibnamefont {Uzawa}}, \bibinfo {author} {\bibfnamefont
  {H.}~\bibnamefont {Terai}}, \bibinfo {author} {\bibfnamefont
  {Z.}~\bibnamefont {Wang}}, \bibinfo {author} {\bibfnamefont {H.}~\bibnamefont
  {Aoki}},\ and\ \bibinfo {author} {\bibfnamefont {R.}~\bibnamefont
  {Shimano}},\ }\bibfield  {title} {\bibinfo {title} {{Light-induced collective
  pseudospin precession resonating with Higgs mode in a superconductor }},\
  }\href@noop {} {\bibfield  {journal} {\bibinfo  {journal}
  {\href{https://www.science.org/doi/10.1126/science.1254697}{Science}}\
  }\textbf {\bibinfo {volume} {345}} (\bibinfo {year} {2014})}\BibitemShut
  {NoStop}%
\bibitem [{\citenamefont {Matsunaga}\ \emph {et~al.}(2017)\citenamefont
  {Matsunaga}, \citenamefont {Tsuji}, \citenamefont {Makise}, \citenamefont
  {Terai}, \citenamefont {Aoki},\ and\ \citenamefont
  {Shimano}}]{Intro:Ex_PumpProbeHiggsMode_Matsunaga2017}%
  \BibitemOpen
  \bibfield  {author} {\bibinfo {author} {\bibfnamefont {R.}~\bibnamefont
  {Matsunaga}}, \bibinfo {author} {\bibfnamefont {N.}~\bibnamefont {Tsuji}},
  \bibinfo {author} {\bibfnamefont {K.}~\bibnamefont {Makise}}, \bibinfo
  {author} {\bibfnamefont {H.}~\bibnamefont {Terai}}, \bibinfo {author}
  {\bibfnamefont {H.}~\bibnamefont {Aoki}},\ and\ \bibinfo {author}
  {\bibfnamefont {R.}~\bibnamefont {Shimano}},\ }\bibfield  {title} {\bibinfo
  {title} {{Polarization-resolved terahertz third-harmonic generation in a
  single-crystal superconductor NbN: Dominance of the Higgs mode beyond the BCS
  approximation}},\ }\href {https://doi.org/10.1103/PhysRevB.96.020505}
  {\bibfield  {journal} {\bibinfo  {journal} {Phys. Rev. B}\ }\textbf {\bibinfo
  {volume} {96}},\ \bibinfo {pages} {020505} (\bibinfo {year}
  {2017})}\BibitemShut {NoStop}%
\bibitem [{\citenamefont {Katsumi}\ \emph {et~al.}(2018)\citenamefont
  {Katsumi}, \citenamefont {Tsuji}, \citenamefont {Hamada}, \citenamefont
  {Matsunaga}, \citenamefont {Schneeloch}, \citenamefont {Zhong}, \citenamefont
  {Gu}, \citenamefont {Aoki}, \citenamefont {Gallais},\ and\ \citenamefont
  {Shimano}}]{Intro:Ex_PumpProbe_HiggsMode2}%
  \BibitemOpen
  \bibfield  {author} {\bibinfo {author} {\bibfnamefont {K.}~\bibnamefont
  {Katsumi}}, \bibinfo {author} {\bibfnamefont {N.}~\bibnamefont {Tsuji}},
  \bibinfo {author} {\bibfnamefont {Y.~I.}\ \bibnamefont {Hamada}}, \bibinfo
  {author} {\bibfnamefont {R.}~\bibnamefont {Matsunaga}}, \bibinfo {author}
  {\bibfnamefont {J.}~\bibnamefont {Schneeloch}}, \bibinfo {author}
  {\bibfnamefont {R.~D.}\ \bibnamefont {Zhong}}, \bibinfo {author}
  {\bibfnamefont {G.~D.}\ \bibnamefont {Gu}}, \bibinfo {author} {\bibfnamefont
  {H.}~\bibnamefont {Aoki}}, \bibinfo {author} {\bibfnamefont {Y.}~\bibnamefont
  {Gallais}},\ and\ \bibinfo {author} {\bibfnamefont {R.}~\bibnamefont
  {Shimano}},\ }\bibfield  {title} {\bibinfo {title} {{Higgs Mode in the
  $d$-Wave Superconductor
  $\mathrm{Bi}_{2}\mathrm{Sr}_{2}\mathrm{CaCu}_{2}\mathrm{O}_{8+x}$ Driven by
  an Intense Terahertz Pulse}},\ }\href
  {https://doi.org/10.1103/PhysRevLett.120.117001} {\bibfield  {journal}
  {\bibinfo  {journal} {Phys. Rev. Lett.}\ }\textbf {\bibinfo {volume} {120}},\
  \bibinfo {pages} {117001} (\bibinfo {year} {2018})}\BibitemShut {NoStop}%
\bibitem [{\citenamefont {Chu}\ \emph {et~al.}(2020)\citenamefont {Chu},
  \citenamefont {Kim}, \citenamefont {Katsumi}, \citenamefont {Kovalev},
  \citenamefont {Dawson}, \citenamefont {Schwarz}, \citenamefont {Yoshikawa},
  \citenamefont {Kim}, \citenamefont {Putzky}, \citenamefont {Li},
  \citenamefont {Raffy}, \citenamefont {Germanskiy}, \citenamefont {Deinert},
  \citenamefont {Awari}, \citenamefont {Ilyakov}, \citenamefont {Green},
  \citenamefont {Chen}, \citenamefont {Bawatna}, \citenamefont {Cristiani},
  \citenamefont {Logvenov}, \citenamefont {Gallais}, \citenamefont {Boris},
  \citenamefont {Keimer}, \citenamefont {Schnyder}, \citenamefont {Manske},
  \citenamefont {Gensch}, \citenamefont {Wang}, \citenamefont {Shimano},\ and\
  \citenamefont {Kaiser}}]{Intro:Ex_THG_HiggsMode2}%
  \BibitemOpen
  \bibfield  {author} {\bibinfo {author} {\bibfnamefont {H.}~\bibnamefont
  {Chu}}, \bibinfo {author} {\bibfnamefont {M.-J.}\ \bibnamefont {Kim}},
  \bibinfo {author} {\bibfnamefont {K.}~\bibnamefont {Katsumi}}, \bibinfo
  {author} {\bibfnamefont {S.}~\bibnamefont {Kovalev}}, \bibinfo {author}
  {\bibfnamefont {R.~D.}\ \bibnamefont {Dawson}}, \bibinfo {author}
  {\bibfnamefont {L.}~\bibnamefont {Schwarz}}, \bibinfo {author} {\bibfnamefont
  {N.}~\bibnamefont {Yoshikawa}}, \bibinfo {author} {\bibfnamefont
  {G.}~\bibnamefont {Kim}}, \bibinfo {author} {\bibfnamefont {D.}~\bibnamefont
  {Putzky}}, \bibinfo {author} {\bibfnamefont {Z.~Z.}\ \bibnamefont {Li}},
  \bibinfo {author} {\bibfnamefont {H.}~\bibnamefont {Raffy}}, \bibinfo
  {author} {\bibfnamefont {S.}~\bibnamefont {Germanskiy}}, \bibinfo {author}
  {\bibfnamefont {J.-C.}\ \bibnamefont {Deinert}}, \bibinfo {author}
  {\bibfnamefont {N.}~\bibnamefont {Awari}}, \bibinfo {author} {\bibfnamefont
  {I.}~\bibnamefont {Ilyakov}}, \bibinfo {author} {\bibfnamefont
  {B.}~\bibnamefont {Green}}, \bibinfo {author} {\bibfnamefont
  {M.}~\bibnamefont {Chen}}, \bibinfo {author} {\bibfnamefont {M.}~\bibnamefont
  {Bawatna}}, \bibinfo {author} {\bibfnamefont {G.}~\bibnamefont {Cristiani}},
  \bibinfo {author} {\bibfnamefont {G.}~\bibnamefont {Logvenov}}, \bibinfo
  {author} {\bibfnamefont {Y.}~\bibnamefont {Gallais}}, \bibinfo {author}
  {\bibfnamefont {A.~V.}\ \bibnamefont {Boris}}, \bibinfo {author}
  {\bibfnamefont {B.}~\bibnamefont {Keimer}}, \bibinfo {author} {\bibfnamefont
  {A.~P.}\ \bibnamefont {Schnyder}}, \bibinfo {author} {\bibfnamefont
  {D.}~\bibnamefont {Manske}}, \bibinfo {author} {\bibfnamefont
  {M.}~\bibnamefont {Gensch}}, \bibinfo {author} {\bibfnamefont
  {Z.}~\bibnamefont {Wang}}, \bibinfo {author} {\bibfnamefont {R.}~\bibnamefont
  {Shimano}},\ and\ \bibinfo {author} {\bibfnamefont {S.}~\bibnamefont
  {Kaiser}},\ }\bibfield  {title} {\bibinfo {title} {{Phase-resolved Higgs
  response in superconducting cuprates }},\ }\href@noop {} {\bibfield
  {journal} {\bibinfo  {journal}
  {\href{https://www.nature.com/articles/s41467-020-15613-1}{Nat. Commun.}}\
  }\textbf {\bibinfo {volume} {11}} (\bibinfo {year} {2020})}\BibitemShut
  {NoStop}%
\bibitem [{\citenamefont {Sooryakumar}\ and\ \citenamefont
  {Klein}(1980)}]{Intro:Ex_Raman1}%
  \BibitemOpen
  \bibfield  {author} {\bibinfo {author} {\bibfnamefont {R.}~\bibnamefont
  {Sooryakumar}}\ and\ \bibinfo {author} {\bibfnamefont {M.~V.}\ \bibnamefont
  {Klein}},\ }\bibfield  {title} {\bibinfo {title} {{Raman Scattering by
  Superconducting-Gap Excitations and Their Coupling to Charge-Density
  Waves}},\ }\href {https://doi.org/10.1103/PhysRevLett.45.660} {\bibfield
  {journal} {\bibinfo  {journal} {Phys. Rev. Lett.}\ }\textbf {\bibinfo
  {volume} {45}},\ \bibinfo {pages} {660} (\bibinfo {year} {1980})}\BibitemShut
  {NoStop}%
\bibitem [{\citenamefont {Sooryakumar}\ and\ \citenamefont
  {Klein}(1981)}]{Intro:Ex_Raman2}%
  \BibitemOpen
  \bibfield  {author} {\bibinfo {author} {\bibfnamefont {R.}~\bibnamefont
  {Sooryakumar}}\ and\ \bibinfo {author} {\bibfnamefont {M.~V.}\ \bibnamefont
  {Klein}},\ }\bibfield  {title} {\bibinfo {title} {{Raman scattering from
  superconducting gap excitations in the presence of a magnetic field}},\
  }\href {https://doi.org/10.1103/PhysRevB.23.3213} {\bibfield  {journal}
  {\bibinfo  {journal} {Phys. Rev. B}\ }\textbf {\bibinfo {volume} {23}},\
  \bibinfo {pages} {3213} (\bibinfo {year} {1981})}\BibitemShut {NoStop}%
\bibitem [{\citenamefont {M\'easson}\ \emph {et~al.}(2014)\citenamefont
  {M\'easson}, \citenamefont {Gallais}, \citenamefont {Cazayous}, \citenamefont
  {Clair}, \citenamefont {Rodi\`ere}, \citenamefont {Cario},\ and\
  \citenamefont {Sacuto}}]{Intro:Ex_Raman3}%
  \BibitemOpen
  \bibfield  {author} {\bibinfo {author} {\bibfnamefont {M.-A.}\ \bibnamefont
  {M\'easson}}, \bibinfo {author} {\bibfnamefont {Y.}~\bibnamefont {Gallais}},
  \bibinfo {author} {\bibfnamefont {M.}~\bibnamefont {Cazayous}}, \bibinfo
  {author} {\bibfnamefont {B.}~\bibnamefont {Clair}}, \bibinfo {author}
  {\bibfnamefont {P.}~\bibnamefont {Rodi\`ere}}, \bibinfo {author}
  {\bibfnamefont {L.}~\bibnamefont {Cario}},\ and\ \bibinfo {author}
  {\bibfnamefont {A.}~\bibnamefont {Sacuto}},\ }\bibfield  {title} {\bibinfo
  {title} {{Amplitude Higgs mode in the $2\mathit{H}$-$\mathrm{NbSe}_{2}$
  superconductor}},\ }\href {https://doi.org/10.1103/PhysRevB.89.060503}
  {\bibfield  {journal} {\bibinfo  {journal} {Phys. Rev. B}\ }\textbf {\bibinfo
  {volume} {89}},\ \bibinfo {pages} {060503} (\bibinfo {year}
  {2014})}\BibitemShut {NoStop}%
\bibitem [{\citenamefont {Grasset}\ \emph {et~al.}(2018)\citenamefont
  {Grasset}, \citenamefont {Cea}, \citenamefont {Gallais}, \citenamefont
  {Cazayous}, \citenamefont {Sacuto}, \citenamefont {Cario}, \citenamefont
  {Benfatto},\ and\ \citenamefont {M\'easson}}]{Intro:Ex_Raman4}%
  \BibitemOpen
  \bibfield  {author} {\bibinfo {author} {\bibfnamefont {R.}~\bibnamefont
  {Grasset}}, \bibinfo {author} {\bibfnamefont {T.}~\bibnamefont {Cea}},
  \bibinfo {author} {\bibfnamefont {Y.}~\bibnamefont {Gallais}}, \bibinfo
  {author} {\bibfnamefont {M.}~\bibnamefont {Cazayous}}, \bibinfo {author}
  {\bibfnamefont {A.}~\bibnamefont {Sacuto}}, \bibinfo {author} {\bibfnamefont
  {L.}~\bibnamefont {Cario}}, \bibinfo {author} {\bibfnamefont
  {L.}~\bibnamefont {Benfatto}},\ and\ \bibinfo {author} {\bibfnamefont
  {M.-A.}\ \bibnamefont {M\'easson}},\ }\bibfield  {title} {\bibinfo {title}
  {{Higgs-mode radiance and charge-density-wave order in
  $2\mathit{H}$-$\mathrm{NbSe}_{2}$}},\ }\href
  {https://doi.org/10.1103/PhysRevB.97.094502} {\bibfield  {journal} {\bibinfo
  {journal} {Phys. Rev. B}\ }\textbf {\bibinfo {volume} {97}},\ \bibinfo
  {pages} {094502} (\bibinfo {year} {2018})}\BibitemShut {NoStop}%
\bibitem [{\citenamefont {Grasset}\ \emph {et~al.}(2019)\citenamefont
  {Grasset}, \citenamefont {Gallais}, \citenamefont {Sacuto}, \citenamefont
  {Cazayous}, \citenamefont {Ma\~nas Valero}, \citenamefont {Coronado},\ and\
  \citenamefont {M\'easson}}]{Intro:Ex_Raman5}%
  \BibitemOpen
  \bibfield  {author} {\bibinfo {author} {\bibfnamefont {R.}~\bibnamefont
  {Grasset}}, \bibinfo {author} {\bibfnamefont {Y.}~\bibnamefont {Gallais}},
  \bibinfo {author} {\bibfnamefont {A.}~\bibnamefont {Sacuto}}, \bibinfo
  {author} {\bibfnamefont {M.}~\bibnamefont {Cazayous}}, \bibinfo {author}
  {\bibfnamefont {S.}~\bibnamefont {Ma\~nas Valero}}, \bibinfo {author}
  {\bibfnamefont {E.}~\bibnamefont {Coronado}},\ and\ \bibinfo {author}
  {\bibfnamefont {M.-A.}\ \bibnamefont {M\'easson}},\ }\bibfield  {title}
  {\bibinfo {title} {{Pressure-Induced Collapse of the Charge Density Wave and
  Higgs Mode Visibility in $2\mathit{H}$-$\mathrm{TaS}_{2}$}},\ }\href
  {https://doi.org/10.1103/PhysRevLett.122.127001} {\bibfield  {journal}
  {\bibinfo  {journal} {Phys. Rev. Lett.}\ }\textbf {\bibinfo {volume} {122}},\
  \bibinfo {pages} {127001} (\bibinfo {year} {2019})}\BibitemShut {NoStop}%
\bibitem [{\citenamefont {Majumdar}\ \emph {et~al.}(2020)\citenamefont
  {Majumdar}, \citenamefont {VanGennep}, \citenamefont {Brisbois},
  \citenamefont {Chareev}, \citenamefont {Sadakov}, \citenamefont {Usoltsev},
  \citenamefont {Mito}, \citenamefont {Silhanek}, \citenamefont {Sarkar},
  \citenamefont {Hassan}, \citenamefont {Karis}, \citenamefont {Ahuja},\ and\
  \citenamefont {Abdel-Hafiez}}]{Intro:RamanRelated}%
  \BibitemOpen
  \bibfield  {author} {\bibinfo {author} {\bibfnamefont {A.}~\bibnamefont
  {Majumdar}}, \bibinfo {author} {\bibfnamefont {D.}~\bibnamefont {VanGennep}},
  \bibinfo {author} {\bibfnamefont {J.}~\bibnamefont {Brisbois}}, \bibinfo
  {author} {\bibfnamefont {D.}~\bibnamefont {Chareev}}, \bibinfo {author}
  {\bibfnamefont {A.~V.}\ \bibnamefont {Sadakov}}, \bibinfo {author}
  {\bibfnamefont {A.~S.}\ \bibnamefont {Usoltsev}}, \bibinfo {author}
  {\bibfnamefont {M.}~\bibnamefont {Mito}}, \bibinfo {author} {\bibfnamefont
  {A.~V.}\ \bibnamefont {Silhanek}}, \bibinfo {author} {\bibfnamefont
  {T.}~\bibnamefont {Sarkar}}, \bibinfo {author} {\bibfnamefont
  {A.}~\bibnamefont {Hassan}}, \bibinfo {author} {\bibfnamefont
  {O.}~\bibnamefont {Karis}}, \bibinfo {author} {\bibfnamefont
  {R.}~\bibnamefont {Ahuja}},\ and\ \bibinfo {author} {\bibfnamefont
  {M.}~\bibnamefont {Abdel-Hafiez}},\ }\bibfield  {title} {\bibinfo {title}
  {{Interplay of charge density wave and multiband superconductivity in layered
  quasi-two-dimensional materials: The case of
  $2\mathit{H}\text{\ensuremath{-}}\mathrm{Nb}\mathrm{S}_{2}$ and
  $2\mathit{H}\text{\ensuremath{-}}\mathrm{Nb}\mathrm{Se}_{2}$}},\ }\href
  {https://doi.org/10.1103/PhysRevMaterials.4.084005} {\bibfield  {journal}
  {\bibinfo  {journal} {Phys. Rev. Mater.}\ }\textbf {\bibinfo {volume} {4}},\
  \bibinfo {pages} {084005} (\bibinfo {year} {2020})}\BibitemShut {NoStop}%
\bibitem [{\citenamefont {Mizuguchi}\ and\ \citenamefont
  {Takano}(2010)}]{Intro:Fe_SC_review}%
  \BibitemOpen
  \bibfield  {author} {\bibinfo {author} {\bibfnamefont {Y.}~\bibnamefont
  {Mizuguchi}}\ and\ \bibinfo {author} {\bibfnamefont {Y.}~\bibnamefont
  {Takano}},\ }\bibfield  {title} {\bibinfo {title} {{Review of Fe
  Chalcogenides as the Simplest Fe-Based Superconductor }},\ }\href@noop {}
  {\bibfield  {journal} {\bibinfo  {journal}
  {\href{https://journals.jps.jp/doi/10.1143/JPSJ.79.102001}{J. Phys. Soc.
  Japan}}\ }\textbf {\bibinfo {volume} {79}} (\bibinfo {year}
  {2010})}\BibitemShut {NoStop}%
\bibitem [{\citenamefont {Buzea}\ and\ \citenamefont
  {Yamashita}(2001)}]{Intro:MgB2_SC_review}%
  \BibitemOpen
  \bibfield  {author} {\bibinfo {author} {\bibfnamefont {C.}~\bibnamefont
  {Buzea}}\ and\ \bibinfo {author} {\bibfnamefont {T.}~\bibnamefont
  {Yamashita}},\ }\bibfield  {title} {\bibinfo {title} {{Review of the
  superconducting properties of $\mathrm{MgB}_{2}$ }},\ }\href@noop {}
  {\bibfield  {journal} {\bibinfo  {journal}
  {\href{https://iopscience.iop.org/article/10.1088/0953-2048/14/11/201}{Supercond.
  Sci. Technol.}}\ }\textbf {\bibinfo {volume} {14}} (\bibinfo {year}
  {2001})}\BibitemShut {NoStop}%
\bibitem [{\citenamefont {Xu}(2017)}]{Intro:Niobium_SC_review}%
  \BibitemOpen
  \bibfield  {author} {\bibinfo {author} {\bibfnamefont {X.}~\bibnamefont
  {Xu}},\ }\bibfield  {title} {\bibinfo {title} {{A review and prospects for
  $\mathrm{Nb}_{3}\mathrm{Sn}$ superconductor development }},\ }\href@noop {}
  {\bibfield  {journal} {\bibinfo  {journal}
  {\href{https://iopscience.iop.org/article/10.1088/1361-6668/aa7976}{Supercond.
  Sci. Technol.}}\ }\textbf {\bibinfo {volume} {30}} (\bibinfo {year}
  {2017})}\BibitemShut {NoStop}%
\bibitem [{\citenamefont {Chao}\ \emph {et~al.}(2021)\citenamefont {Chao},
  \citenamefont {Qiangwei}, \citenamefont {Zhijun}, \citenamefont {Chunsheng},
  \citenamefont {Hechang}, \citenamefont {Zheng},\ and\ \citenamefont
  {Jianlin}}]{Kagome_s_wave}%
  \BibitemOpen
  \bibfield  {author} {\bibinfo {author} {\bibfnamefont {M.}~\bibnamefont
  {Chao}}, \bibinfo {author} {\bibfnamefont {Y.}~\bibnamefont {Qiangwei}},
  \bibinfo {author} {\bibfnamefont {T.}~\bibnamefont {Zhijun}}, \bibinfo
  {author} {\bibfnamefont {G.}~\bibnamefont {Chunsheng}}, \bibinfo {author}
  {\bibfnamefont {L.}~\bibnamefont {Hechang}}, \bibinfo {author} {\bibfnamefont
  {L.}~\bibnamefont {Zheng}},\ and\ \bibinfo {author} {\bibfnamefont
  {L.}~\bibnamefont {Jianlin}},\ }\bibfield  {title} {\bibinfo {title} {{S-Wave
  Superconductivity in Kagome Metal $\mathrm{CsV}_{3}\mathrm{Sb}_{5}$ Revealed
  by ${}^{121/123}$Sb NQR and ${}^{51}$V NMR Measurements}},\ }\bibfield
  {journal} {\bibinfo  {journal}
  {\href{https://iopscience.iop.org/article/10.1088/0256-307X/38/7/077402}{Chin.
  Phys. Lett.}}\ }\textbf {\bibinfo {volume} {38}},\ \href
  {https://doi.org/10.1088/0256-307X/38/7/077402}
  {10.1088/0256-307X/38/7/077402} (\bibinfo {year} {2021})\BibitemShut
  {NoStop}%
\bibitem [{\citenamefont {Ortiz}\ \emph {et~al.}(2021)\citenamefont {Ortiz},
  \citenamefont {Teicher}, \citenamefont {Kautzsch}, \citenamefont {Sarte},
  \citenamefont {Ratcliff}, \citenamefont {Harter}, \citenamefont {Ruff},
  \citenamefont {Seshadri},\ and\ \citenamefont {Wilson}}]{Kagome_StarOfDavid}%
  \BibitemOpen
  \bibfield  {author} {\bibinfo {author} {\bibfnamefont {B.~R.}\ \bibnamefont
  {Ortiz}}, \bibinfo {author} {\bibfnamefont {S.~M.~L.}\ \bibnamefont
  {Teicher}}, \bibinfo {author} {\bibfnamefont {L.}~\bibnamefont {Kautzsch}},
  \bibinfo {author} {\bibfnamefont {P.~M.}\ \bibnamefont {Sarte}}, \bibinfo
  {author} {\bibfnamefont {N.}~\bibnamefont {Ratcliff}}, \bibinfo {author}
  {\bibfnamefont {J.}~\bibnamefont {Harter}}, \bibinfo {author} {\bibfnamefont
  {J.~P.~C.}\ \bibnamefont {Ruff}}, \bibinfo {author} {\bibfnamefont
  {R.}~\bibnamefont {Seshadri}},\ and\ \bibinfo {author} {\bibfnamefont
  {S.~D.}\ \bibnamefont {Wilson}},\ }\bibfield  {title} {\bibinfo {title}
  {{Fermi Surface Mapping and the Nature of Charge-Density-Wave Order in the
  Kagome Superconductor $\mathrm{CsV}_{3}\mathrm{Sb}_{5}$}},\ }\href
  {https://doi.org/10.1103/PhysRevX.11.041030} {\bibfield  {journal} {\bibinfo
  {journal} {Phys. Rev. X}\ }\textbf {\bibinfo {volume} {11}},\ \bibinfo
  {pages} {041030} (\bibinfo {year} {2021})}\BibitemShut {NoStop}%
\bibitem [{\citenamefont {Jiang}\ \emph {et~al.}(2022)\citenamefont {Jiang},
  \citenamefont {Wu}, \citenamefont {Yin}, \citenamefont {Wang}, \citenamefont
  {Hasan}, \citenamefont {Wilson}, \citenamefont {Chen},\ and\ \citenamefont
  {Hu}}]{Intro:Kagome_SC_review}%
  \BibitemOpen
  \bibfield  {author} {\bibinfo {author} {\bibfnamefont {K.}~\bibnamefont
  {Jiang}}, \bibinfo {author} {\bibfnamefont {T.}~\bibnamefont {Wu}}, \bibinfo
  {author} {\bibfnamefont {J.-X.}\ \bibnamefont {Yin}}, \bibinfo {author}
  {\bibfnamefont {Z.}~\bibnamefont {Wang}}, \bibinfo {author} {\bibfnamefont
  {M.~Z.}\ \bibnamefont {Hasan}}, \bibinfo {author} {\bibfnamefont {S.~D.}\
  \bibnamefont {Wilson}}, \bibinfo {author} {\bibfnamefont {X.}~\bibnamefont
  {Chen}},\ and\ \bibinfo {author} {\bibfnamefont {J.}~\bibnamefont {Hu}},\
  }\bibfield  {title} {\bibinfo {title} {{Kagome superconductors
  $\mathrm{AV}_{3}\mathrm{Sb}_{5}$ (A=K, Rb, Cs) }},\ }\href@noop {} {\bibfield
   {journal} {\bibinfo  {journal}
  {\href{https://academic.oup.com/nsr/article/10/2/nwac199/6724256?login=true}{Natl.
  Sci. Rev.}}\ }\textbf {\bibinfo {volume} {nwac199}} (\bibinfo {year}
  {2022})}\BibitemShut {NoStop}%
\bibitem [{\citenamefont {Luo}\ \emph {et~al.}(2022)\citenamefont {Luo},
  \citenamefont {Zhao}, \citenamefont {Zhou}, \citenamefont {Yang},
  \citenamefont {Fang}, \citenamefont {Yang}, \citenamefont {Gao},
  \citenamefont {Zhou},\ and\ \citenamefont {Zheng}}]{Kagome_Possible_CDW}%
  \BibitemOpen
  \bibfield  {author} {\bibinfo {author} {\bibfnamefont {J.}~\bibnamefont
  {Luo}}, \bibinfo {author} {\bibfnamefont {Z.}~\bibnamefont {Zhao}}, \bibinfo
  {author} {\bibfnamefont {Y.~Z.}\ \bibnamefont {Zhou}}, \bibinfo {author}
  {\bibfnamefont {J.}~\bibnamefont {Yang}}, \bibinfo {author} {\bibfnamefont
  {A.~F.}\ \bibnamefont {Fang}}, \bibinfo {author} {\bibfnamefont {H.~T.}\
  \bibnamefont {Yang}}, \bibinfo {author} {\bibfnamefont {H.~J.}\ \bibnamefont
  {Gao}}, \bibinfo {author} {\bibfnamefont {R.}~\bibnamefont {Zhou}},\ and\
  \bibinfo {author} {\bibfnamefont {G.-q.}\ \bibnamefont {Zheng}},\ }\bibfield
  {title} {\bibinfo {title} {{Possible star-of-David pattern charge density
  wave with additional modulation in the kagome superconductor
  $\mathrm{CsV}_{3}\mathrm{Sb}_{5}$}},\ }\href@noop {} {\bibfield  {journal}
  {\bibinfo  {journal} {\href{https://doi.org/10.1038/s41535-022-00437-7}{npj
  Quantum Mater.}}\ }\textbf {\bibinfo {volume} {7}} (\bibinfo {year}
  {2022})}\BibitemShut {NoStop}%
\bibitem [{\citenamefont {Zheng}\ \emph {et~al.}(2022)\citenamefont {Zheng},
  \citenamefont {Wu}, \citenamefont {Yang}, \citenamefont {Nie}, \citenamefont
  {Shan}, \citenamefont {Sun}, \citenamefont {Song}, \citenamefont {Yu},
  \citenamefont {Li}, \citenamefont {Zhao}, \citenamefont {Li}, \citenamefont
  {Kang}, \citenamefont {Zhou}, \citenamefont {Liu}, \citenamefont {Xiang},
  \citenamefont {Ying}, \citenamefont {Wang}, \citenamefont {Wu},\ and\
  \citenamefont {Chen}}]{Kagome_Possible_CDW_3Q}%
  \BibitemOpen
  \bibfield  {author} {\bibinfo {author} {\bibfnamefont {L.}~\bibnamefont
  {Zheng}}, \bibinfo {author} {\bibfnamefont {Z.}~\bibnamefont {Wu}}, \bibinfo
  {author} {\bibfnamefont {Y.}~\bibnamefont {Yang}}, \bibinfo {author}
  {\bibfnamefont {L.}~\bibnamefont {Nie}}, \bibinfo {author} {\bibfnamefont
  {M.}~\bibnamefont {Shan}}, \bibinfo {author} {\bibfnamefont {K.}~\bibnamefont
  {Sun}}, \bibinfo {author} {\bibfnamefont {D.}~\bibnamefont {Song}}, \bibinfo
  {author} {\bibfnamefont {F.}~\bibnamefont {Yu}}, \bibinfo {author}
  {\bibfnamefont {J.}~\bibnamefont {Li}}, \bibinfo {author} {\bibfnamefont
  {D.}~\bibnamefont {Zhao}}, \bibinfo {author} {\bibfnamefont {S.}~\bibnamefont
  {Li}}, \bibinfo {author} {\bibfnamefont {B.}~\bibnamefont {Kang}}, \bibinfo
  {author} {\bibfnamefont {Y.}~\bibnamefont {Zhou}}, \bibinfo {author}
  {\bibfnamefont {K.}~\bibnamefont {Liu}}, \bibinfo {author} {\bibfnamefont
  {Z.}~\bibnamefont {Xiang}}, \bibinfo {author} {\bibfnamefont
  {J.}~\bibnamefont {Ying}}, \bibinfo {author} {\bibfnamefont {Z.}~\bibnamefont
  {Wang}}, \bibinfo {author} {\bibfnamefont {T.}~\bibnamefont {Wu}},\ and\
  \bibinfo {author} {\bibfnamefont {X.}~\bibnamefont {Chen}},\ }\bibfield
  {title} {\bibinfo {title} {{Emergent charge order in pressurized kagome
  superconductor $\mathrm{CsV}_{3}\mathrm{Sb}_{5}$}},\ }\href@noop {}
  {\bibfield  {journal} {\bibinfo  {journal}
  {\href{https://doi.org/10.1038/s41586-022-05351-3}{Nature}}\ }\textbf
  {\bibinfo {volume} {611}} (\bibinfo {year} {2022})}\BibitemShut {NoStop}%
\bibitem [{\citenamefont {Roppongi}\ \emph {et~al.}(2023)\citenamefont
  {Roppongi}, \citenamefont {Ishihara}, \citenamefont {Tanaka}, \citenamefont
  {Ogawa}, \citenamefont {Okada}, \citenamefont {Liu}, \citenamefont {Mukasa},
  \citenamefont {Mizukami}, \citenamefont {Uwamoto}, \citenamefont {Grasset},
  \citenamefont {Konczykowski}, \citenamefont {Ortiz}, \citenamefont {Wilson},
  \citenamefont {Hashimoto},\ and\ \citenamefont
  {Shibauchi}}]{Kagome_Anisotropic_s_wave}%
  \BibitemOpen
  \bibfield  {author} {\bibinfo {author} {\bibfnamefont {M.}~\bibnamefont
  {Roppongi}}, \bibinfo {author} {\bibfnamefont {K.}~\bibnamefont {Ishihara}},
  \bibinfo {author} {\bibfnamefont {Y.}~\bibnamefont {Tanaka}}, \bibinfo
  {author} {\bibfnamefont {K.}~\bibnamefont {Ogawa}}, \bibinfo {author}
  {\bibfnamefont {K.}~\bibnamefont {Okada}}, \bibinfo {author} {\bibfnamefont
  {S.}~\bibnamefont {Liu}}, \bibinfo {author} {\bibfnamefont {K.}~\bibnamefont
  {Mukasa}}, \bibinfo {author} {\bibfnamefont {Y.}~\bibnamefont {Mizukami}},
  \bibinfo {author} {\bibfnamefont {Y.}~\bibnamefont {Uwamoto}}, \bibinfo
  {author} {\bibfnamefont {R.}~\bibnamefont {Grasset}}, \bibinfo {author}
  {\bibfnamefont {M.}~\bibnamefont {Konczykowski}}, \bibinfo {author}
  {\bibfnamefont {B.~R.}\ \bibnamefont {Ortiz}}, \bibinfo {author}
  {\bibfnamefont {S.~D.}\ \bibnamefont {Wilson}}, \bibinfo {author}
  {\bibfnamefont {K.}~\bibnamefont {Hashimoto}},\ and\ \bibinfo {author}
  {\bibfnamefont {T.}~\bibnamefont {Shibauchi}},\ }\bibfield  {title} {\bibinfo
  {title} {{Bulk evidence of anisotropic s-wave pairing with no sign change in
  the kagome superconductor $\mathrm{CsV}_{3}\mathrm{Sb}_{5}$}},\ }\bibfield
  {journal} {\bibinfo  {journal} {Nat. Commun.}\ }\textbf {\bibinfo {volume}
  {14}},\ \href {https://doi.org/https://doi.org/10.1038/s41467-023-36273-x}
  {https://doi.org/10.1038/s41467-023-36273-x} (\bibinfo {year}
  {2023})\BibitemShut {NoStop}%
\bibitem [{\citenamefont {Guo}\ \emph {et~al.}(2023)\citenamefont {Guo},
  \citenamefont {Wagner}, \citenamefont {Putzke}, \citenamefont {Chen},
  \citenamefont {Wang}, \citenamefont {Zhang}, \citenamefont {Gutierrez-Amigo},
  \citenamefont {Errea}, \citenamefont {G.~Vergniory}, \citenamefont {Felser},
  \citenamefont {Fischer}, \citenamefont {Neupert},\ and\ \citenamefont
  {Moll}}]{Kagome_Preserved_TRS}%
  \BibitemOpen
  \bibfield  {author} {\bibinfo {author} {\bibfnamefont {C.}~\bibnamefont
  {Guo}}, \bibinfo {author} {\bibfnamefont {G.}~\bibnamefont {Wagner}},
  \bibinfo {author} {\bibfnamefont {C.}~\bibnamefont {Putzke}}, \bibinfo
  {author} {\bibfnamefont {D.}~\bibnamefont {Chen}}, \bibinfo {author}
  {\bibfnamefont {K.}~\bibnamefont {Wang}}, \bibinfo {author} {\bibfnamefont
  {L.}~\bibnamefont {Zhang}}, \bibinfo {author} {\bibfnamefont
  {M.}~\bibnamefont {Gutierrez-Amigo}}, \bibinfo {author} {\bibfnamefont
  {I.}~\bibnamefont {Errea}}, \bibinfo {author} {\bibfnamefont
  {M.}~\bibnamefont {G.~Vergniory}}, \bibinfo {author} {\bibfnamefont
  {C.}~\bibnamefont {Felser}}, \bibinfo {author} {\bibfnamefont {M.~H.}\
  \bibnamefont {Fischer}}, \bibinfo {author} {\bibfnamefont {T.}~\bibnamefont
  {Neupert}},\ and\ \bibinfo {author} {\bibfnamefont {P.~J.~W.}\ \bibnamefont
  {Moll}},\ }\bibfield  {title} {\bibinfo {title} {{Correlated order at the
  tipping point in the kagome metal $\mathrm{CsV}_{3}\mathrm{Sb}_{5}$}},\
  }\href@noop {} {\bibfield  {journal} {\bibinfo  {journal}
  {\href{https://doi.org/10.48550/arXiv.2304.00972}{arXiv:2304.00972}}\ }
  (\bibinfo {year} {2023})}\BibitemShut {NoStop}%
\bibitem [{\citenamefont {Saykin}\ \emph {et~al.}(2023)\citenamefont {Saykin},
  \citenamefont {Farhang}, \citenamefont {Kountz}, \citenamefont {Chen},
  \citenamefont {Ortiz}, \citenamefont {Shekhar}, \citenamefont {Felser},
  \citenamefont {Wilson}, \citenamefont {Thomale}, \citenamefont {Xia},\ and\
  \citenamefont {Kapitulnik}}]{Kagome_NoTRSB}%
  \BibitemOpen
  \bibfield  {author} {\bibinfo {author} {\bibfnamefont {D.~R.}\ \bibnamefont
  {Saykin}}, \bibinfo {author} {\bibfnamefont {C.}~\bibnamefont {Farhang}},
  \bibinfo {author} {\bibfnamefont {E.~D.}\ \bibnamefont {Kountz}}, \bibinfo
  {author} {\bibfnamefont {D.}~\bibnamefont {Chen}}, \bibinfo {author}
  {\bibfnamefont {B.~R.}\ \bibnamefont {Ortiz}}, \bibinfo {author}
  {\bibfnamefont {C.}~\bibnamefont {Shekhar}}, \bibinfo {author} {\bibfnamefont
  {C.}~\bibnamefont {Felser}}, \bibinfo {author} {\bibfnamefont {S.~D.}\
  \bibnamefont {Wilson}}, \bibinfo {author} {\bibfnamefont {R.}~\bibnamefont
  {Thomale}}, \bibinfo {author} {\bibfnamefont {J.}~\bibnamefont {Xia}},\ and\
  \bibinfo {author} {\bibfnamefont {A.}~\bibnamefont {Kapitulnik}},\ }\bibfield
   {title} {\bibinfo {title} {{High Resolution Polar Kerr Effect Studies of
  $\mathrm{CsV}_{3}\mathrm{Sb}_{5}$: Tests for Time-Reversal Symmetry Breaking
  below the Charge-Order Transition}},\ }\href
  {https://doi.org/10.1103/PhysRevLett.131.016901} {\bibfield  {journal}
  {\bibinfo  {journal} {Phys. Rev. Lett.}\ }\textbf {\bibinfo {volume} {131}},\
  \bibinfo {pages} {016901} (\bibinfo {year} {2023})}\BibitemShut {NoStop}%
\bibitem [{\citenamefont {Leggett}(1966)}]{Intro:Leggett_original}%
  \BibitemOpen
  \bibfield  {author} {\bibinfo {author} {\bibfnamefont {A.~J.}\ \bibnamefont
  {Leggett}},\ }\bibfield  {title} {\bibinfo {title} {{Number-Phase
  Fluctuations in Two-Band Superconductors }},\ }\href@noop {} {\bibfield
  {journal} {\bibinfo  {journal}
  {\href{https://academic.oup.com/ptp/article/36/5/901/1857809}{Prog. Theor.
  Phys.}}\ }\textbf {\bibinfo {volume} {36}} (\bibinfo {year}
  {1966})}\BibitemShut {NoStop}%
\bibitem [{\citenamefont {Blumberg}\ \emph {et~al.}(2007)\citenamefont
  {Blumberg}, \citenamefont {Mialitsin}, \citenamefont {Dennis}, \citenamefont
  {Klein}, \citenamefont {Zhigadlo},\ and\ \citenamefont
  {Karpinski}}]{Intro:MgB2RamanLeggett}%
  \BibitemOpen
  \bibfield  {author} {\bibinfo {author} {\bibfnamefont {G.}~\bibnamefont
  {Blumberg}}, \bibinfo {author} {\bibfnamefont {A.}~\bibnamefont {Mialitsin}},
  \bibinfo {author} {\bibfnamefont {B.~S.}\ \bibnamefont {Dennis}}, \bibinfo
  {author} {\bibfnamefont {M.~V.}\ \bibnamefont {Klein}}, \bibinfo {author}
  {\bibfnamefont {N.~D.}\ \bibnamefont {Zhigadlo}},\ and\ \bibinfo {author}
  {\bibfnamefont {J.}~\bibnamefont {Karpinski}},\ }\bibfield  {title} {\bibinfo
  {title} {{Observation of Leggett's Collective Mode in a Multiband
  $\mathrm{MgB}_{2}$ Superconductor}},\ }\href
  {https://doi.org/10.1103/PhysRevLett.99.227002} {\bibfield  {journal}
  {\bibinfo  {journal} {Phys. Rev. Lett.}\ }\textbf {\bibinfo {volume} {99}},\
  \bibinfo {pages} {227002} (\bibinfo {year} {2007})}\BibitemShut {NoStop}%
\bibitem [{\citenamefont {Krull}\ \emph {et~al.}(2016)\citenamefont {Krull},
  \citenamefont {Bittner}, \citenamefont {Uhrig}, \citenamefont {Manske},\ and\
  \citenamefont {Schnyder}}]{Intro:Higgs_Leggett1}%
  \BibitemOpen
  \bibfield  {author} {\bibinfo {author} {\bibfnamefont {H.}~\bibnamefont
  {Krull}}, \bibinfo {author} {\bibfnamefont {N.}~\bibnamefont {Bittner}},
  \bibinfo {author} {\bibfnamefont {G.~S.}\ \bibnamefont {Uhrig}}, \bibinfo
  {author} {\bibfnamefont {D.}~\bibnamefont {Manske}},\ and\ \bibinfo {author}
  {\bibfnamefont {A.~P.}\ \bibnamefont {Schnyder}},\ }\bibfield  {title}
  {\bibinfo {title} {{Coupling of Higgs and Leggett modes in non-equilibrium
  superconductors }},\ }\href@noop {} {\bibfield  {journal} {\bibinfo
  {journal} {\href{https://www.nature.com/articles/ncomms11921}{Nat. Commun.}}\
  }\textbf {\bibinfo {volume} {7}} (\bibinfo {year} {2016})}\BibitemShut
  {NoStop}%
\bibitem [{\citenamefont {Murotani}\ \emph {et~al.}(2017)\citenamefont
  {Murotani}, \citenamefont {Tsuji},\ and\ \citenamefont
  {Aoki}}]{Intro:Higgs_Leggett2}%
  \BibitemOpen
  \bibfield  {author} {\bibinfo {author} {\bibfnamefont {Y.}~\bibnamefont
  {Murotani}}, \bibinfo {author} {\bibfnamefont {N.}~\bibnamefont {Tsuji}},\
  and\ \bibinfo {author} {\bibfnamefont {H.}~\bibnamefont {Aoki}},\ }\bibfield
  {title} {\bibinfo {title} {{Theory of light-induced resonances with
  collective Higgs and Leggett modes in multiband superconductors}},\ }\href
  {https://doi.org/10.1103/PhysRevB.95.104503} {\bibfield  {journal} {\bibinfo
  {journal} {Phys. Rev. B}\ }\textbf {\bibinfo {volume} {95}},\ \bibinfo
  {pages} {104503} (\bibinfo {year} {2017})}\BibitemShut {NoStop}%
\bibitem [{\citenamefont {Murotani}\ and\ \citenamefont
  {Shimano}(2019)}]{Intro:Higgs_Leggett3}%
  \BibitemOpen
  \bibfield  {author} {\bibinfo {author} {\bibfnamefont {Y.}~\bibnamefont
  {Murotani}}\ and\ \bibinfo {author} {\bibfnamefont {R.}~\bibnamefont
  {Shimano}},\ }\bibfield  {title} {\bibinfo {title} {{Nonlinear optical
  response of collective modes in multiband superconductors assisted by
  nonmagnetic impurities}},\ }\href
  {https://doi.org/10.1103/PhysRevB.99.224510} {\bibfield  {journal} {\bibinfo
  {journal} {Phys. Rev. B}\ }\textbf {\bibinfo {volume} {99}},\ \bibinfo
  {pages} {224510} (\bibinfo {year} {2019})}\BibitemShut {NoStop}%
\bibitem [{\citenamefont {Burnell}\ \emph {et~al.}(2010)\citenamefont
  {Burnell}, \citenamefont {Hu}, \citenamefont {Parish},\ and\ \citenamefont
  {Bernevig}}]{Intro:Leggett_Nonlinear1}%
  \BibitemOpen
  \bibfield  {author} {\bibinfo {author} {\bibfnamefont {F.~J.}\ \bibnamefont
  {Burnell}}, \bibinfo {author} {\bibfnamefont {J.}~\bibnamefont {Hu}},
  \bibinfo {author} {\bibfnamefont {M.~M.}\ \bibnamefont {Parish}},\ and\
  \bibinfo {author} {\bibfnamefont {B.~A.}\ \bibnamefont {Bernevig}},\
  }\bibfield  {title} {\bibinfo {title} {{Leggett mode in a strong-coupling
  model of iron arsenide superconductors}},\ }\href
  {https://doi.org/10.1103/PhysRevB.82.144506} {\bibfield  {journal} {\bibinfo
  {journal} {Phys. Rev. B}\ }\textbf {\bibinfo {volume} {82}},\ \bibinfo
  {pages} {144506} (\bibinfo {year} {2010})}\BibitemShut {NoStop}%
\bibitem [{\citenamefont {Lin}\ and\ \citenamefont
  {Hu}(2012)}]{Intro:Leggett_Massless}%
  \BibitemOpen
  \bibfield  {author} {\bibinfo {author} {\bibfnamefont {S.-Z.}\ \bibnamefont
  {Lin}}\ and\ \bibinfo {author} {\bibfnamefont {X.}~\bibnamefont {Hu}},\
  }\bibfield  {title} {\bibinfo {title} {Massless leggett mode in three-band
  superconductors with time-reversal-symmetry breaking},\ }\href
  {https://doi.org/10.1103/PhysRevLett.108.177005} {\bibfield  {journal}
  {\bibinfo  {journal} {Phys. Rev. Lett.}\ }\textbf {\bibinfo {volume} {108}},\
  \bibinfo {pages} {177005} (\bibinfo {year} {2012})}\BibitemShut {NoStop}%
\bibitem [{\citenamefont {Balatsky}\ \emph {et~al.}(2000)\citenamefont
  {Balatsky}, \citenamefont {Kumar},\ and\ \citenamefont
  {Schrieffer}}]{Intro:Leggett_Nonlinear2}%
  \BibitemOpen
  \bibfield  {author} {\bibinfo {author} {\bibfnamefont {A.~V.}\ \bibnamefont
  {Balatsky}}, \bibinfo {author} {\bibfnamefont {P.}~\bibnamefont {Kumar}},\
  and\ \bibinfo {author} {\bibfnamefont {J.~R.}\ \bibnamefont {Schrieffer}},\
  }\bibfield  {title} {\bibinfo {title} {{Collective Mode in a Superconductor
  with Mixed-Symmetry Order Parameter Components}},\ }\href
  {https://doi.org/10.1103/PhysRevLett.84.4445} {\bibfield  {journal} {\bibinfo
   {journal} {Phys. Rev. Lett.}\ }\textbf {\bibinfo {volume} {84}},\ \bibinfo
  {pages} {4445} (\bibinfo {year} {2000})}\BibitemShut {NoStop}%
\bibitem [{\citenamefont {Bittner}\ \emph {et~al.}(2015)\citenamefont
  {Bittner}, \citenamefont {Einzel}, \citenamefont {Klam},\ and\ \citenamefont
  {Manske}}]{Intro:Leggett_Nonlinear3}%
  \BibitemOpen
  \bibfield  {author} {\bibinfo {author} {\bibfnamefont {N.}~\bibnamefont
  {Bittner}}, \bibinfo {author} {\bibfnamefont {D.}~\bibnamefont {Einzel}},
  \bibinfo {author} {\bibfnamefont {L.}~\bibnamefont {Klam}},\ and\ \bibinfo
  {author} {\bibfnamefont {D.}~\bibnamefont {Manske}},\ }\bibfield  {title}
  {\bibinfo {title} {{Leggett Modes and the Anderson-Higgs Mechanism in
  Superconductors without Inversion Symmetry}},\ }\href
  {https://doi.org/10.1103/PhysRevLett.115.227002} {\bibfield  {journal}
  {\bibinfo  {journal} {Phys. Rev. Lett.}\ }\textbf {\bibinfo {volume} {115}},\
  \bibinfo {pages} {227002} (\bibinfo {year} {2015})}\BibitemShut {NoStop}%
\bibitem [{\citenamefont {Ota}\ \emph {et~al.}(2011)\citenamefont {Ota},
  \citenamefont {Machida}, \citenamefont {Koyama},\ and\ \citenamefont
  {Aoki}}]{Intro:Leggett_Nonlinear4}%
  \BibitemOpen
  \bibfield  {author} {\bibinfo {author} {\bibfnamefont {Y.}~\bibnamefont
  {Ota}}, \bibinfo {author} {\bibfnamefont {M.}~\bibnamefont {Machida}},
  \bibinfo {author} {\bibfnamefont {T.}~\bibnamefont {Koyama}},\ and\ \bibinfo
  {author} {\bibfnamefont {H.}~\bibnamefont {Aoki}},\ }\bibfield  {title}
  {\bibinfo {title} {{Collective modes in multiband superfluids and
  superconductors: Multiple dynamical classes}},\ }\href
  {https://doi.org/10.1103/PhysRevB.83.060507} {\bibfield  {journal} {\bibinfo
  {journal} {Phys. Rev. B}\ }\textbf {\bibinfo {volume} {83}},\ \bibinfo
  {pages} {060507} (\bibinfo {year} {2011})}\BibitemShut {NoStop}%
\bibitem [{\citenamefont {Marciani}\ \emph {et~al.}(2013)\citenamefont
  {Marciani}, \citenamefont {Fanfarillo}, \citenamefont {Castellani},\ and\
  \citenamefont {Benfatto}}]{Intro:Leggett_Nonlinear5}%
  \BibitemOpen
  \bibfield  {author} {\bibinfo {author} {\bibfnamefont {M.}~\bibnamefont
  {Marciani}}, \bibinfo {author} {\bibfnamefont {L.}~\bibnamefont
  {Fanfarillo}}, \bibinfo {author} {\bibfnamefont {C.}~\bibnamefont
  {Castellani}},\ and\ \bibinfo {author} {\bibfnamefont {L.}~\bibnamefont
  {Benfatto}},\ }\bibfield  {title} {\bibinfo {title} {{Leggett modes in
  iron-based superconductors as a probe of time-reversal symmetry breaking}},\
  }\href {https://doi.org/10.1103/PhysRevB.88.214508} {\bibfield  {journal}
  {\bibinfo  {journal} {Phys. Rev. B}\ }\textbf {\bibinfo {volume} {88}},\
  \bibinfo {pages} {214508} (\bibinfo {year} {2013})}\BibitemShut {NoStop}%
\bibitem [{\citenamefont {Cea}\ and\ \citenamefont
  {Benfatto}(2016)}]{Intro:Leggett_Nonlinear6}%
  \BibitemOpen
  \bibfield  {author} {\bibinfo {author} {\bibfnamefont {T.}~\bibnamefont
  {Cea}}\ and\ \bibinfo {author} {\bibfnamefont {L.}~\bibnamefont {Benfatto}},\
  }\bibfield  {title} {\bibinfo {title} {{Signature of the Leggett mode in the
  ${A}_{1g}$ Raman response: From $\mathrm{MgB}_{2}$ to iron-based
  superconductors}},\ }\href {https://doi.org/10.1103/PhysRevB.94.064512}
  {\bibfield  {journal} {\bibinfo  {journal} {Phys. Rev. B}\ }\textbf {\bibinfo
  {volume} {94}},\ \bibinfo {pages} {064512} (\bibinfo {year}
  {2016})}\BibitemShut {NoStop}%
\bibitem [{\citenamefont {Tanaka}(2001)}]{PhaseSoliton_Tanaka}%
  \BibitemOpen
  \bibfield  {author} {\bibinfo {author} {\bibfnamefont {Y.}~\bibnamefont
  {Tanaka}},\ }\bibfield  {title} {\bibinfo {title} {Soliton in two-band
  superconductor},\ }\href {https://doi.org/10.1103/PhysRevLett.88.017002}
  {\bibfield  {journal} {\bibinfo  {journal} {Phys. Rev. Lett.}\ }\textbf
  {\bibinfo {volume} {88}},\ \bibinfo {pages} {017002} (\bibinfo {year}
  {2001})}\BibitemShut {NoStop}%
\bibitem [{\citenamefont {Yerin}\ and\ \citenamefont
  {Drechsler}(2021)}]{PhaseSoliton_Yerin}%
  \BibitemOpen
  \bibfield  {author} {\bibinfo {author} {\bibfnamefont {Y.}~\bibnamefont
  {Yerin}}\ and\ \bibinfo {author} {\bibfnamefont {S.-L.}\ \bibnamefont
  {Drechsler}},\ }\bibfield  {title} {\bibinfo {title} {Phase solitons in a
  weakly coupled three-component superconductor},\ }\href
  {https://doi.org/10.1103/PhysRevB.104.014518} {\bibfield  {journal} {\bibinfo
   {journal} {Phys. Rev. B}\ }\textbf {\bibinfo {volume} {104}},\ \bibinfo
  {pages} {014518} (\bibinfo {year} {2021})}\BibitemShut {NoStop}%
\bibitem [{\citenamefont {Kamatani}\ \emph {et~al.}(2022)\citenamefont
  {Kamatani}, \citenamefont {Kitamura}, \citenamefont {Tsuji}, \citenamefont
  {Shimano},\ and\ \citenamefont {Morimoto}}]{Leggett_Kamatani}%
  \BibitemOpen
  \bibfield  {author} {\bibinfo {author} {\bibfnamefont {T.}~\bibnamefont
  {Kamatani}}, \bibinfo {author} {\bibfnamefont {S.}~\bibnamefont {Kitamura}},
  \bibinfo {author} {\bibfnamefont {N.}~\bibnamefont {Tsuji}}, \bibinfo
  {author} {\bibfnamefont {R.}~\bibnamefont {Shimano}},\ and\ \bibinfo {author}
  {\bibfnamefont {T.}~\bibnamefont {Morimoto}},\ }\bibfield  {title} {\bibinfo
  {title} {{Optical response of the Leggett mode in multiband superconductors
  in the linear response regime}},\ }\href
  {https://doi.org/10.1103/PhysRevB.105.094520} {\bibfield  {journal} {\bibinfo
   {journal} {Phys. Rev. B}\ }\textbf {\bibinfo {volume} {105}},\ \bibinfo
  {pages} {094520} (\bibinfo {year} {2022})}\BibitemShut {NoStop}%
\bibitem [{\citenamefont {Landau}\ and\ \citenamefont
  {Lifshitz}(1969)}]{LandauLifshitz}%
  \BibitemOpen
  \bibfield  {author} {\bibinfo {author} {\bibfnamefont {L.~D.}\ \bibnamefont
  {Landau}}\ and\ \bibinfo {author} {\bibfnamefont {E.~M.}\ \bibnamefont
  {Lifshitz}},\ }\href@noop {} {\emph {\bibinfo {title} {{Statistical Physics
  }}}}\ (\bibinfo  {publisher} {Pergamon Press},\ \bibinfo {address} {Oxford},\
  \bibinfo {year} {1969})\BibitemShut {NoStop}%
\bibitem [{\citenamefont {Dzyaloshinskii}(1964)}]{LI_SymmetryDiscusstion}%
  \BibitemOpen
  \bibfield  {author} {\bibinfo {author} {\bibfnamefont {I.}~\bibnamefont
  {Dzyaloshinskii}},\ }\bibfield  {title} {\bibinfo {title} {{Theory of
  Helicoidal Structures in Antiferromagnets. I. Nonmetals }},\ }\href@noop {}
  {\bibfield  {journal} {\bibinfo  {journal}
  {\href{http://jetp.ras.ru/cgi-bin/dn/e_019_04_0960.pdf}{JETP}}\ }\textbf
  {\bibinfo {volume} {19}} (\bibinfo {year} {1964})}\BibitemShut {NoStop}%
\bibitem [{\citenamefont {Mineev}\ and\ \citenamefont
  {Samokhin}(1994)}]{II:LI_NCSC_1st}%
  \BibitemOpen
  \bibfield  {author} {\bibinfo {author} {\bibfnamefont {V.~P.}\ \bibnamefont
  {Mineev}}\ and\ \bibinfo {author} {\bibfnamefont {K.~V.}\ \bibnamefont
  {Samokhin}},\ }\bibfield  {title} {\bibinfo {title} {{Helical phases in
  superconductors }},\ }\href@noop {} {\bibfield  {journal} {\bibinfo
  {journal} {\href{http://www.jetp.ras.ru/cgi-bin/dn/e_078_03_0401.pdf}{JETP}}\
  }\textbf {\bibinfo {volume} {105}} (\bibinfo {year} {1994})}\BibitemShut
  {NoStop}%
\bibitem [{\citenamefont {Mineev}\ and\ \citenamefont
  {Samokhin}(2008)}]{II:LI_NCSC_2nd}%
  \BibitemOpen
  \bibfield  {author} {\bibinfo {author} {\bibfnamefont {V.~P.}\ \bibnamefont
  {Mineev}}\ and\ \bibinfo {author} {\bibfnamefont {K.~V.}\ \bibnamefont
  {Samokhin}},\ }\bibfield  {title} {\bibinfo {title} {{Nonuniform states in
  noncentrosymmetric superconductors: Derivation of Lifshitz invariants from
  microscopic theory}},\ }\href {https://doi.org/10.1103/PhysRevB.78.144503}
  {\bibfield  {journal} {\bibinfo  {journal} {Phys. Rev. B}\ }\textbf {\bibinfo
  {volume} {78}},\ \bibinfo {pages} {144503} (\bibinfo {year}
  {2008})}\BibitemShut {NoStop}%
\bibitem [{\citenamefont {Samokhin}(2013)}]{II:LI_NCSC_3rd}%
  \BibitemOpen
  \bibfield  {author} {\bibinfo {author} {\bibfnamefont {K.~V.}\ \bibnamefont
  {Samokhin}},\ }\bibfield  {title} {\bibinfo {title} {{Gradient energy of
  superconductors without inversion symmetry}},\ }\href
  {https://doi.org/10.1016/j.physc.2013.03.002} {\bibfield  {journal} {\bibinfo
   {journal} {Physica C: Superconductivity}\ }\textbf {\bibinfo {volume}
  {489}},\ \bibinfo {pages} {19} (\bibinfo {year} {2013})}\BibitemShut
  {NoStop}%
\bibitem [{\citenamefont {Kanasugi}\ and\ \citenamefont
  {Yanase}(2022)}]{II:LI_Yanase_1st}%
  \BibitemOpen
  \bibfield  {author} {\bibinfo {author} {\bibfnamefont {S.}~\bibnamefont
  {Kanasugi}}\ and\ \bibinfo {author} {\bibfnamefont {Y.}~\bibnamefont
  {Yanase}},\ }\bibfield  {title} {\bibinfo {title} {{Anapole superconductivity
  from $\mathcal{PT}$-symmetric mixed-parity interband pairing}},\ }\href
  {https://doi.org/10.1038/s42005-022-00804-7} {\bibfield  {journal} {\bibinfo
  {journal} {Commun. Phys.}\ }\textbf {\bibinfo {volume} {5}},\ \bibinfo
  {pages} {39} (\bibinfo {year} {2022})}\BibitemShut {NoStop}%
\bibitem [{\citenamefont {Kitamura}\ \emph {et~al.}(2023)\citenamefont
  {Kitamura}, \citenamefont {Kanasugi}, \citenamefont {Chazono},\ and\
  \citenamefont {Yanase}}]{II:LI_Yanase_2nd}%
  \BibitemOpen
  \bibfield  {author} {\bibinfo {author} {\bibfnamefont {T.}~\bibnamefont
  {Kitamura}}, \bibinfo {author} {\bibfnamefont {S.}~\bibnamefont {Kanasugi}},
  \bibinfo {author} {\bibfnamefont {M.}~\bibnamefont {Chazono}},\ and\ \bibinfo
  {author} {\bibfnamefont {Y.}~\bibnamefont {Yanase}},\ }\bibfield  {title}
  {\bibinfo {title} {{Quantum geometry induced anapole superconductivity}},\
  }\href {https://doi.org/10.1103/PhysRevB.107.214513} {\bibfield  {journal}
  {\bibinfo  {journal} {Phys. Rev. B}\ }\textbf {\bibinfo {volume} {107}},\
  \bibinfo {pages} {214513} (\bibinfo {year} {2023})}\BibitemShut {NoStop}%
\bibitem [{\citenamefont {Kopsky}\ and\ \citenamefont
  {Sannikov}(1977)}]{II:LI_com_incom_1st}%
  \BibitemOpen
  \bibfield  {author} {\bibinfo {author} {\bibfnamefont {V.}~\bibnamefont
  {Kopsky}}\ and\ \bibinfo {author} {\bibfnamefont {D.~G.}\ \bibnamefont
  {Sannikov}},\ }\bibfield  {title} {\bibinfo {title} {{Gradient invariants and
  incommensurate phase transitions }},\ }\href@noop {} {\bibfield  {journal}
  {\bibinfo  {journal}
  {\href{https://iopscience.iop.org/article/10.1088/0022-3719/10/21/021}{J.
  Phys. C: Solid State Phys.}}\ }\textbf {\bibinfo {volume} {10}} (\bibinfo
  {year} {1977})}\BibitemShut {NoStop}%
\bibitem [{\citenamefont {Ishibashi}\ and\ \citenamefont
  {Dvo\v{r}\'{a}k}(1978)}]{II:LI_com_incom_2nd}%
  \BibitemOpen
  \bibfield  {author} {\bibinfo {author} {\bibfnamefont {Y.}~\bibnamefont
  {Ishibashi}}\ and\ \bibinfo {author} {\bibfnamefont {V.}~\bibnamefont
  {Dvo\v{r}\'{a}k}},\ }\bibfield  {title} {\bibinfo {title} {{Incommensurate
  Phase Transitions under the Existence of the Lifshitz Invariant }},\
  }\href@noop {} {\bibfield  {journal} {\bibinfo  {journal}
  {\href{https://journals.jps.jp/doi/10.1143/JPSJ.44.32}{J. Phys. Soc. Jpn.}}\
  }\textbf {\bibinfo {volume} {44}},\ \bibinfo {pages} {32} (\bibinfo {year}
  {1978})}\BibitemShut {NoStop}%
\bibitem [{\citenamefont {Sparavigna}(2009)}]{II:LI_LiquidCrystal}%
  \BibitemOpen
  \bibfield  {author} {\bibinfo {author} {\bibfnamefont {A.}~\bibnamefont
  {Sparavigna}},\ }\bibfield  {title} {\bibinfo {title} {{Role of Lifshitz
  Invariants in Liquid Crystals }},\ }\href@noop {} {\bibfield  {journal}
  {\bibinfo  {journal} {\href{http://dx.doi.org/10.3390/ma2020674}{Materials}}\
  }\textbf {\bibinfo {volume} {2}} (\bibinfo {year} {2009})}\BibitemShut
  {NoStop}%
\bibitem [{\citenamefont {Dzyaloshinskii}(1958)}]{II:LI_DM_1st}%
  \BibitemOpen
  \bibfield  {author} {\bibinfo {author} {\bibfnamefont {I.}~\bibnamefont
  {Dzyaloshinskii}},\ }\bibfield  {title} {\bibinfo {title} {{A thermodynamic
  theory of “weak” ferromagnetism of antiferromagnetics }},\ }\href@noop {}
  {\bibfield  {journal} {\bibinfo  {journal}
  {\href{https://www.sciencedirect.com/science/article/pii/0022369758900763}{J.
  Phys. Chem. Solids.}}\ }\textbf {\bibinfo {volume} {4}} (\bibinfo {year}
  {1958})}\BibitemShut {NoStop}%
\bibitem [{\citenamefont {Moriya}(1960)}]{II:LI_DM_2nd}%
  \BibitemOpen
  \bibfield  {author} {\bibinfo {author} {\bibfnamefont {T.}~\bibnamefont
  {Moriya}},\ }\bibfield  {title} {\bibinfo {title} {{Anisotropic Superexchange
  Interaction and Weak Ferromagnetism}},\ }\href
  {https://doi.org/10.1103/PhysRev.120.91} {\bibfield  {journal} {\bibinfo
  {journal} {Phys. Rev.}\ }\textbf {\bibinfo {volume} {120}},\ \bibinfo {pages}
  {91} (\bibinfo {year} {1960})}\BibitemShut {NoStop}%
\bibitem [{\citenamefont {Fert}\ \emph {et~al.}(2017)\citenamefont {Fert},
  \citenamefont {Reyren},\ and\ \citenamefont {Cros}}]{MagneticSkyrmion}%
  \BibitemOpen
  \bibfield  {author} {\bibinfo {author} {\bibfnamefont {A.}~\bibnamefont
  {Fert}}, \bibinfo {author} {\bibfnamefont {N.}~\bibnamefont {Reyren}},\ and\
  \bibinfo {author} {\bibfnamefont {V.}~\bibnamefont {Cros}},\ }\bibfield
  {title} {\bibinfo {title} {{Magnetic skyrmions: advances in physics and
  potential applications}},\ }\bibfield  {journal} {\bibinfo  {journal} {Nat.
  Rev. Mat.}\ }\textbf {\bibinfo {volume} {2}},\ \href
  {https://doi.org/10.1038/natrevmats.2017.31} {10.1038/natrevmats.2017.31}
  (\bibinfo {year} {2017})\BibitemShut {NoStop}%
\bibitem [{\citenamefont {Stokes}\ \emph {et~al.}(1993)\citenamefont {Stokes},
  \citenamefont {Hatch},\ and\ \citenamefont
  {Nelson}}]{LI_Classification_solids}%
  \BibitemOpen
  \bibfield  {author} {\bibinfo {author} {\bibfnamefont {H.~T.}\ \bibnamefont
  {Stokes}}, \bibinfo {author} {\bibfnamefont {D.~M.}\ \bibnamefont {Hatch}},\
  and\ \bibinfo {author} {\bibfnamefont {H.~M.}\ \bibnamefont {Nelson}},\
  }\bibfield  {title} {\bibinfo {title} {{Landau, Lifshitz, and weak Lifshitz
  conditions in the Landau theory of phase transitions in solids}},\ }\href
  {https://doi.org/10.1103/PhysRevB.47.9080} {\bibfield  {journal} {\bibinfo
  {journal} {Phys. Rev. B}\ }\textbf {\bibinfo {volume} {47}},\ \bibinfo
  {pages} {9080} (\bibinfo {year} {1993})}\BibitemShut {NoStop}%
\bibitem [{\citenamefont {Balents}(2010)}]{Intro:Kagome_SpinLiquid1}%
  \BibitemOpen
  \bibfield  {author} {\bibinfo {author} {\bibfnamefont {L.}~\bibnamefont
  {Balents}},\ }\bibfield  {title} {\bibinfo {title} {{Spin liquids in
  frustrated magnets }},\ }\href@noop {} {\bibfield  {journal} {\bibinfo
  {journal} {\href{https://www.nature.com/articles/nature08917}{Nature}}\
  }\textbf {\bibinfo {volume} {464}} (\bibinfo {year} {2010})}\BibitemShut
  {NoStop}%
\bibitem [{\citenamefont {Yan}\ \emph {et~al.}(2011)\citenamefont {Yan},
  \citenamefont {Huse},\ and\ \citenamefont
  {White}}]{Intro:Kagome_SpinLiquid2}%
  \BibitemOpen
  \bibfield  {author} {\bibinfo {author} {\bibfnamefont {S.}~\bibnamefont
  {Yan}}, \bibinfo {author} {\bibfnamefont {D.~A.}\ \bibnamefont {Huse}},\ and\
  \bibinfo {author} {\bibfnamefont {S.~R.}\ \bibnamefont {White}},\ }\bibfield
  {title} {\bibinfo {title} {{Spin-Liquid Ground State of the $S$ = 1/2 Kagome
  Heisenberg Antiferromagnet }},\ }\href@noop {} {\bibfield  {journal}
  {\bibinfo  {journal}
  {\href{https://www.science.org/doi/10.1126/science.1201080}{Science}}\
  }\textbf {\bibinfo {volume} {332}} (\bibinfo {year} {2011})}\BibitemShut
  {NoStop}%
\bibitem [{\citenamefont {Han}\ \emph {et~al.}(2012)\citenamefont {Han},
  \citenamefont {Helton}, \citenamefont {Chu}, \citenamefont {Nocera},
  \citenamefont {Rodriguez-Rivera}, \citenamefont {Broholm},\ and\
  \citenamefont {Lee}}]{Intro:Kagome_SpinLiquid3}%
  \BibitemOpen
  \bibfield  {author} {\bibinfo {author} {\bibfnamefont {T.-H.}\ \bibnamefont
  {Han}}, \bibinfo {author} {\bibfnamefont {J.~S.}\ \bibnamefont {Helton}},
  \bibinfo {author} {\bibfnamefont {S.}~\bibnamefont {Chu}}, \bibinfo {author}
  {\bibfnamefont {D.~G.}\ \bibnamefont {Nocera}}, \bibinfo {author}
  {\bibfnamefont {J.~A.}\ \bibnamefont {Rodriguez-Rivera}}, \bibinfo {author}
  {\bibfnamefont {C.}~\bibnamefont {Broholm}},\ and\ \bibinfo {author}
  {\bibfnamefont {Y.~S.}\ \bibnamefont {Lee}},\ }\bibfield  {title} {\bibinfo
  {title} {{Fractionalized excitations in the spin-liquid state of a
  kagome-lattice antiferromagnet }},\ }\href@noop {} {\bibfield  {journal}
  {\bibinfo  {journal}
  {\href{https://www.nature.com/articles/nature11659}{Nature}}\ }\textbf
  {\bibinfo {volume} {492}} (\bibinfo {year} {2012})}\BibitemShut {NoStop}%
\bibitem [{\citenamefont {Sachdev}(1992)}]{Intro:Kagome_antiferro}%
  \BibitemOpen
  \bibfield  {author} {\bibinfo {author} {\bibfnamefont {S.}~\bibnamefont
  {Sachdev}},\ }\bibfield  {title} {\bibinfo {title} {{Kagom\'e- and
  triangular-lattice Heisenberg antiferromagnets: Ordering from quantum
  fluctuations and quantum-disordered ground states with unconfined bosonic
  spinons}},\ }\href {https://doi.org/10.1103/PhysRevB.45.12377} {\bibfield
  {journal} {\bibinfo  {journal} {Phys. Rev. B}\ }\textbf {\bibinfo {volume}
  {45}},\ \bibinfo {pages} {12377} (\bibinfo {year} {1992})}\BibitemShut
  {NoStop}%
\bibitem [{\citenamefont {Hirschberger}\ \emph {et~al.}(2019)\citenamefont
  {Hirschberger}, \citenamefont {Nakajima}, \citenamefont {Gao}, \citenamefont
  {Peng}, \citenamefont {Kikkawa}, \citenamefont {Kurumaji}, \citenamefont
  {Kriener}, \citenamefont {Yamasaki}, \citenamefont {Sagayama}, \citenamefont
  {Nakao}, \citenamefont {Ohishi}, \citenamefont {Kakurai}, \citenamefont
  {Taguchi}, \citenamefont {Yu}, \citenamefont {Arima},\ and\ \citenamefont
  {Tokura}}]{Intro:Kagome_Breathing}%
  \BibitemOpen
  \bibfield  {author} {\bibinfo {author} {\bibfnamefont {M.}~\bibnamefont
  {Hirschberger}}, \bibinfo {author} {\bibfnamefont {T.}~\bibnamefont
  {Nakajima}}, \bibinfo {author} {\bibfnamefont {S.}~\bibnamefont {Gao}},
  \bibinfo {author} {\bibfnamefont {L.}~\bibnamefont {Peng}}, \bibinfo {author}
  {\bibfnamefont {A.}~\bibnamefont {Kikkawa}}, \bibinfo {author} {\bibfnamefont
  {T.}~\bibnamefont {Kurumaji}}, \bibinfo {author} {\bibfnamefont
  {M.}~\bibnamefont {Kriener}}, \bibinfo {author} {\bibfnamefont
  {Y.}~\bibnamefont {Yamasaki}}, \bibinfo {author} {\bibfnamefont
  {H.}~\bibnamefont {Sagayama}}, \bibinfo {author} {\bibfnamefont
  {H.}~\bibnamefont {Nakao}}, \bibinfo {author} {\bibfnamefont
  {K.}~\bibnamefont {Ohishi}}, \bibinfo {author} {\bibfnamefont
  {K.}~\bibnamefont {Kakurai}}, \bibinfo {author} {\bibfnamefont
  {Y.}~\bibnamefont {Taguchi}}, \bibinfo {author} {\bibfnamefont
  {X.}~\bibnamefont {Yu}}, \bibinfo {author} {\bibfnamefont {T.-h.}\
  \bibnamefont {Arima}},\ and\ \bibinfo {author} {\bibfnamefont
  {Y.}~\bibnamefont {Tokura}},\ }\bibfield  {title} {\bibinfo {title}
  {{Skyrmion phase and competing magnetic orders on a breathing kagomé lattice
  }},\ }\href@noop {} {\bibfield  {journal} {\bibinfo  {journal}
  {\href{https://www.nature.com/articles/s41467-019-13675-4}{Nat. Commun.}}\
  }\textbf {\bibinfo {volume} {10}} (\bibinfo {year} {2019})}\BibitemShut
  {NoStop}%
\bibitem [{\citenamefont {Sun}\ \emph {et~al.}(2011)\citenamefont {Sun},
  \citenamefont {Gu}, \citenamefont {Katsura},\ and\ \citenamefont
  {Das~Sarma}}]{Intro:Kagome_FlatBand1}%
  \BibitemOpen
  \bibfield  {author} {\bibinfo {author} {\bibfnamefont {K.}~\bibnamefont
  {Sun}}, \bibinfo {author} {\bibfnamefont {Z.}~\bibnamefont {Gu}}, \bibinfo
  {author} {\bibfnamefont {H.}~\bibnamefont {Katsura}},\ and\ \bibinfo {author}
  {\bibfnamefont {S.}~\bibnamefont {Das~Sarma}},\ }\bibfield  {title} {\bibinfo
  {title} {{Nearly Flatbands with Nontrivial Topology}},\ }\href
  {https://doi.org/10.1103/PhysRevLett.106.236803} {\bibfield  {journal}
  {\bibinfo  {journal} {Phys. Rev. Lett.}\ }\textbf {\bibinfo {volume} {106}},\
  \bibinfo {pages} {236803} (\bibinfo {year} {2011})}\BibitemShut {NoStop}%
\bibitem [{\citenamefont {Li}\ \emph {et~al.}(2018)\citenamefont {Li},
  \citenamefont {Zhuang}, \citenamefont {Wang}, \citenamefont {Feng},
  \citenamefont {Gao}, \citenamefont {Xu}, \citenamefont {Hao}, \citenamefont
  {Wang}, \citenamefont {Zhang}, \citenamefont {Wu}, \citenamefont {Dou},
  \citenamefont {Chen}, \citenamefont {Hu},\ and\ \citenamefont
  {Du}}]{Intro:Kagome_FlatBand2}%
  \BibitemOpen
  \bibfield  {author} {\bibinfo {author} {\bibfnamefont {Z.}~\bibnamefont
  {Li}}, \bibinfo {author} {\bibfnamefont {J.}~\bibnamefont {Zhuang}}, \bibinfo
  {author} {\bibfnamefont {L.}~\bibnamefont {Wang}}, \bibinfo {author}
  {\bibfnamefont {H.}~\bibnamefont {Feng}}, \bibinfo {author} {\bibfnamefont
  {Q.}~\bibnamefont {Gao}}, \bibinfo {author} {\bibfnamefont {X.}~\bibnamefont
  {Xu}}, \bibinfo {author} {\bibfnamefont {W.}~\bibnamefont {Hao}}, \bibinfo
  {author} {\bibfnamefont {X.}~\bibnamefont {Wang}}, \bibinfo {author}
  {\bibfnamefont {C.}~\bibnamefont {Zhang}}, \bibinfo {author} {\bibfnamefont
  {K.}~\bibnamefont {Wu}}, \bibinfo {author} {\bibfnamefont {S.~X.}\
  \bibnamefont {Dou}}, \bibinfo {author} {\bibfnamefont {L.}~\bibnamefont
  {Chen}}, \bibinfo {author} {\bibfnamefont {Z.}~\bibnamefont {Hu}},\ and\
  \bibinfo {author} {\bibfnamefont {Y.}~\bibnamefont {Du}},\ }\bibfield
  {title} {\bibinfo {title} {{Realization of flat band with possible nontrivial
  topology in electronic Kagome lattice }},\ }\href@noop {} {\bibfield
  {journal} {\bibinfo  {journal}
  {\href{https://www.science.org/doi/10.1126/sciadv.aau4511}{Sci. Adv.}}\
  }\textbf {\bibinfo {volume} {4}} (\bibinfo {year} {2018})}\BibitemShut
  {NoStop}%
\bibitem [{\citenamefont {Kang}\ \emph
  {et~al.}(2020{\natexlab{a}})\citenamefont {Kang}, \citenamefont {Fang},
  \citenamefont {Ye}, \citenamefont {Po}, \citenamefont {Denlinger},
  \citenamefont {Jozwiak}, \citenamefont {Bostwick}, \citenamefont {Rotenberg},
  \citenamefont {Kaxiras}, \citenamefont {Checkelsky},\ and\ \citenamefont
  {Comin}}]{Intro:Kagome_FlatBand3}%
  \BibitemOpen
  \bibfield  {author} {\bibinfo {author} {\bibfnamefont {M.}~\bibnamefont
  {Kang}}, \bibinfo {author} {\bibfnamefont {S.}~\bibnamefont {Fang}}, \bibinfo
  {author} {\bibfnamefont {L.}~\bibnamefont {Ye}}, \bibinfo {author}
  {\bibfnamefont {H.~C.}\ \bibnamefont {Po}}, \bibinfo {author} {\bibfnamefont
  {J.}~\bibnamefont {Denlinger}}, \bibinfo {author} {\bibfnamefont
  {C.}~\bibnamefont {Jozwiak}}, \bibinfo {author} {\bibfnamefont
  {A.}~\bibnamefont {Bostwick}}, \bibinfo {author} {\bibfnamefont
  {E.}~\bibnamefont {Rotenberg}}, \bibinfo {author} {\bibfnamefont
  {E.}~\bibnamefont {Kaxiras}}, \bibinfo {author} {\bibfnamefont {J.~G.}\
  \bibnamefont {Checkelsky}},\ and\ \bibinfo {author} {\bibfnamefont
  {R.}~\bibnamefont {Comin}},\ }\bibfield  {title} {\bibinfo {title}
  {{Topological flat bands in frustrated kagome lattice CoSn }},\ }\href@noop
  {} {\bibfield  {journal} {\bibinfo  {journal}
  {\href{https://www.nature.com/articles/s41467-020-17465-1}{Nat. Commun.}}\
  }\textbf {\bibinfo {volume} {11}} (\bibinfo {year}
  {2020}{\natexlab{a}})}\BibitemShut {NoStop}%
\bibitem [{\citenamefont {Sun}\ \emph {et~al.}(2022)\citenamefont {Sun},
  \citenamefont {Zhou}, \citenamefont {Wang}, \citenamefont {Kumar},
  \citenamefont {Geng}, \citenamefont {Yue}, \citenamefont {Han}, \citenamefont
  {Haraguchi}, \citenamefont {Shimada}, \citenamefont {Cheng}, \citenamefont
  {Chen}, \citenamefont {Shi}, \citenamefont {Wu}, \citenamefont {Meng},\ and\
  \citenamefont {Feng}}]{Intro:Kagome_FlatBand4}%
  \BibitemOpen
  \bibfield  {author} {\bibinfo {author} {\bibfnamefont {Z.}~\bibnamefont
  {Sun}}, \bibinfo {author} {\bibfnamefont {H.}~\bibnamefont {Zhou}}, \bibinfo
  {author} {\bibfnamefont {C.}~\bibnamefont {Wang}}, \bibinfo {author}
  {\bibfnamefont {S.}~\bibnamefont {Kumar}}, \bibinfo {author} {\bibfnamefont
  {D.}~\bibnamefont {Geng}}, \bibinfo {author} {\bibfnamefont {S.}~\bibnamefont
  {Yue}}, \bibinfo {author} {\bibfnamefont {X.}~\bibnamefont {Han}}, \bibinfo
  {author} {\bibfnamefont {Y.}~\bibnamefont {Haraguchi}}, \bibinfo {author}
  {\bibfnamefont {K.}~\bibnamefont {Shimada}}, \bibinfo {author} {\bibfnamefont
  {P.}~\bibnamefont {Cheng}}, \bibinfo {author} {\bibfnamefont
  {L.}~\bibnamefont {Chen}}, \bibinfo {author} {\bibfnamefont {Y.}~\bibnamefont
  {Shi}}, \bibinfo {author} {\bibfnamefont {K.}~\bibnamefont {Wu}}, \bibinfo
  {author} {\bibfnamefont {S.}~\bibnamefont {Meng}},\ and\ \bibinfo {author}
  {\bibfnamefont {B.}~\bibnamefont {Feng}},\ }\bibfield  {title} {\bibinfo
  {title} {{Observation of Topological Flat Bands in the Kagome Semiconductor
  $\mathrm{Nb}_{3}\mathrm{Cl}_{8}$ }},\ }\href@noop {} {\bibfield  {journal}
  {\bibinfo  {journal}
  {\href{https://pubs.acs.org/doi/10.1021/acs.nanolett.2c00778}{Nano Lett.}}\
  }\textbf {\bibinfo {volume} {22}} (\bibinfo {year} {2022})}\BibitemShut
  {NoStop}%
\bibitem [{\citenamefont {Guo}\ and\ \citenamefont
  {Franz}(2009)}]{Intro:Kagome_Topology1}%
  \BibitemOpen
  \bibfield  {author} {\bibinfo {author} {\bibfnamefont {H.-M.}\ \bibnamefont
  {Guo}}\ and\ \bibinfo {author} {\bibfnamefont {M.}~\bibnamefont {Franz}},\
  }\bibfield  {title} {\bibinfo {title} {{Topological insulator on the kagome
  lattice}},\ }\href {https://doi.org/10.1103/PhysRevB.80.113102} {\bibfield
  {journal} {\bibinfo  {journal} {Phys. Rev. B}\ }\textbf {\bibinfo {volume}
  {80}},\ \bibinfo {pages} {113102} (\bibinfo {year} {2009})}\BibitemShut
  {NoStop}%
\bibitem [{\citenamefont {Tang}\ \emph {et~al.}(2011)\citenamefont {Tang},
  \citenamefont {Mei},\ and\ \citenamefont {Wen}}]{Intro:Kagome_Topology2}%
  \BibitemOpen
  \bibfield  {author} {\bibinfo {author} {\bibfnamefont {E.}~\bibnamefont
  {Tang}}, \bibinfo {author} {\bibfnamefont {J.-W.}\ \bibnamefont {Mei}},\ and\
  \bibinfo {author} {\bibfnamefont {X.-G.}\ \bibnamefont {Wen}},\ }\bibfield
  {title} {\bibinfo {title} {{High-Temperature Fractional Quantum Hall
  States}},\ }\href {https://doi.org/10.1103/PhysRevLett.106.236802} {\bibfield
   {journal} {\bibinfo  {journal} {Phys. Rev. Lett.}\ }\textbf {\bibinfo
  {volume} {106}},\ \bibinfo {pages} {236802} (\bibinfo {year}
  {2011})}\BibitemShut {NoStop}%
\bibitem [{\citenamefont {He}\ \emph {et~al.}(2017)\citenamefont {He},
  \citenamefont {Zaletel}, \citenamefont {Oshikawa},\ and\ \citenamefont
  {Pollmann}}]{Intro:Kagome_Topology3}%
  \BibitemOpen
  \bibfield  {author} {\bibinfo {author} {\bibfnamefont {Y.-C.}\ \bibnamefont
  {He}}, \bibinfo {author} {\bibfnamefont {M.~P.}\ \bibnamefont {Zaletel}},
  \bibinfo {author} {\bibfnamefont {M.}~\bibnamefont {Oshikawa}},\ and\
  \bibinfo {author} {\bibfnamefont {F.}~\bibnamefont {Pollmann}},\ }\bibfield
  {title} {\bibinfo {title} {{Signatures of Dirac Cones in a DMRG Study of the
  Kagome Heisenberg Model}},\ }\href
  {https://doi.org/10.1103/PhysRevX.7.031020} {\bibfield  {journal} {\bibinfo
  {journal} {Phys. Rev. X}\ }\textbf {\bibinfo {volume} {7}},\ \bibinfo {pages}
  {031020} (\bibinfo {year} {2017})}\BibitemShut {NoStop}%
\bibitem [{\citenamefont {Ye}\ \emph {et~al.}(2018)\citenamefont {Ye},
  \citenamefont {Kang}, \citenamefont {Liu}, \citenamefont {von Cube},
  \citenamefont {Wicker}, \citenamefont {Suzuki}, \citenamefont {Jozwiak},
  \citenamefont {Bostwick}, \citenamefont {Rotenberg}, \citenamefont {Bell},
  \citenamefont {Fu}, \citenamefont {Comin},\ and\ \citenamefont
  {Checkelsky}}]{Intro:Kagome_Topology4}%
  \BibitemOpen
  \bibfield  {author} {\bibinfo {author} {\bibfnamefont {L.}~\bibnamefont
  {Ye}}, \bibinfo {author} {\bibfnamefont {M.}~\bibnamefont {Kang}}, \bibinfo
  {author} {\bibfnamefont {J.}~\bibnamefont {Liu}}, \bibinfo {author}
  {\bibfnamefont {F.}~\bibnamefont {von Cube}}, \bibinfo {author}
  {\bibfnamefont {C.~R.}\ \bibnamefont {Wicker}}, \bibinfo {author}
  {\bibfnamefont {T.}~\bibnamefont {Suzuki}}, \bibinfo {author} {\bibfnamefont
  {C.}~\bibnamefont {Jozwiak}}, \bibinfo {author} {\bibfnamefont
  {A.}~\bibnamefont {Bostwick}}, \bibinfo {author} {\bibfnamefont
  {E.}~\bibnamefont {Rotenberg}}, \bibinfo {author} {\bibfnamefont {D.~C.}\
  \bibnamefont {Bell}}, \bibinfo {author} {\bibfnamefont {L.}~\bibnamefont
  {Fu}}, \bibinfo {author} {\bibfnamefont {R.}~\bibnamefont {Comin}},\ and\
  \bibinfo {author} {\bibfnamefont {J.~G.}\ \bibnamefont {Checkelsky}},\
  }\bibfield  {title} {\bibinfo {title} {{Massive Dirac fermions in a
  ferromagnetic kagome metal }},\ }\href@noop {} {\bibfield  {journal}
  {\bibinfo  {journal}
  {\href{https://www.nature.com/articles/nature25987}{Nature}}\ }\textbf
  {\bibinfo {volume} {555}} (\bibinfo {year} {2018})}\BibitemShut {NoStop}%
\bibitem [{\citenamefont {Kang}\ \emph
  {et~al.}(2020{\natexlab{b}})\citenamefont {Kang}, \citenamefont {Ye},
  \citenamefont {Fang}, \citenamefont {You}, \citenamefont {Levitan},
  \citenamefont {Han}, \citenamefont {Facio}, \citenamefont {Jozwiak},
  \citenamefont {Bostwick}, \citenamefont {Rotenberg}, \citenamefont {Chan},
  \citenamefont {McDonald}, \citenamefont {Graf}, \citenamefont {Kaznatcheev},
  \citenamefont {Vescovo}, \citenamefont {Bell}, \citenamefont {Kaxiras},
  \citenamefont {van~den Brink}, \citenamefont {Richter}, \citenamefont
  {Prasad~Ghimire}, \citenamefont {Checkelsky},\ and\ \citenamefont
  {Comin}}]{Intro:Kagome_Topology5}%
  \BibitemOpen
  \bibfield  {author} {\bibinfo {author} {\bibfnamefont {M.}~\bibnamefont
  {Kang}}, \bibinfo {author} {\bibfnamefont {L.}~\bibnamefont {Ye}}, \bibinfo
  {author} {\bibfnamefont {S.}~\bibnamefont {Fang}}, \bibinfo {author}
  {\bibfnamefont {J.-S.}\ \bibnamefont {You}}, \bibinfo {author} {\bibfnamefont
  {A.}~\bibnamefont {Levitan}}, \bibinfo {author} {\bibfnamefont
  {M.}~\bibnamefont {Han}}, \bibinfo {author} {\bibfnamefont {J.~I.}\
  \bibnamefont {Facio}}, \bibinfo {author} {\bibfnamefont {C.}~\bibnamefont
  {Jozwiak}}, \bibinfo {author} {\bibfnamefont {A.}~\bibnamefont {Bostwick}},
  \bibinfo {author} {\bibfnamefont {E.}~\bibnamefont {Rotenberg}}, \bibinfo
  {author} {\bibfnamefont {M.~K.}\ \bibnamefont {Chan}}, \bibinfo {author}
  {\bibfnamefont {R.~D.}\ \bibnamefont {McDonald}}, \bibinfo {author}
  {\bibfnamefont {D.}~\bibnamefont {Graf}}, \bibinfo {author} {\bibfnamefont
  {K.}~\bibnamefont {Kaznatcheev}}, \bibinfo {author} {\bibfnamefont
  {E.}~\bibnamefont {Vescovo}}, \bibinfo {author} {\bibfnamefont {D.~C.}\
  \bibnamefont {Bell}}, \bibinfo {author} {\bibfnamefont {E.}~\bibnamefont
  {Kaxiras}}, \bibinfo {author} {\bibfnamefont {J.}~\bibnamefont {van~den
  Brink}}, \bibinfo {author} {\bibfnamefont {M.}~\bibnamefont {Richter}},
  \bibinfo {author} {\bibfnamefont {M.}~\bibnamefont {Prasad~Ghimire}},
  \bibinfo {author} {\bibfnamefont {J.~G.}\ \bibnamefont {Checkelsky}},\ and\
  \bibinfo {author} {\bibfnamefont {R.}~\bibnamefont {Comin}},\ }\bibfield
  {title} {\bibinfo {title} {{Dirac fermions and flat bands in the ideal kagome
  metal FeSn }},\ }\href@noop {} {\bibfield  {journal} {\bibinfo  {journal}
  {\href{https://www.nature.com/articles/s41563-019-0531-0}{Nat. Mater.}}\
  }\textbf {\bibinfo {volume} {19}} (\bibinfo {year}
  {2020}{\natexlab{b}})}\BibitemShut {NoStop}%
\bibitem [{\citenamefont {Grigorishin}(2016)}]{II:FreeEnergy}%
  \BibitemOpen
  \bibfield  {author} {\bibinfo {author} {\bibfnamefont {K.~V.}\ \bibnamefont
  {Grigorishin}},\ }\bibfield  {title} {\bibinfo {title} {{Effective
  Ginzburg–Landau free energy functional for multi-band isotropic
  superconductors }},\ }\href@noop {} {\bibfield  {journal} {\bibinfo
  {journal}
  {\href{https://www.sciencedirect.com/science/article/abs/pii/S0375960116300020}{Phys.
  Lett. A}}\ }\textbf {\bibinfo {volume} {380}} (\bibinfo {year}
  {2016})}\BibitemShut {NoStop}%
\bibitem [{\citenamefont {Doh}\ \emph {et~al.}(1999)\citenamefont {Doh},
  \citenamefont {Sigrist}, \citenamefont {Cho},\ and\ \citenamefont
  {Lee}}]{DragEffect_Doh}%
  \BibitemOpen
  \bibfield  {author} {\bibinfo {author} {\bibfnamefont {H.}~\bibnamefont
  {Doh}}, \bibinfo {author} {\bibfnamefont {M.}~\bibnamefont {Sigrist}},
  \bibinfo {author} {\bibfnamefont {B.~K.}\ \bibnamefont {Cho}},\ and\ \bibinfo
  {author} {\bibfnamefont {S.-I.}\ \bibnamefont {Lee}},\ }\bibfield  {title}
  {\bibinfo {title} {{Phenomenological Theory of Superconductivity and
  Magnetism in
  ${\mathrm{Ho}}_{1-x}$${\mathrm{Dy}}_{\mathit{x}}{\mathrm{Ni}}_{2}{\mathrm{B}}_{2}\mathrm{C}$}},\
  }\href {https://doi.org/10.1103/PhysRevLett.83.5350} {\bibfield  {journal}
  {\bibinfo  {journal} {Phys. Rev. Lett.}\ }\textbf {\bibinfo {volume} {83}},\
  \bibinfo {pages} {5350} (\bibinfo {year} {1999})}\BibitemShut {NoStop}%
\bibitem [{\citenamefont {Yerin}\ \emph {et~al.}(2022)\citenamefont {Yerin},
  \citenamefont {Drechsler}, \citenamefont {Cuoco},\ and\ \citenamefont
  {Petrillo}}]{DragEffect_Yerin}%
  \BibitemOpen
  \bibfield  {author} {\bibinfo {author} {\bibfnamefont {Y.}~\bibnamefont
  {Yerin}}, \bibinfo {author} {\bibfnamefont {S.-L.}\ \bibnamefont
  {Drechsler}}, \bibinfo {author} {\bibfnamefont {M.}~\bibnamefont {Cuoco}},\
  and\ \bibinfo {author} {\bibfnamefont {C.}~\bibnamefont {Petrillo}},\
  }\bibfield  {title} {\bibinfo {title} {Magneto-topological transitions in
  multicomponent superconductors},\ }\href
  {https://doi.org/10.1103/PhysRevB.106.054517} {\bibfield  {journal} {\bibinfo
   {journal} {Phys. Rev. B}\ }\textbf {\bibinfo {volume} {106}},\ \bibinfo
  {pages} {054517} (\bibinfo {year} {2022})}\BibitemShut {NoStop}%
\bibitem [{\citenamefont {Sigrist}\ and\ \citenamefont
  {Ueda}(1991)}]{Sigrist_Ueda}%
  \BibitemOpen
  \bibfield  {author} {\bibinfo {author} {\bibfnamefont {M.}~\bibnamefont
  {Sigrist}}\ and\ \bibinfo {author} {\bibfnamefont {K.}~\bibnamefont {Ueda}},\
  }\bibfield  {title} {\bibinfo {title} {{Phenomenological theory of
  unconventional superconductivity}},\ }\href
  {https://doi.org/10.1103/RevModPhys.63.239} {\bibfield  {journal} {\bibinfo
  {journal} {Rev. Mod. Phys.}\ }\textbf {\bibinfo {volume} {63}},\ \bibinfo
  {pages} {239} (\bibinfo {year} {1991})}\BibitemShut {NoStop}%
\bibitem [{\citenamefont {Serre}(1977)}]{Serre}%
  \BibitemOpen
  \bibfield  {author} {\bibinfo {author} {\bibfnamefont {J.~P.}\ \bibnamefont
  {Serre}},\ }\href@noop {} {\emph {\bibinfo {title} {{Linear Representations
  of Finite Groups }}}}\ (\bibinfo  {publisher} {Springer},\ \bibinfo {address}
  {New York},\ \bibinfo {year} {1977})\BibitemShut {NoStop}%
\bibitem [{\citenamefont {{P. W. Atkins and M. S. Child and C. S. G.
  Phillips}}(1970)}]{TableGroupTheory}%
  \BibitemOpen
  \bibfield  {author} {\bibinfo {author} {\bibnamefont {{P. W. Atkins and M. S.
  Child and C. S. G. Phillips}}},\ }\href@noop {} {\emph {\bibinfo {title}
  {{Tables for group theory }}}}\ (\bibinfo  {publisher} {Oxford University
  Press},\ \bibinfo {address} {Oxford},\ \bibinfo {year} {1970})\BibitemShut
  {NoStop}%
\bibitem [{\citenamefont {Rice}\ and\ \citenamefont {Mele}(1982)}]{Rice-Mele}%
  \BibitemOpen
  \bibfield  {author} {\bibinfo {author} {\bibfnamefont {M.~J.}\ \bibnamefont
  {Rice}}\ and\ \bibinfo {author} {\bibfnamefont {E.~J.}\ \bibnamefont
  {Mele}},\ }\bibfield  {title} {\bibinfo {title} {{Elementary Excitations of a
  Linearly Conjugated Diatomic Polymer}},\ }\href
  {https://doi.org/10.1103/PhysRevLett.49.1455} {\bibfield  {journal} {\bibinfo
   {journal} {Phys. Rev. Lett.}\ }\textbf {\bibinfo {volume} {49}},\ \bibinfo
  {pages} {1455} (\bibinfo {year} {1982})}\BibitemShut {NoStop}%
\bibitem [{\citenamefont {Su}\ \emph {et~al.}(1980)\citenamefont {Su},
  \citenamefont {Schrieffer},\ and\ \citenamefont {Heeger}}]{SSH_model}%
  \BibitemOpen
  \bibfield  {author} {\bibinfo {author} {\bibfnamefont {W.~P.}\ \bibnamefont
  {Su}}, \bibinfo {author} {\bibfnamefont {J.~R.}\ \bibnamefont {Schrieffer}},\
  and\ \bibinfo {author} {\bibfnamefont {A.~J.}\ \bibnamefont {Heeger}},\
  }\bibfield  {title} {\bibinfo {title} {{Soliton excitations in
  polyacetylene}},\ }\href {https://doi.org/10.1103/PhysRevB.22.2099}
  {\bibfield  {journal} {\bibinfo  {journal} {Phys. Rev. B}\ }\textbf {\bibinfo
  {volume} {22}},\ \bibinfo {pages} {2099} (\bibinfo {year}
  {1980})}\BibitemShut {NoStop}%
\bibitem [{\citenamefont {van Otterlo}\ \emph {et~al.}(1999)\citenamefont {van
  Otterlo}, \citenamefont {Golubev}, \citenamefont {Zaikin},\ and\
  \citenamefont {Blatter}}]{IV:EffectiveAction_1st}%
  \BibitemOpen
  \bibfield  {author} {\bibinfo {author} {\bibfnamefont {A.}~\bibnamefont {van
  Otterlo}}, \bibinfo {author} {\bibfnamefont {D.~S.}\ \bibnamefont {Golubev}},
  \bibinfo {author} {\bibfnamefont {A.~D.}\ \bibnamefont {Zaikin}},\ and\
  \bibinfo {author} {\bibfnamefont {G.}~\bibnamefont {Blatter}},\ }\bibfield
  {title} {\bibinfo {title} {{Dynamics and effective actions of BCS
  superconductors }},\ }\href@noop {} {\bibfield  {journal} {\bibinfo
  {journal} {\href{https://doi.org/10.1007/s100510050836}{Eur. Phys. J. B.}}\
  }\textbf {\bibinfo {volume} {10}} (\bibinfo {year} {1999})}\BibitemShut
  {NoStop}%
\bibitem [{\citenamefont {Sharapov}\ \emph {et~al.}(2002)\citenamefont
  {Sharapov}, \citenamefont {Gusynin},\ and\ \citenamefont
  {Beck}}]{IV:EffectiveAction_2nd}%
  \BibitemOpen
  \bibfield  {author} {\bibinfo {author} {\bibfnamefont {S.~G.}\ \bibnamefont
  {Sharapov}}, \bibinfo {author} {\bibfnamefont {V.~P.}\ \bibnamefont
  {Gusynin}},\ and\ \bibinfo {author} {\bibfnamefont {H.}~\bibnamefont
  {Beck}},\ }\bibfield  {title} {\bibinfo {title} {{Effective action approach
  to the Leggett's mode in two-band superconductors }},\ }\href@noop {}
  {\bibfield  {journal} {\bibinfo  {journal}
  {\href{https://doi.org/10.1140/epjb/e2002-00356-9}{Eur. Phys. J. B.}}\
  }\textbf {\bibinfo {volume} {30}} (\bibinfo {year} {2002})}\BibitemShut
  {NoStop}%
\bibitem [{\citenamefont {Benfatto}\ \emph {et~al.}(2004)\citenamefont
  {Benfatto}, \citenamefont {Toschi},\ and\ \citenamefont
  {Caprara}}]{IV:EffectiveAction_3rd}%
  \BibitemOpen
  \bibfield  {author} {\bibinfo {author} {\bibfnamefont {L.}~\bibnamefont
  {Benfatto}}, \bibinfo {author} {\bibfnamefont {A.}~\bibnamefont {Toschi}},\
  and\ \bibinfo {author} {\bibfnamefont {S.}~\bibnamefont {Caprara}},\
  }\bibfield  {title} {\bibinfo {title} {{Low-energy phase-only action in a
  superconductor: A comparison with the $\mathrm{XY}$ model}},\ }\href
  {https://doi.org/10.1103/PhysRevB.69.184510} {\bibfield  {journal} {\bibinfo
  {journal} {Phys. Rev. B}\ }\textbf {\bibinfo {volume} {69}},\ \bibinfo
  {pages} {184510} (\bibinfo {year} {2004})}\BibitemShut {NoStop}%
\bibitem [{\citenamefont {Cea}\ \emph {et~al.}(2018)\citenamefont {Cea},
  \citenamefont {Barone}, \citenamefont {Castellani},\ and\ \citenamefont
  {Benfatto}}]{IV:EffectiveAction_4th}%
  \BibitemOpen
  \bibfield  {author} {\bibinfo {author} {\bibfnamefont {T.}~\bibnamefont
  {Cea}}, \bibinfo {author} {\bibfnamefont {P.}~\bibnamefont {Barone}},
  \bibinfo {author} {\bibfnamefont {C.}~\bibnamefont {Castellani}},\ and\
  \bibinfo {author} {\bibfnamefont {L.}~\bibnamefont {Benfatto}},\ }\bibfield
  {title} {\bibinfo {title} {{Polarization dependence of the third-harmonic
  generation in multiband superconductors}},\ }\href
  {https://doi.org/10.1103/PhysRevB.97.094516} {\bibfield  {journal} {\bibinfo
  {journal} {Phys. Rev. B}\ }\textbf {\bibinfo {volume} {97}},\ \bibinfo
  {pages} {094516} (\bibinfo {year} {2018})}\BibitemShut {NoStop}%
\end{thebibliography}%

\end{document}